\documentclass[11pt]{article}
\usepackage{afterpage}
\usepackage{pdflscape}
\usepackage{bm}
\usepackage{booktabs}
\usepackage{caption}
\usepackage{amsfonts,amssymb}
\usepackage{mathrsfs}
\usepackage[utf8]{inputenc}
\usepackage[no-natbib-sort]{jheppub}
\usepackage{mathtools}
\usepackage{hyperref}
\usepackage{tensor}

\usepackage[nameinlink,capitalize]{cleveref}

\DeclarePairedDelimiter\floor{\lfloor}{\rfloor}

\crefname{figure}{Figure}{Figures}
\crefformat{subequation}{#2(#1)#3}
\crefrangeformat{subequation}{#3(#1)#4 to #5(#2)#6}

\makeatletter{}
\g@addto@macro\bfseries{\boldmath}
\renewcommand{\@email}[1]{\texttt{#1}}
\def\NAT@sort{\z@}
\makeatother

\let\originalleft\left
\let\originalright\right
\renewcommand*{\left}{\mathopen{}\mathclose\bgroup\originalleft}
\renewcommand*{\right}{\aftergroup\egroup\originalright}

    \DeclareMathOperator{\SU}{SU}
    \DeclareMathOperator{\U}{U}

    \newcommand*{\asu}{\ensuremath\mathfrak{su}}
    \newcommand*{\au}{\ensuremath\mathfrak{u}}

    \newcommand*{\cN}{\mathcal{N}}
    \newcommand*{\cO}{\mathcal{O}}

    \newcommand*{\F}{\mathbb{F}}
    \newcommand*{\bP}{\mathbb{P}}
    \newcommand*{\Z}{\mathbb{Z}}
    
    \newcommand*{\C}{\mathbb{C}}

    \newcommand*{\SM}{\ensuremath(\SU(3) \times \SU(2) \times \U(1)) / \Z_6}

    \DeclareMathOperator{\sign}{sign}

    \newcommand*{\cZ}{\mathcal{Z}}
    \newcommand*{\sL}{\mathscr{L}}
    \newcommand*{\sO}{\mathscr{O}}

    \newcommand*{\chIndex}[1]{\ensuremath\chi_{#1}}
    \newcommand*{\httv}{\ensuremath H_{2,2}^{\text{vert}}}

    \newcommand*{\mr}{M_{\text{red}}}
    \newcommand*{\lv}{{\Lambda}_{\rm vert}}
    \newcommand*{\blv}{\bar{\Lambda}_{\rm vert}}
    \newcommand*{\hfz}{H^4 (X_5,\Z)}

\newcommand{\remove}[1]{}
\title{
Chiral spectrum of the universal
tuned $\SM$ 4D F-theory model\\}
\author[\dag]{Patrick Jefferson,}
\author[\dag]{Washington Taylor,}
\author[\ddag]{and Andrew P. Turner}

\affiliation[\dag]{Center for Theoretical Physics \\
      Department of Physics \\
      Massachusetts Institute of Technology \\
      77 Massachusetts Avenue \\
      Cambridge, MA 02139, USA}

\affiliation[\ddag]{Department of Physics and Astronomy \\
      University of Pennsylvania \\
      Philadelphia, PA 19104, USA}

\emailAdd{\tt{pjeffers} \rm{at} \tt{mit.edu}}
\emailAdd{\tt{wati} \rm{at} \tt{mit.edu}}
\emailAdd{\tt{turnerap} \rm{at} \tt{sas.upenn.edu}}

\preprint{MIT-CTP-5421}

\abstract{

We use the recently developed methods of
\href{https://arxiv.org/abs/2108.07810}{\texttt{2108.07810}} to
analyze vertical flux backgrounds and associated chiral matter spectra
in the 4D universal $\SM$ model introduced in
\href{https://arxiv.org/abs/1912.10991}{\texttt{1912.10991}}, which is
believed to describe the most %general
generic family of F-theory vacua with
tuned $\SM$ gauge symmetry. Our analysis focuses on a resolution of a
particular presentation of the $\SM$ model in which the elliptic fiber
is realized as a cubic in $\bP^2$ fibered over an arbitrary
smooth threefold base. We show that vertical fluxes can produce
nonzero multiplicities for all chiral matter families that satisfy 4D
anomaly cancellation, which include as a special case the chiral
matter families of the Minimal Supersymmetric Standard Model.
}

\begin{document}

\maketitle

\flushbottom

%%%%%%%%%%%%%%%%%%%%%%%%%%%%%%%%%%%%%%%%%%%%%%%%%%%%%%%%%%%%%%%%%%%%%%%%%%%%%%
%%%%%%%%%%%%%%%%%%%%%%%%%%%%%%%%%%%%%%%%%%%%%%%%%%%%%%%%%%%%%%%%%%%%%%%%%%%%%%
%%%%%%%%%%%%%%%%%%%%%%%%%%%%%%%%%%%%%%%%%%%%%%%%%%%%%%%%%%%%%%%%%%%%%%%%%%%%%%
\section{Introduction}
\label{sec:intro}

%It is well established that string theory contains an enormous number
%of vacua whose kinematics are consistent with low-energy
%experimental observations.\patrick{Modified this sentence a bit.} One set of such features, which are distinct
%characteristics of the Standard Model and have been realized in
%various types of string constructions (see \cite{Cvetic:2022fnv}
%for
%a review of recent developments),
%are the presence of a gauge group with Lie algebra $\sm$
%and the existence of chiral
%fermions transforming in representations of the gauge symmetry group
%$(\SU(3) \times \SU(2) \times \U(1) )/ \Gamma$, where $\Gamma \in \{
%1, \Z_2, \Z_3, \Z_6\}$ is a possibly trivial
%discrete subgroup of the center of $\SU(3) \times \SU(2) \times
%\U(1)$. This paper focuses on a generic class of models with
%these features; in particular we analyze the chiral matter spectrum
%and aspects of the Yukawa interactions of the universal tuned F-theory model
%with gauge group $\SM$ introduced in \cite{Raghuram:2019efb}.

One of the hallmarks of the Standard Model is the presence of chiral fermions transforming in representations of the gauge symmetry group
$(\SU(3) \times \SU(2) \times \U(1) )/ \Gamma$, where $\Gamma \in \{
1, \Z_2, \Z_3, \Z_6\}$ is a possibly trivial
discrete subgroup of the center of $\SU(3) \times \SU(2) \times
\U(1)$. Chiral fermions charged under the Standard Model gauge group have been realized in numerous string theory constructions---see \cite{Cvetic:2022fnv}
for a review of recent developments. The purpose of this paper is to analyze the chiral matter spectrum
and aspects of the Yukawa interactions of the universal tuned F-theory model
with gauge group $\SM$ introduced in \cite{Raghuram:2019efb}.
%which is believed to describe the most generic class of 4D F-theory vacua with tuned Standard Model gauge group (we clarify the precise definition of ``tuned'' gauge symmetry below).\patrick{Rewrote this first paragraph.}

There are obvious motivations for finding specific, dynamically
stable solutions of string theory that reproduce all observed aspects
of the Standard Model.  However, rather than aiming for a complete string-theoretic construction of the Standard Model, we instead take a top-down
 approach and focus
on sets of solutions in the space of string vacua that  exhibit
the gross features of the Standard Model, specifically the gauge group and
allowed representations in the chiral matter spectrum.
 One reason for this approach is that taking a bird's
eye view can give us a comprehensive picture of the set of candidate
string vacua, which may help to address broad questions about how the
fundamental interactions of the Standard Model can be reproduced by
a self-consistent theory of quantum gravity (many examples of which we
expect to come from string theory).
Some obvious questions of this sort are how
common or uncommon various realizations of the Standard Model are in
the space of string theory solutions with given supersymmetry, and
what kinds of  physics
beyond the Standard Model are associated with
these different realizations.
Another, related,
question is what, if any, constraints are imposed on the
kinematics of
Standard Model-like theories, and whether or not these constraints can
teach us anything about more general constraints imposed by
consistency with quantum gravity that we might expect to be manifest
at low energies.
% In some contexts
  The set of
string theory vacua with prescribed low-energy data is
sometimes referred to as part of
the string \emph{landscape}, while low-energy effective theories with
similar kinematic structure that do not admit a UV completion in
string theory (and therefore, perhaps not in quantum gravity more
generally) are said to belong to the \emph{swampland} \cite{VafaSwamp,OoguriVafaSwamp}.
Adopting this
terminology, the questions described above can be rephrased as
questions about, respectively, the prevalence of Standard Model-like vacua within the
string landscape, and identification of the ``boundary'' between the landscape
and the swampland as viewed from the perspective of Standard
Model-like constructions. Taking the broad
approach of constructing large classes of Standard Model-like theories
also provides the more pragmatic opportunity of comparing how natural
or phenomenologically-relevant different classes of such constructions
may be, and for identifying more specific features of observed physics
that may arise in subsets of the classes of constructed models.

The questions described above could in principle be explored in the context of various different types of constructions belonging to different branches of string theory. In this paper, however, we focus exclusively on Standard Model-like constructions in F-theory \cite{VafaF-theory, MorrisonVafaI, MorrisonVafaII}. F-theory provides a
remarkably effective set of tools for exploring the landscape of 4D $\cN=1$ string vacua.
%supersymmetric string
%vacua that are described at low energies by 4D $\cN=1$ chiral
%gauge theories coupled to gravity.
The reason for this is that F-theory relates nonperturbative type IIB flux compactifications on compact K\"ahler
threefolds with 7-branes to the geometry of singular elliptically-fibered Calabi--Yau (CY) fourfolds, and this relationship
%between type IIB string theory compactifications and CY geometry has
has in turn led to the development of a systematic procedure for
constructing 4D $\cN=1$ string vacua with desired kinematics simply by tuning the mathematical properties of the elliptic CY singularities using celebrated results in algebraic geometry. An enormous number of string vacua have been constructed following this procedure, and although F-theory is dual in certain regimes to other constructions such
as heterotic string compactifications, F-theory is believed to give
the broadest global picture
currently available of the supersymmetric string landscape in terms
of a unified space of elliptic CY fourfolds that are
connected through various topology-changing transitions (see, e.g., \cite{TaylorWangMC,HalversonLongSungAlg,TaylorWangLandscape,TaylorWangVacua}).

An extensive literature on 4D F-theory flux compactifications has been
produced over the past fifteen years, with most papers on standard
model-like F-theory vacua focusing on GUT models whose gauge group is
broken, giving the Standard Model gauge group at low energies, see,
e.g.,
\cite{Donagi:2008ca,BeasleyHeckmanVafaI,BeasleyHeckmanVafaII,DonagiWijnholtGUTs,Chen:2009me,Heckman:2009mn,Blumenhagen:2009yv,Dudas:2009hu,Grimm:2009yu,Marsano:2009wr,Leontaris:2011wtt,Callaghan:2012rv,Mayrhofer:2013ara,Callaghan:2013kaa,Braun:2014pva,Li:2021eyn,Li:2022aek},
and \cite{HeckmanReview} for a review of the extensive literature on
SU(5) F-theory GUT constructions. The topic of this paper is a
somewhat less extensively explored type of construction, namely
F-theory vacua with exact $(\SU(3) \times \SU(2) \times \U(1) )/
\Gamma$ gauge symmetry and no larger geometric GUT
symmetry. Specifically, we study the family of F-theory
vacua engineered by the universal tuned $\SM$ model described in
\cite{Raghuram:2019efb}.  The $\SM$ model
describes a
universal class of F-theory Weierstrass models with tuned\footnote{A
  tuned gauge group is one that is directly tuned in the Weierstrass
  model defining the F-theory compactification, corresponding to
  fixing some specific complex structure moduli, as opposed to one
  that arises on a rigid divisor as a generic feature of the F-theory
  base geometry, or which arises from breaking a larger such rigid
  group.  In this sense, the standard SU(5) F-theory GUT constructions
  reviewed in \cite{HeckmanReview} also rely on tuning of the SU(5)
  structure, while the more recent flux-breaking GUT constructions of
  \cite{Li:2021eyn,Li:2022aek} and the direct SM construction of
  \cite{GrassiHalversonShanesonTaylor} use rigid gauge factors, which
  may be more prevalent in the landscape.  } $\SM$ gauge symmetry that
has geometrically generic matter for this
gauge group,\footnote{The notion of genericity for matter
  representations in 6D and 4D F-theory was defined and explored in
  \cite{TaylorTurnerGeneric}.  Note that while universal tuned
  Weierstrass models will exist for other choices of global gauge
  group structure associated with other quotient factors $\Gamma$, the
  Standard Model chiral matter is only generic for $\Gamma =\Z_6$.
  Note also that generic matter types are associated with the geometry
  of general Weierstrass tuning of any given gauge group; while in 6D
  this definition can be naturally understood as a feature of the
  low-energy theory (where generic matter is associated with the
  largest-dimensional branch of the moduli space for fixed gauge
  group), this distinction is less transparent in 4D, and other
  constructions such as that of \cite{Li:2022aek} do not seem to share
  the same natural structure of generic matter in 4D.  }
which includes the matter representations of the Minimal
Supersymmetric Standard Model (MSSM) and three additional exotic
matter representations; these representations can combine into three
independent anomaly-free families (one of which is that of the
MSSM). Note that a subclass of these models, which arise naturally
through a toric fiber (``$F_{11}$'') construction
\cite{KleversEtAlToric}, contains the exact matter spectrum of the
MSSM with no exotics; relatedly, chiral matter in $F_{11}$ models was
investigated in \cite{CveticEtAlThreeParam,CveticEtAlQuadrillion}, and vector-like matter
in these models (which we do not address here) has been explored in
\cite{Bies:2014sra,Bies:2021nje,Bies:2021xfh,Bies:2022wvj}.

One of the primary questions we address in this paper is whether the
tuned $\SM$ model over an arbitrary
base that allows this tuning
naturally contains chiral matter in the other two allowed
families, or whether the other two families are for some reason
forbidden by F-theory geometry and hence belong to the swampland. To
this end we use the recently developed approach of
\cite{Jefferson:2021bid} to determine the lattice of vertical fluxes
of the $\SM$ model defined over an arbitrary smooth base and in the
presence of a non-trivial flux background. By computing the full set
of linear dependences of the vertical fluxes, we show that all three
independent families of anomaly-free chiral matter are realized, and
hence that the linear constraints imposed by
F-theory geometry on chiral multiplicities precisely
match those imposed by 4D anomaly cancellation. Furthermore, we
 examine aspects of the quantization of the fluxes (and hence, of the
 chiral indices) both in general and
 for specific choices of F-theory base, and
comment on the implications for the numbers of allowed chiral
families.

More precisely,
we compute %the first known
an explicit resolution of a presentation of the $\SM$
model in which the elliptic fiber is realized as a general cubic in
$\bP^2$. We use this resolution to confirm that the gauge
symmetries, matter representations, and Yukawa interactions
anticipated in \cite{Raghuram:2019efb}
are realized geometrically by the singular elliptic CY fourfold
associated to the $\SM$ model. We then use the methods of
\cite{Jefferson:2021bid} to show that the lattice of vertical fluxes
preserving the 4D local Lorentz and gauge symmetry,
which gives at least a subset of the full set of chiral multiplicities realized by the F-theory model, spans the full linear space of anomaly-free chiral spectra transforming
under $\SM$.
The analysis carried out here is performed in a (mostly) base-independent
fashion so that the results can be applied to a large number of
F-theory vacua with a wide range of bases.
Note, however, that not all base geometries admit tuning of this gauge
structure.  In particular, at large Hodge numbers, there are many
rigid gauge factors and it can become difficult to find room in the
geometry for tuning additional features.

The remainder of this paper is structured as follows:
In \cref{sec:chiral}, we give a brief summary of the main results of
the paper.  In \cref{sec:resolution}, we outline the basic setup and
describe explicitly the resolution of the universal tuned $\SM$ model.
In \cref{sec:gauge,codimtwogeo,sec:Yukawa}, we study the explicit
resolutions of singularities at codimension one, two, three,
corresponding to the gauge factors, matter representations, and Yukawa
interactions.  In \cref{sec:intersection}, we analyze the chiral
matter multiplicities in the presence of (vertical) fluxes using the
formalism of \cite{Jefferson:2021bid}, and in \cref{sec:specific} we
consider explicit examples.
In \cref{sec:quantization} we discuss some more detailed aspects of the quantization of chiral multiplicities.
We conclude in \cref{sec:discussion} with a discussion of the
results and further directions for future work.

\section{Summary of results}

\label{sec:chiral}

\begin{table}
\begin{center}
\includegraphics{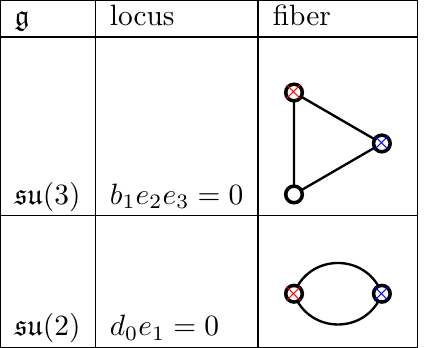}
\includegraphics{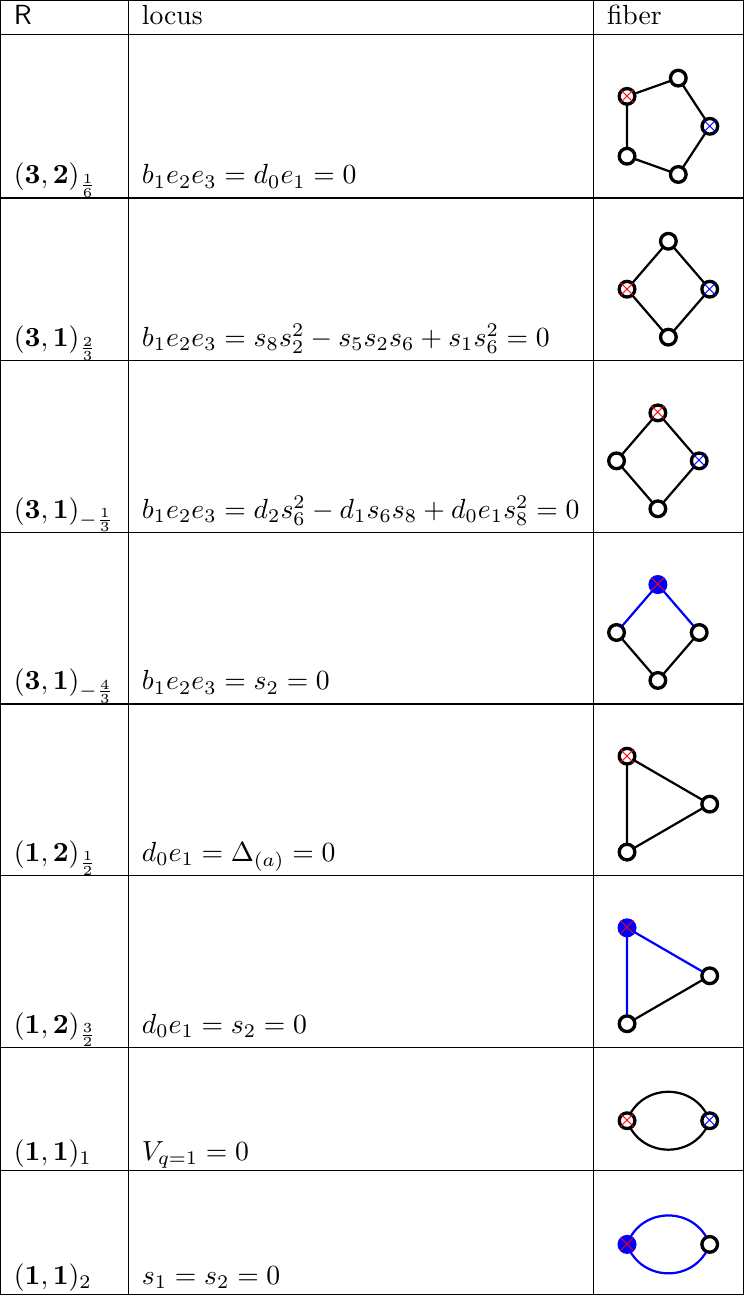}
\end{center}
\caption{Singular elliptic fibers of the resolution $X_5 \rightarrow
  X_0$ (see \cref{eq:321res}) of the $\SM$ model over codimension-one and
  codimension-two components of the discriminant locus $\Delta = 0$;
  these singular fibers correspond to (resp.) gauge symmetries and
  charged matter representations. The nodes of the graphs in the
  right-most columns represent irreducible components of the singular
  elliptic fibers, while the edges of the graphs correspond to points
  of intersection between pairs of irreducible components. Each
  irreducible component is birational to $\bP^1$. A blue (red)
  `$\times$' in the center of a node indicates that the curve
  represented by that node intersects the zero section (generating
  section) at a point away from the points of intersection with other
  $\bP^1$ components. A blue node is an exceptional $\bP^1$ wrapped by the zero section; such curves correspond in the
  M-theory frame to primitive BPS particles whose KK central charges
  are approximately the same scale as their Coulomb branch central
  charges. (Note that we express the loci in terms of the parameters
  of the resolution $X_5$, so that $d_0 e_1=0,b_1 e_2 e_3 =0$ are,
  respectively, the $\SU(2),\SU(3)$ gauge divisors $\Sigma_2,\Sigma_3
  \subset B$.)}
	\label{tab:codim12}
\end{table}

\begin{table}
\begin{center}
\includegraphics{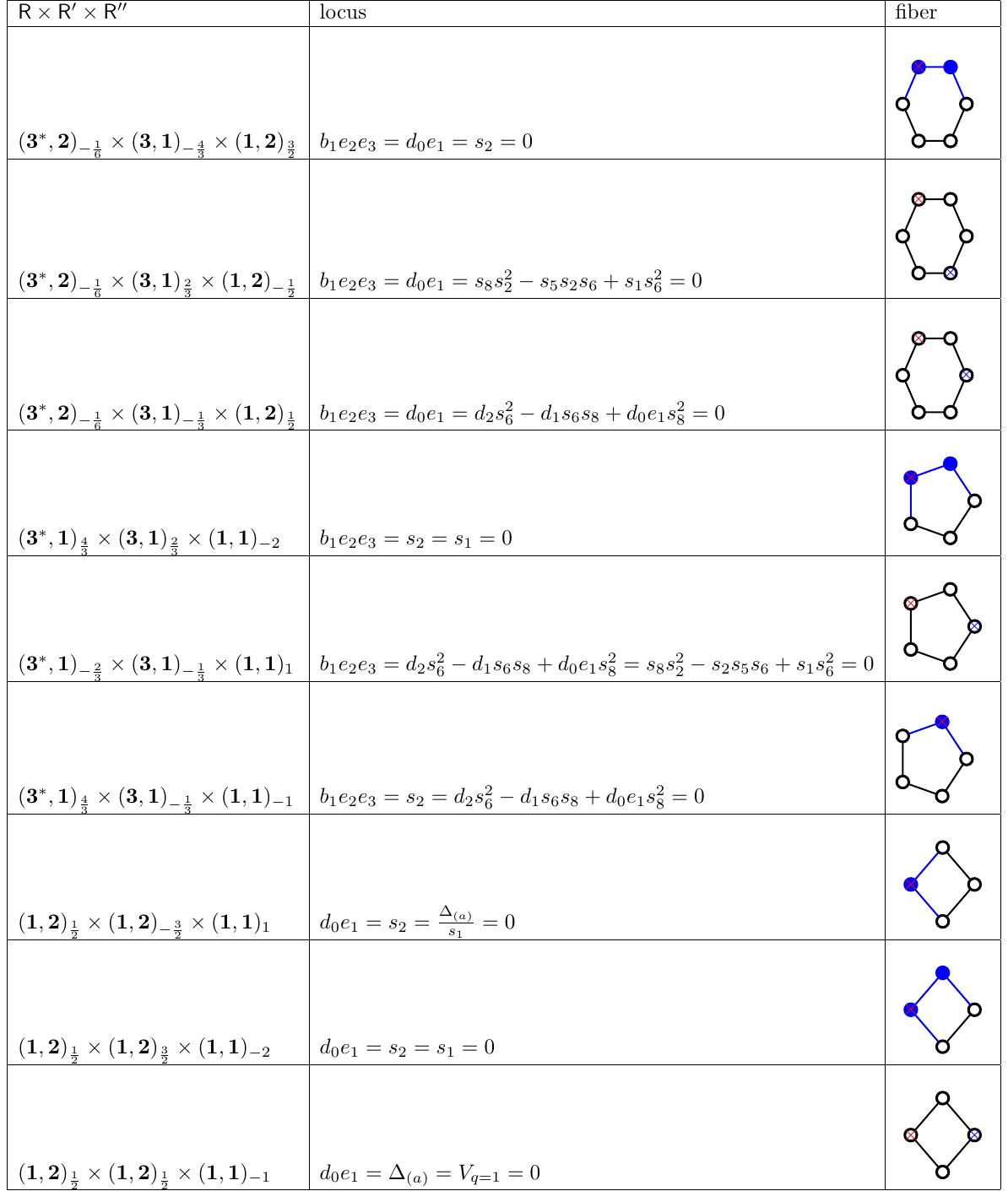}
\end{center}
\caption{Table of singular elliptic fibers of the resolution $X_5 \rightarrow X_0$ of the $\SM$ model over codimension-three  components of the discriminant locus $\Delta = 0$; these singular fibers correspond to Yukawa interactions. The left-most column describes the triples of chiral representations that are contracted to gauge-singlets in each Yukawa interaction. See the caption of \cref{tab:codim12} for an explanation of the graphs in the right-most column and the definitions of the singular loci.
}
\label{tab:codim3}
\end{table}

Our analysis of the chiral spectrum of the tuned $G_{\rm SM}=\SM$ model is divided into
two parts. First, we resolve the singular F-theory background $X_0$ in
order to study the kinematics of the corresponding low-energy
effective 4D theory for an arbitrary
choice of base and characteristic data where the tuned $G_{\rm SM}$
structure is possible. Second, we switch on a generic
vertical flux background and use the geometry of the resolved model to
compute the corresponding chiral excesses induced in the 4D
spectrum. We summarize the results of these two parts of the analysis
in the subsections below.

\subsection{Resolution of singularities and 4D kinematics}

Let $X_0 \rightarrow B$ be the singular elliptic CY variety defining the $\SM$
model, where $B$ is the base of the elliptic fibration. In \cref{sec:resolution}, we use well-established
methods\footnote{A distinctive feature of our analysis is the fact
  that the singular $\SM$ model is constructed using a general cubic
  in $\bP^2$ as opposed to the more standard Weierstrass
  equation. As a result, the resolution $X_5 \rightarrow X_0$ exhibits
  a rational, rather than holomorphic, zero section, which indicates
  the presence of primitive BPS particles with non-trivial KK charge
  in the spectrum of the low-energy 3D theory corresponding to
  $X_5$. The presence of non-trivial KK charges changes the way in
  which the 4D spectrum is ``imprinted'' on the spectrum of its 3D KK
  reduction, and thus some care is required in recovering the 4D
  kinematics from the 3D KK theory. We stress that the methods we use
  in our analysis here
are not novel; other models sharing this feature
  have been similarly analyzed in the literature, see,
  e.g.,
  \cite{Grimm:2013oga,KleversEtAlToric,LawrieEtAlRational}.} to study the
geometry of a particular resolution $X_5 \rightarrow X_0$ in order to
explore the kinematic structure of the low-energy effective 4D
theory. Although for the purposes of computing the chiral spectrum following the approach of \cite{Jefferson:2021bid} it
is sufficient to simply resolve the singularities of $X_0$ and use the
structure of the codimension-one singular fibers of $X_5$ to write the
intersection numbers of $X_5$ in terms of a basis of Cartan divisors,
we nevertheless provide a comprehensive description of the singular
fibers over the codimension-one, codimension-two, and
codimension-three components of the discriminant locus in $B$. Our
analysis of the singular fibers of $X_5$ shows that the 4D effective
theory exhibits the gauge symmetries and charged matter spectrum listed in the left-hand columns of
\cref{tab:codim12}.  The gauge symmetries and charged
matter spectrum we find agree with the analysis of the singular limit
$X_0$ in \cite{Raghuram:2019efb}, and hence our analysis provides a
check of the predictions made in \cite{Raghuram:2019efb}
regarding the
gauge symmetries and matter spectrum of the $\SM$ model. Our analysis also includes detailed geometric information relevant for further investigation
of structures such as the Yukawa interactions of the
theory---see \cref{tab:codim3}.

\subsection{Vertical flux backgrounds and chiral matter spectrum}

We next use the resolution $X_5 \rightarrow X_0$ to compute the
multiplicities of 4D $\cN=1$ chiral multiplets in the $\SM$ model induced by a non-trivial vertical flux background. The multiplicities of 4D $\cN=1$ chiral multiplets can be expressed in terms
of the chiral indices
\begin{equation}
\label{chiralindices}
\chi_{\mathsf{r}} =
  n_{\mathsf{r}} - n_{\mathsf{r}^*}\,,
\end{equation}
where $n_{\mathsf r}, n_{\mathsf r^*}$ are the numbers of chiral and anti-chiral multiplets transforming in the complex representations $\mathsf r$, $\mathsf r^*$. We direct the reader to \cref{ftheoryfluxes} for
a detailed review of the terminology and concepts discussed below.

Computing the
chiral multiplicities for a 4D F-theory compactification in terms of the geometry of a resolved elliptic CY is a
well-studied problem in F-theory, reviewed in \cite{WeigandTASI}.
Base-independent analyses of this problem for specific gauge groups
have also been worked out in, e.g.,
\cite{Marsano_2011,Kuntzler:2012bu,Cveti__2014,LinWeigandG4}.  In
\cite{Jefferson:2021bid}, we developed a
base-independent analysis, in which the intersection numbers of a resolved elliptic CY $X$
are used to compute the integral pairing on the vertical homology
subgroup,
	\begin{align}
	M\colon H_{2,2}^{\text{vert}}(X,\Z) \times H_{2,2}^{\text{vert}}(X,\Z) \to \Z\,.
	\end{align}
After homological
equivalences on vertical cycles are modded out, this nondegenerate
intersection pairing $M_{\rm red}$ was found to be independent of the
choice of resolution in many cases, and conjectured to be
resolution-independent in general.  If true, the
resolution-independence of the matrix $M_{\text{red}}$ suggests that
the integral pairing on vertical cohomology is intrinsic to the
singular F-theory limit that directly encodes information about the
low-energy 4D theory; this is physically natural since all physical
features such as the chiral spectrum should be resolution-independent,
but this kind of cohomology of singular spaces is not fully understood
mathematically.
Other recent work aimed at a direct characterization of the physics of
F-theory models in terms of the singular geometry described by type
IIB string theory includes \cite{Grassi:2018wfy,Grassi:2021ptc,Katz:2022vwe}.

  The chiral multiplicities of any theory of interest
can be computed from simple linear algebra using $M_{\rm red}$, where
schematically $\Theta_{IJ} = M_{(IJ)(KL)} \phi^{KL}$ in terms of a
(homologically redundant)
basis of surfaces $S_{IJ} = \hat{D}_I \cap \hat{D}_J$, $\phi^{IJ}$
parameterizes the allowed (vertical) fluxes, and the chiral
multiplicities can be related to certain $\Theta_{IJ}$, while other
$\Theta_{IJ}$ are constrained to vanish by various symmetry
principles.  For example, for an F-theory Tate model with a
nonabelian gauge group (defined over an arbitrary base, and with generic
complex parameters, with no additional Kodaira singularities at higher
codimension), the reduced matrix $\mr$ takes the schematic
form
	\begin{equation}
        M_\text{red} =  \begin{pmatrix}
               D_{\alpha'} \cdot K \cdot D_\alpha  & D_{\alpha'}  \cdot D_{\alpha} \cdot D_\beta  & 0 & 0 \\
              D_{\alpha' } \cdot D_{\beta'} \cdot D_{\alpha }  &0 &0 & * \\
            0 & 0& -\kappa^{ij}\Sigma \cdot D_\alpha \cdot D_{\alpha'}& * \\
        0 & * &  * & *
        \end{pmatrix} \,,
\label{eq:mr-matter}
	\end{equation}
in a reduced basis $S_{0 \alpha},S_{\alpha \beta}, S_{i \alpha},
S_{ij}$ where $0$ represents the zero section, indices $\alpha$
correspond to base divisors, and $i$ correspond to Cartan elements of
the gauge group; the matrix elements denoted by ``$*$'' are undetermined and
the non-zero matrix elements are  given by intersection products of divisors $D_\alpha$ in the base $B$, where
$K$ is the canonical class of the base.  A rational change of basis
removes all the $*$ elements except the bottom right block, which
becomes a matrix $M_{\rm phys}$ relating
chiral multiplicities to
the subset of flux parameters
$\phi^{IJ}$ that %lift
correspond to symmetry-preserving
F-theory fluxes (note that all other flux parameters are forced to
vanish in order to satisfy the constraints $\Theta_{I \alpha} = 0$, which are necessary to preserve 4D Poincar\'{e} and gauge symmetry). This simple
schematic structure becomes more complicated for models with
additional abelian $\U(1)$ factors, as described in
\cite{Jefferson:2021bid}.  In particular, the presence of additional
sections complicates the form of the reduced matrix
\labelcref{eq:mr-matter}, and
makes it harder to describe the general solution of the
constraints as easily.  This is described in some detail for a simpler
model with gauge group $(\SU(2) \times \U(1))/Z_2$ in
\cite{Jefferson:2021bid}; working through the details of this for the
more complicated universal $\SM$ model is a central part of the work in this
paper.  The matrix $M_{\text{red}}$ representing the integral pairing
for the resolved $\SM$ model is displayed in \cref{Mred} and \cref{t:mr-alternate} in  two
particular homologically independent bases for $S_{IJ}$, which are useful in complementary aspects of the analysis.
This reduced matrix is used in \cref{sec:intersection} to describe the chiral
matter content of the model with fluxes.

An interesting question one can ask related to the landscape of
F-theory compactifications is whether or not all linearly independent
anomaly-free
combinations of chiral matter can be realized in F-theory,
i.e., whether F-theory imposes further linear constraints on the
available chiral spectrum beyond those associated with anomalies.
(Note that we expect that all F-theory models will satisfy anomaly constraints;
 the geometric manifestation of these constraints has been studied in,
 e.g., \cite{LinWeigandG4, Bies_2017,
   Corvilain:2017luj,Cheng:2021zjh}).
We address
this question by geometrically computing the possible combinations of
chiral matter multiplicities that can be realized in the $\SM$ model
and comparing them to the full list of possible chiral matter
combinations compatible with 4D anomaly cancellation, namely the set
of all integer values for the chiral indices $\chi_{\mathsf{r}}$
(where $\mathsf{r}$ is one of the eight complex representations
appearing in the left-most column of the right hand table in
\cref{tab:codim12}) subject to the constraints
\begin{equation}
    \label{eq:321-AC-sols}
    \begin{aligned}
        \chi_{(\bm{3}, \bm{1})_{\frac{2}{3}}} &=& -&\chi_{(\bm{1}, \bm{1})_1}& +&& 2 &\chi_{(\bm{1}, \bm{2})_{\frac{3}{2}}}& -&& &\chi_{(\bm{3}, \bm{1})_{-\frac{4}{3}}} \\
        \chi_{(\bm{3}, \bm{1})_{-\frac{1}{3}}} &=& -&\chi_{(\bm{1}, \bm{1})_1}& +&& 2 &\chi_{(\bm{1}, \bm{2})_{\frac{3}{2}}}& -&& 4 &\chi_{(\bm{3}, \bm{1})_{-\frac{4}{3}}}\\
        \chi_{(\bm{1}, \bm{2})_{\frac{1}{2}}} &=& -&\chi_{(\bm{1}, \bm{1})_1}& -&& &\chi_{(\bm{1}, \bm{2})_{\frac{3}{2}}}& -&& 2 &\chi_{(\bm{3}, \bm{1})_{-\frac{4}{3}}} \\
        \chi_{(\bm{3}, \bm{2})_{\frac{1}{6}}} &=& &\chi_{(\bm{1}, \bm{1})_1}& -&& 2 &\chi_{(\bm{1}, \bm{2})_{\frac{3}{2}}}& +&& 2 &\chi_{(\bm{3}, \bm{1})_{-\frac{4}{3}}} \\
        \chi_{(\bm{1}, \bm{1})_2} &=& && -&& &\chi_{(\bm{1}, \bm{2})_{\frac{3}{2}}}& +&& &\chi_{(\bm{3}, \bm{1})_{-\frac{4}{3}}}\,.
    \end{aligned}
\end{equation}
Since there are eight possible chiral matter representations and five
constraint equations (above), this leaves behind a three-dimensional
linear space of anomaly-free chiral matter multiplicities.
%The three
%primitive
One basis for the three
linearly-independent families of anomaly-free matter is
displayed in \cref{t:321-families}. Note that setting the chiral
multiplicities of the exotic\footnote{We refer to charged matter
  transforming in representations that do not belong to the MSSM as
  ``exotic'' matter, although these are still generic features of the
class of  universal tuned $\SM$ F-theory models.} representations $(\bm{1}, \bm{2})_{\frac{3}{2}}$
and $(\bm{3}, \bm{1})_{-\frac{4}{3}}$ equal to zero reduces this
three-dimensional family of anomaly-free chiral multiplicities to a
one-dimensional family corresponding to a single generation of the
MSSM chiral matter spectrum.\footnote{Another specific
  linear combination of the anomaly-free families (with multiplicities
  of each family multiplied by $2, 0, 1$ respectively), has no matter
  charged under the $\SU(3)$ factor and indeed the resulting
  multiplicities correspond precisely to those of the $(\SU(2)
  \times\U(1))/\Z_2$ model \cite{KleversEtAlToric} whose chiral
  spectrum is explored in \cite{Jefferson:2021bid}.}

We compare the  anomaly-free chiral matter spectra  satisfying
\cref{eq:321-AC-sols} to the chiral indices of the resolved $\SM$
model induced by a non-trivial vertical flux background; see
\cref{chiasphi}. Consistent with the findings of
\cite{Jefferson:2021bid}, namely that all anomaly-free linear
combinations of matter can be realized in F-theory (without
considering multiplicities)
for the models considered there, we indeed find that the full set of
three independent linear families
of possible chiral matter multiplicities are realized in the $\SM$
model in the presence of a non-trivial vertical flux background for a
generic choice of base.
 We
comment on the precise quantization of the chiral indices imposed by
the geometry of specific F-theory vacua (i.e., a specific choice of
base and associated topological data) in \cref{sec:quantization}.
In addition to the quantization constraints, tadpole constraints,
which we do not consider here, will
also limit the range of chiral matter multiplicities available in any
given model.

\begin{table}
\begin{center}
\begin{tabular}{|c |ccccc |ccc |}
\hline
&$( \textbf{3}, \textbf{2})_{\frac{1}{6}} $&$ (
  \textbf{3}, \textbf{1})_{\frac{2}{3}}  $&$( \textbf{3},
  \textbf{1})_{-\frac{1}{3}}   $&$
( \textbf{1}, \textbf{2})_{\frac{1}{2}}  $&$ ( \textbf{1},
  \textbf{1})_{1}  $&$
	 ( \textbf{3}, \textbf{1})_{-\frac{4}{3}}
 $&$( \textbf{1}, \textbf{2})_{\frac{3}{2}}  $&$ ( \textbf{1},
  \textbf{1})_{2}  $\\
\hline
(MSSM) & 1& -1& -1& -1& 1& 0& 0& 0\\
(exotic 1) & 2 & -1 & -4 & -2 & 0 & 1 & 0 & 1\\
(exotic 2) & -2 & 2 & 2 & -1 & 0 & 0 & 1 & -1\\
\hline
\end{tabular}
\end{center}
\caption[x]{Multiplicities of the three anomaly-free
  families of generic chiral matter in 4D models with $\SM$ gauge group}
\label{t:321-families}
\end{table}

\section{Basic setup and resolution}

\label{sec:resolution}

The universal tuned $\SM$ model was
identified in \cite{Raghuram:2019efb} by un-Higgsing a U$(1)$ model
with charge $q=4$ matter introduced in \cite{Raghuram34}. The charge
$q=4$ $\U(1)$ model of \cite{Raghuram34} was constructed as a singular
hypersurface of an ambient projective bundle $Y_0$ whose fibers are
isomorphic to $\bP^2$ with homogeneous coordinates $[u:v:w]$. The hypersurface equation is as follows:
 	\begin{align}
  \label{eq:p0hyp}
	\begin{split}
		p_{0}^{q=4} &= (a_1 v+b_1 w) \left(d_0 v^2+d_1 v w+d_2 w^2\right)\\
	&+u \left(s_1 u^2+s_2 u v+s_3 v^2+s_5 u w+s_6 v w+s_8 w^2\right) = 0\,.
	\end{split}
	\end{align}
In terms of the above construction, the un-Higgsing $\U(1) \to \SM$ is achieved by restricting to a special locus in complex structure moduli space on which the sections $a_1 = s_3$ vanish:
	\begin{equation}
%	\label{eq:X0hyp}
		X_0 = \{ p_0^{q=4} =  a_1 = s_3 = 0\} =: \{ p_0=0\} \subset Y_0\,,
	\end{equation}
so that
\begin{equation}
		p_{0} = b_1 w \left(d_0 v^2+d_1 v w+d_2 w^2\right)
	+u \left(s_1 u^2+s_2 u v+s_5 u w+s_6 v w+s_8 w^2\right)
= 0\,.
	\label{eq:X0hyp}
\end{equation}
We now discuss this construction in more detail. First, observe that the above elliptic fibration is equipped with two rational sections:
	\begin{align}
	\begin{split}
	\label{eq:sections}
		\text{zero section:}& \quad \{  s_1 u + s_2 v =w= 0 \}\,, \\
		\text{generating section:}& \quad \{ u = w =0 \}\,.
	\end{split}
	\end{align}
The coefficients of
the cubic (in $u,v,w$) monomials appearing in the homogeneous
polynomial $p_0$ in \cref{eq:X0hyp} are (the pullbacks
to $Y_0$
of) sections of
various line bundles over $B$. We may equivalently express these line bundles in terms of their first Chern classes, each of which can in turn be expressed as a linear combination of the following divisor classes:
	\begin{align}
	\label{eq:chardata}
		K\,, \quad \Sigma_2 = [d_0]\,, \quad \Sigma_3 = [b_1]\,, \quad Y = [s_2]\,.
	\end{align}
We refer to the above four divisor classes as the characteristic data of the elliptic fibration $Y_0 \rightarrow B$. Let us describe these divisor classes in further detail:
	\begin{itemize}
		\item $K$ is the canonical class of $B$;
		\item $\Sigma_2,
\Sigma_3$ are the $\asu(2)$, $\asu(3)$ gauge
divisors wrapped by 7-branes;
	\item and $Y$ is the divisor class of the
locus in $B$ over which the zero section and the
generating section associated with the $\au(1)$
factor
intersect.
\end{itemize}
The classes of the zero loci of the other sections appearing in $p_0$ besides $d_0,b_1,s_2$ are
	\begin{align}
	\begin{split}
	\label{eq:chardata2}
		[s_1] &=-K+Y - \Sigma_3 - \Sigma_2\,,\\
		[d_1] &=-2K - Y - \Sigma_3\,,\\
		[d_2] &=-4 K- 2Y - 2 \Sigma_3 - \Sigma_2\,,\\
		[s_5] &=-2 Y - \Sigma_3 - \Sigma_2\,,\\
		[s_6] &=-K\,,\\
		[s_8] &=- 3 K - Y - \Sigma_3 - \Sigma_2\,.
	\end{split}
	\end{align}
The ambient space of the elliptic fibration, $Y_0$, can be viewed as the projectivization of a rank-two vector bundle:
 	\begin{align}
	\label{canonproj}
		Y_0 =  \bP(\oplus_{i=1}^3 \sL_i) \xrightarrow{\varpi} B,~~~~ \sL_i \rightarrow B\,.
	\end{align}
To ensure that the homogeneous polynomial $p_0$ is a well-defined section of the line bundle
	\begin{align}
		\otimes_i \sL_i
	\end{align}
we assign the following choice (unique up to an overall additive
constant) of divisor classes\footnote{We use bold symbols $\boldsymbol
  D$ to denote divisor classes in the Chow ring of the ambient
  projective bundle $Y_0$; in particular, $\bm{D}_\alpha$ denotes the
  pullback of a divisor $D_\alpha \in B$ (e.g., $\bm K = K^\alpha \bm
  D_\alpha$ is the pullback of $K$ to the Chow ring of $Y_0$).}
$\boldsymbol{L}_i = c_1(\sL_i)$ to the hyperplanes $u,v,w = 0$
in the Chow ring of $Y_0$
	\begin{align}
	\begin{split}
		[u] &=\bm{H} +\bm{L}_1 =\bm{H} -\bm{Y} + \boldsymbol \Sigma_3\\
		[v] &=\bm{H} +\bm{L}_2 = \bm{H}-\bm{K} - \bm{Y} - \boldsymbol \Sigma_2\\
		[w] &=\bm{H} +\bm{L}_3 = \bm{H}-\bm{K} +\boldsymbol  \Sigma_3\,.
	\end{split}
	\end{align}
 Although the hypersurface equation \cref{eq:X0hyp} differs from the
 standard Weierstrass form presented in \cite{Raghuram:2019efb} (with the two equations being related by Nagell's
 algorithm\footnote{See Appendix B of
   \cite{CveticKleversPiraguaMultU1} for a discussion of Nagell's
   algorithm.}), the singular locus of $X_0$ is nevertheless still given
 by the zero locus of the discriminant $\Delta = 4 f^3 + 27 g^2$.
 Explicit expressions for $f,g$ can be found in
 \cite{Raghuram34,Raghuram:2019efb} (for related discussions, see also \cite{CveticKleversPiraguaMultU1,Cvetic:2015ioa}).

We resolve the singularities of $X_0$ through codimension two by means
of the following sequence of blowups\footnote{The notation $f_{i+1} =
  (g_{i+1,1},\dots,g_{i+1,n_{i+1}} |e_{i+1} )$ is shorthand for the
  blowup $f_{i+1}:Y_{i+1} \rightarrow Y_i$ of the ambient space $Y_i$
  along the blowup center $\{ g_{i+1,1} = \cdots = g_{i+1,n_{i+1}} = 0
  \} \subset Y_i$ whose exceptional divisor is the zero locus $e_{i+1} =
  0$. The corresponding blowup $X_{i+1} \rightarrow X_i$ is then given
  by the restriction of $f_{i+1}$ to the complete intersection $X_i =
  \{p_{i,1} = \cdots = p_{i,m_i} =0\} \subset Y_i$. We abuse notation
  and implement the $i+1$th blowup by making the substitution
  $g_{i+1,j} \rightarrow e_{i+1} g_{i+1,j}$ whenever $g_{i+1,j}$ is a local coordinate, and we denote the proper
  transform of $X_i \subset Y_i$ by $X_{i+1} \subset Y_{i+1}$.}
	\begin{align}
	\begin{array}{c}
	\label{eq:321res}
			X_5  \overset{(w,t|e_5)}{\longrightarrow} X_4\overset{(u,w|e_4)}{\longrightarrow}X_3 \overset{(e_2,w|e_3)}{\longrightarrow}X_2 \overset{(b_1,u|e_2)}{\longrightarrow} X_1 \overset{ (d_0, u ,w |e_1)}{\longrightarrow} X_0.
	\end{array}
	\end{align}
In this sequence of blowups, the variable $t$ appearing in the defining equation for $X_4$ is defined as follows:
	\begin{align}
		t := s_2 v + e_1 e_2 e_3 e_4 s_1 u\,.
	\end{align}
The first three blowups appearing in \cref{eq:321res} (counting from
the right)
are
conceptually straightforward, as they are restricted to
codimension-one components of the discriminant locus in $B$. However,
the fourth and fifth blowups appearing in \cref{eq:321res}, whose
corresponding blowdown maps are (respectively) $X_4 \rightarrow X_3,
X_5 \rightarrow X_4$, are somewhat unintuitive in that they appear to
blow up the entire base $B$. However, it turns out that these choices
of blowups produce new divisors with precisely the desired geometry:
the proper transform of each blowup is topologically a small blowup of
the base $B$ at the points over which the elliptic fibers become
singular, and hence is a smooth birational representative of a
rational section (see \cref{eq:sections}).\footnote{To see why the
  fourth and fifth blowups in \cref{eq:321res} have precisely the
  intended effect, let us study blowups of this form in the context of
  a toy model. Let $Y_0$ be the total space of a $\mathbb P^2$
  fibration over a base $B$. Suppose that the fibers of $Y_0$ have
  homogeneous coordinates $[x:y:z]$ and let us restrict to a $\mathbb
  C^2$ open set of $B$ with local affine coordinates $(a,b)$. Let $X_0
  \subset Y_0$ be the hypersurface $ax+by =0$. The hypersurface $X_0$
  has a section $x =y =0$. There is a conifold singularity at $a = b =
  x = y = 0$. Suppose we resolve this singularity by means of a blowup
  along $x = y = 0$, i.e. we make the substitution $x \rightarrow e_1
  x,y\rightarrow e_1 y$ with $e_1=0$ a local equation for the
  exceptional divisor in the ambient space. The proper transform $X_1$
  is the hypersurface $a x+ by = 0$ in the new ambient space $Y_1$
  with an affine open set described by the homogeneous coordinates
  $[x:y][e_1x:e_1y:z](a,b)$ (and some unspecified $\mathbb C^*$
  action). Having blown up the section, we can study its proper
  transform by restricting to the exceptional locus $e_1=0$. Away from
  the ``discriminant locus'' $a=b=0$ in $B$, we find that the
  exceptional divisor is isomorphic to the original section, and
  corresponds to the set of points $[-b:a][0:0:z](a,b)$. When
  restricted to the discriminant locus, we find that the point
  $[0:0:z]$ of the fiber has been replaced with a $\mathbb P^1$,
  i.e. the set of points $[x:y][0:0:z](0,0)$. Thus, we see that the
  proper transform of the section has the topology of a small
  blowup of the base $B$ at the point $(0,0)$. An analogous story
  holds for the proper transforms of the sections \cref{eq:sections}
  under the fourth and fifth blowups in \cref{eq:321res}.}

The proper transform of $X_0$ under the above composition of blowups, $X_5$, is the complete intersection
	\begin{align}
	\begin{split}
	\label{eq:X5hyp}
		X_5 = &~\{p_{5,1} =p_{5,2} =0 \} \subset Y_5 \\
		 p_{5,1} = &~b_1 d_1 e_3 e_4 e_5 v w^2+b_1 d_2 e_1 e_3^2 e_4^2 e_5^2 w^3+b_1 d_0 v^2 w+e_2 e_4 t u^2+e_1 e_2 e_3 e_4^2 s_5 u^2 w\\
		 &+e_4 s_6 u v w+e_1 e_3 e_4^2 e_5 s_8 u w^2\\
		 p_{5,2}=&~-te_5 +e_1 e_2 e_3 e_4 s_1 u+s_2 v
\end{split}
	\end{align}
and the ambient space $Y_5$ is equipped with homogeneous coordinates
	\begin{align}
	\begin{split}
	\label{eq:X5coord}
		&[e_1 e_2 e_3 e_4 u:v:e_1 e_3 e_4 e_5 w] ~[e_2 e_3 e_4 u:e_3 e_4 e_5 w:d_0] \\
		~&[e_4 u:b_1]~ [e_2:e_4 e_5 w]~[u:e_5 w]~[w:t]
	\end{split}
	\end{align}
The above homogeneous coordinates are identified under the rescalings
	\begin{align}
	\begin{split}
	\label{eq:321scale}
		&u\to \lambda _0 \lambda _1 \lambda _2 \lambda _4 u,~~v\to \lambda _0 v,~~w\to \lambda _0 \lambda _1 \lambda _3 \lambda _4 \lambda_5 w,~~d_0\to d_0 \lambda _1,~~b_1\to b_1 \lambda _2\,,\\
		&e_1\to \frac{e_1}{\lambda _1},~~e_2\to \frac{e_2 \lambda _3}{\lambda _2},~~e_3\to \frac{e_3}{\lambda _3},~~e_4\to \frac{e_4}{\lambda _4},~~e_5\to \frac{e_5}{\lambda _5},~~t\to \lambda _0 \lambda _5 t
	\end{split}
	\end{align}
        for $\lambda_{i} \in \C^{*}$. One can verify via explicit
        computation in the affine open sets of $X_5$ that the rank of
        the Jacobian $ [[\partial_i p_{5,j}]] $ does not (in general) reduce over
        any codimension-two loci in $B$, indicating that $X_5$ is
        smooth through codimension-two in $B$ without additional tunings of the parameters defining the $\SM$ Weierstrass model.\footnote{Note that certain choices of base $B$ can impose additional specializations on the Weierstrass model parameters that can in principle lead to additional singularities appearing over higher codimension components of the discriminant locus.}
We abuse terminology
        and refer to $X_5$ as ``smooth'', a ``resolution'', etc.,
        keeping in mind that $X_5$ may contain singular fibers over
        certain codimension-three loci; these codimension-three
        singularities do not affect the main results of this paper.

We next analyze the geometry of $X_5$ to identify the kinematics of
the low-energy effective 4D $\cN=1$ supergravity theory describing the
$\SM$ model at long distances. Unless otherwise stated, throughout our
analysis we work over an arbitrary smooth base $B$.

\section{Codimension one: gauge symmetry}
\label{sec:gauge}

In this section, we analyze the geometry of the singular elliptic fibers of $X_5$ over codimension one loci of the discriminant locus in $B$, in order to verify that our resolution correctly exhibits the expected gauge algebra
	\begin{equation}
		\mathfrak{g} = \oplus_s \mathfrak{g}_s = \asu(3) \oplus \asu(2) \oplus \au(1)
	\end{equation}
characterizing the 4D $\cN=1$ theory describing F-theory compactified on the singular fourfold $X_0$.

\subsection{\texorpdfstring{$\au(1)$}{}}
\label{sec:U1gauge}

In F-theory compactifications pure $\au(1)$ gauge symmetries are
identified with the Mordell--Weil (MW) group of rational sections of
the elliptic fibration
\cite{MorrisonVafaII,Aspinwall:2000kf,AspinwallMorrisonNonsimply,Grimm:2010ez},
where the zero section (i.e., a choice of zero element of the MW
group, extended fiber-wise over $B$) corresponds to the Kaluza--Klein
(KK) $\au(1)$ and all other sections generating the
free part of the MW group
correspond to additional $\au(1)$ factors.\footnote{More precisely, in
  the dual M-theory picture, in the vicinity of a divisor of a smooth
  elliptic CY, it is possible to locally expand the M-theory 3-form in
  a basis of harmonic forms $\omega_i$ as $C_3 = A^I \wedge \omega_I +
  \cdots$, where $A^i$ are abelian gauge fields and the harmonic forms
  $\omega_i$ are Poincar\'e dual to the divisors $\hat D_i$. See
  Section 4.3 of \cite{WeigandTASI} and references therein for further
  discussion.} Therefore, pure abelian gauge symmetries are not
determined by codimension one singularities in $B$, although they are
nonetheless associated to smooth divisors $\hat D_{\bar a}$ of $X_5$;
see \cref{eq:ratsections} for a description of the proper transforms
of the zero and generating sections $\hat D_{ 0}, \hat D_{ 1}$ under
the composition of blowups leading to $X_5$.

Note that since $\hat D_{ 0}$ is rational, rather than holomorphic, the resolution $X_5$ describes a phase of M-theory in which some of the primitive BPS particles in the 3D spectrum carry non-trivial $\au(1)_{\text{KK}}$ charge; this point is explored further in \cref{KKcharges}.\footnote{See, e.g., \cite{Grimm:2013oga} for a discussion of the interplay between rational sections and KK charges in the context of 6D F-theory compactifications.} Since 3D BPS particles arise from M2-branes wrapping holomorphic curves $C$, these quantized $\au(1)_{\text{KK}}$ charges can be computed in terms of the intersection pairing $\hat D_{\bar 0} \cdot C$. Similarly, the addtional $\au(1)$ charges of BPS particles can be determined by computing the intersection pairings $\hat D_{\bar 1} \cdot C$.

\subsection{\texorpdfstring{$\asu(3) \oplus \asu(2)$}{}}
\label{sec:SUgauge}

In contrast to $\au(1)$ gauge symmetries, simple nonabelian gauge
symmetries $\mathfrak{g}_s$ are associated to singularities of the
elliptic fibers over codimension one components of the discriminant
locus $\Delta =0$ in $B$. Since the discriminant $\Delta$ of the \emph{singular} Weierstrass model \cref{eq:X0hyp} exhibits the
following behavior,
	\begin{align}
	\begin{split}
		f&=\cO(\varepsilon^0),~~~~g=\cO(\varepsilon^0),~~~~	\Delta= \varepsilon^5 {b_1}^3 {d_0}^2 + \cO( \varepsilon^6)\,,
	\end{split}
	\end{align}
we can see there is an I$_2$ singularity overe $d_0 = 0$ and an
I$_3^{}$ singularity over $b_1 = 0$ in $X_0$ (see, e.g.,
\cite{KatzEtAlTate}). We use our explicit description of the resolution $X_5 \rightarrow X_0$ in \cref{eq:X5hyp,eq:X5coord} to verify that
both singular fibers are split in the sense of arithmetic geometry and
indicate the presence of (resp.) $\asu(2)$ and $\asu(3)$ gauge
symmetries. We describe the singular fibers in more detail below.

In the following discussion, for convenience, we study the singular fibers and their intersections in the affine open set $e_4 e_5 \ne 0$, away from possible points of intersection with the rational sections $\hat D_0, \hat D_1$. The singular elliptic fibers $F$ over the locus $d_0 e_1 = 0$ split into the following irreducible rational curves:
	\begin{align}
	\label{eq:su2comp}
		\asu(2) ~~:~~F |_{d_0 e_1 = 0} = F_0+ F_1\,.
	\end{align}
 Setting $d_0=0$, we learn that the irreducible curve $F_0$ is the normalization of the singular cubic curve in $\bP^2$ with coordinates $[u:v:w]$ (after appropriate coordinate redefinitions), where the singularity $u = w =0$ has been replaced with the exceptional curve $F_1$. Setting $e_1 =0$, one can see that $F_1$ is a smooth conic in $\bP^2$ with coordinates $[u:w:d_0]$. The intersection $F_0 \cap F_1$ consists of two distinct points, hence the generic fibers are type I$_2^{}$ and we associate $F_0, F_1$ to, respectively, the affine and non-affine nodes of the affine $\asu(2)$ Dynkin diagram.
% Setting $\lambda_0 = 1/v, \lambda_{i=2,\dots,5} = e_i$ and making the redefinitions $\tilde u= e_2 e_3 e_4 u/v, \tilde w = e_3 e_4 e_5 w/v$, we find
%	\begin{align}
%	\begin{split}
%	\label{F1}
%		F_1  ~:~ &\{ d_1 (b_1e_2 e_3) \tilde w^2+(b_1  e_2 e_3) d_0 \tilde w+s_2 \tilde u^2+s_6 \tilde u  \tilde w= 0\}~ \subset ~\bP^2_{[\tilde u  :\tilde w :d_0]} ,
%	\end{split}
%	\end{align}
%in the affine open complementary to the point $\{ [0:0:1] \}$; it follows $F_1$ is a smooth conic.
%Setting $d_0 = 0$ forces the quadratic polynomial in \cref{F1} to factor, hence
In codimension one, $\hat D_0$ intersects $F_0$ in a point, whereas $\hat D_1$ intersects $F_1$ in a point.
%	\begin{align}
%		\hat D_0 \cdot F_0 =1,~~~~ \hat D_1 \cdot F_1 =1.
%	\end{align}

Next we turn our attention to the fibers over the locus $b_1 e_2 e_3 =0$:
	\begin{align}
	\begin{split}
		\asu(3) ~~:~~F|_{b_1 e_2 e_3 =0} = F'_0 + F'_2 + F'_3\,.
	\end{split}
	\end{align}
Analogous to $F_0$, we associate $F_0'$ to the affine node and $F'_2,F'_3$ to the other two nodes of the affine $\asu(3)$ Dynkin diagram. %Setting $\lambda_{i=1,3,4,5}=e_i, \lambda_0 = 1/(e_1 e_4 u \lambda_2), \lambda_2 = 1/\lambda_2$ and
Making the redefinitions $w \to uw/(e_3 e_5), e_2 \to  e_2/ e_3, v \to e_1 e_4 u v$, we learn that $F_0'$ can be described locally as a smooth conic in a subspace birational to $\bP^2$ with coordinates $[e_2 : v : w]$ with the points $[0:1:0]$ and $[s_2 v:s_1 e_2 :0]$ removed. Next, we consider the curve $F_2'$.
%Setting $\lambda_{i=1,4,5} = e_i, \lambda_3= 1/(e_1 e_4 e_5 w \lambda_0 )$ and
Making coordinate redefinitions $ u \to u/(e_1 e_4),e_3\to   e_3/(e_1 e_4 e_5 w)$, we find that $F_2'$ is the sum of a base and fiber curve in  $\F_1$ with homogeneous coordinates
%	\begin{align}
%	\label{F2prime}
%		F_2' ~:~ \{(d_1\tilde e_3   \tilde v +d_2 \tilde e_3^2  +\tilde d_0 \tilde v^2) b_1+(s_6 \tilde v+s_8\tilde e_3 ) \tilde u  = 0\} ~~\subset ~~ {\F_{1}}_{[\tilde v:\tilde e_3] [\tilde u:b_1]},
%	\end{align}
$[ v: e_3][ u: b_1] \cong [\lambda_0 v: \lambda_0 e_3] [ \lambda_0  \lambda_2 u: \lambda_2b_1 ]$. Finally, in the case of the curve $F_3'$,
%Setting $\lambda_{i=4,5} = e_i, \lambda_0 = 1/v, \lambda_1 = 1/d_0$ and
after making the coordinate redefinitions $ w \to v d_0 w/( e_4 e_5)$, $u \to v d_0 u /e_4$ we learn that
%	\begin{align}
%	\label{F3prime}
%		F_3' ~:~ \{ s_2 e_2 \tilde u^2 + b_1 \tilde w + s_6 \tilde u \tilde w = 0 \} ~~ \subset~~ {\F_{-1}}_{[\tilde u :b_1][e_2:\tilde w]} \setminus \{ [0:1][1:0] \},
%	\end{align}
the class of $F_3'$ is a sum of the base and fiber curve classes in $\F_{-1}$ with homogeneous coordinates $[  u:b_1][e_2 : w] \cong [\lambda_2 u : \lambda_2 b_1] [ \lambda_2^{-1} \lambda_3 e_2: \lambda_3  w]$.
%Using notation analogous to the case of $F_2'$, the $\F_{-1}$ divisor class $F_3'$ is a sum of the base and fiber curve classes, thus it follows from adjunction that $F_3'$ is a rational curve.
It is straightforward to verify that the three pairwise intersections $F_0' \cap F_2', F_0' \cap F_3', F_2' \cap F_3'$ are all points, and hence the fibers over $b_1 e_2 e_3 =0$ are of Kodaira type I$_3^\text{split}$. In codimension one $\hat D_0$ intersects $F_0'$ in a point, whereas $\hat D_1$ intersects $F_3$ in a point.

\begin{figure}
	\begin{center}
	\includegraphics{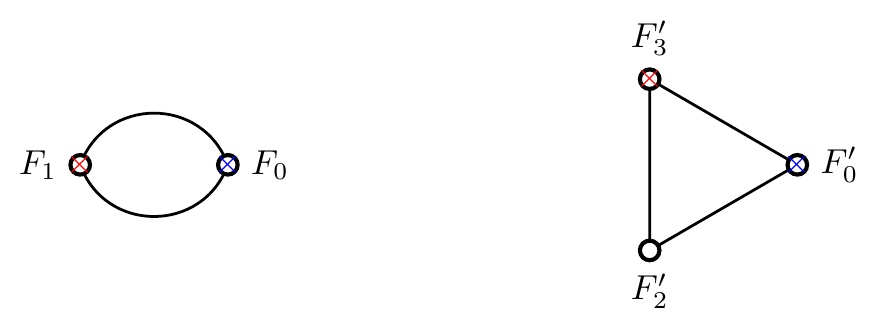}
	\end{center}
	\caption{Schematic diagram of the singular elliptic fibers over the codimension-one $\asu(2)$ (left) and $\asu(3)$ (right) loci of the discriminant locus in $B$. Over the $\asu(2)$ locus $d_0 e_1 = 0$, the singular elliptic fiber $F$ is Kodaira type $\text{I}_2$, and admits the decomposition $F = F_0 + F_1$; over the $\asu(3)$ locus $b_1 e_2 e_3 =0$, the singular elliptic fiber is Kodaira type $\text{I}_3^{\text{split}}$, and admits the decomposition $F = F'_0 + F'_2 + F'_3$. Note that all irreducible components $F_a, F'_b$ are rational curves. The blue $\times$ indicates that $F_0, F'_0$ intersect the zero section $\hat D_0$ in a point, whereas the red $\times$ symbol indicates that $F_1,F'_3$ intersect the generating section $\hat D_1$ in a point.}
	\label{I2andI3}
\end{figure}

See \cref{I2andI3} for a schematic depiction of the codimension-one singular fibers.

\section{Codimension two: local matter}
\label{codimtwogeo}

In this section we study the geometry of the codimension-two singular
fibers (i.e., collisions of codimension-one singular fibers including those described in \cref{sec:gauge}) associated to local matter representations
	\begin{align}
		(\mathsf R_3, \mathsf R_2)_{w_{\bar 1}},
	\end{align}
where
	\begin{equation}
	\mathsf R_{s} := \mathsf r_s \oplus \mathsf r^*_s
	\end{equation}
 is an irreducible quaternionic representation of the nonabelian gauge algebra $\mathfrak{g}_s$ and $w_{\bar 1} = q$ is the $\au(1)$ charge. We show, in particular, that the codimension-two singularities are consistent with the matter spectrum discussed in \cite{Raghuram:2019efb}.

\subsection{Kaluza--Klein charges}
\label{KKcharges}

As was mentioned briefly in \cref{sec:U1gauge}, a special feature of the resolution $X_5$ is that the zero section $\hat D_0$ is rational, rather than holomorphic, which is closely related to our realization of the F-theory vacuum $X_0$ as a hypersurface of a general $\bP^2$ fibration as opposed to a Weierstrass model $y^2z= x^3 + f xz^2 + gz^3$ (the Weierstrass model, by contrast, is naturally equipped with a holomorphic zero section $ x=z=0$). It turns out that the zero section $\hat D_0$ is isomorphic to a small blowup of the base $B$ at various points and is hence only birational to $B$. This can be seen by noting that the locus $s_2=0$, over which the zero and generating sections intersect (and hence over which our choice of zero section fails to be holomorphic), has non-trivial intersection with the discriminant locus. Thus, the singular elliptic fibers of $X_0$ over the codimension-two locus $\Delta = s_2=0$ have singular points that are blown up as part of the resolution $X_5 \rightarrow X_0$; the proper transforms of these points are irreducible $\bP^1$ components of $\text{I}_2$ singular fibers over $\Delta = s_2 =0$. Since these exceptional $\bP^1$'s are by definition extremal generators of the Mori cone of $X_5$, they necessarily have non-trivial intersection with $\hat D_{0} \cong \text{Bl}_{\{ \Delta = s_2 =0\}} B$, and thus it follows that the central charges of the primitive 3D BPS particles corresponding to M2 branes wrapping these exceptional $\bP^1$'s carry non-zero $\au(1)_{\text{KK}}$ charge.
% $\lambda_4 = 1/((\lambda_0 \lambda_1 \lambda_2 u), \lambda_3 = \lambda_2/e_2, \lambda_2 = v/(e_4 u ),  \lambda_0 = 1/v$ and defining $ \tilde t = t/v, \tilde w = w/ (e_2 u ) , \tilde e_3 = e_4 u e_2 e_3/v , \tilde b_1 = v b_1 /( e_4 u )$, we learn that $\hat D_0$ wraps some components of exceptional curves:
%	\begin{align}
%		\hat D_0~:~ \{ \tilde t+  (d_0 \tilde b_1 + e_1 s_5 \tilde e_3 + s_6 ) \tilde w =0 \}  ~~ \subset ~~ \bP^1_{[\tilde e_3:d_0]} \times \bP^1_{[\tilde w:\tilde t]}.
%	\end{align}
These 3D BPS particles' KK central charges are in fact small in comparison to their Coulomb branch central charges (see \cref{eqn:KKcharge}), so there is no mass hierarchy between BPS particles with zero KK charge and BPS particles with non-zero KK charge. While the presence of light particles with non-trivial KK charge clearly does not alter the 4D spectrum, this does imply that we cannot simply read off the 4D kinematics from the naive 3D BPS spectrum in the usual manner.\footnote{For example, in this situation the one-loop Chern--Simons couplings receive contributions from the KK central charges. Thus, it is not possible to determine the form of the Chern--Simons couplings from purely 3D field-theoretic considerations, as the KK tower now participates in the light spectrum \cite{Cvetic:2012xn,Grimm:2013oga}.} Thus, we need to be able to use the geometry of $X_5$ to precisely specify the contribution of the KK charges to the 3D Coulomb branch dynamics, so that we can disentangle the KK spectrum from our analysis. We discuss how to do this below.

We first focus on the zero section. Rescaling the homogeneous coordinates appropriately and defining $w \rightarrow  e_2 u w, e_4 \rightarrow e_4/ u$, we can describe the restriction of the zero section to the locus $s_2=0$ by the complete intersection
	\begin{align}
	\label{s2families}
		\hat D_0|_{s_2=0}~:~ s_1 e_1 e_2 e_3  e_4 =   e_4 t + (s_6 v + b_1 d_0 v^2 )  w + s_5 e_1 e_2 e_3  e_4^2  w =0\,.
	\end{align}
In the above equation, the monomial $s_1 e_1 e_2 e_3 e_4$ vanishes on the union of three components of the discriminant locus restricted to $s_2 =0$, namely the union of the loci $s_1 = 0$, $d_0e_1=0$, and $b_1e_2 e_3=0$:
	\begin{align}
		 f|_{s_2=0}&=\cO(\varepsilon^0),~~~~g|_{s_2=0}=\cO(\varepsilon^0),~~~~	\Delta|_{s_2=0}= \varepsilon^9 (b_1 e_2 e_3)^4 (d_0 e_1)^3 s_1^2 + \cO( \varepsilon^{10})\,.
	\end{align}
We thus expect the intersection of $\hat D_0 \cap \hat Y$ with each of these codimension one loci to describe the following collection of exceptional $\bP^1$s:

Setting $e_1 =0$ and working in the affine open set $s_1 e_2 e_3  \ne 0$, after appropriately rescaling and defining new variables $e_4 \rightarrow v e_4/(e_2 e_3), t \rightarrow v t$ we find
	\begin{align}
		\hat D_0|_{s_2 =e_1 =0} ~:~e_4 t + b_1 d_0 w + s_6 e_4 w =0
	\end{align}
in a subspace birational to $\bP^1 \times \bP^1$ with homogeneous coordinates $[e_4: d_0] [ w: t ]$. Similarly, by setting $e_3 =0$ and working in the open set $s_1 e_1 e_2\ne 0$, after making some appropriate coordinate redefinitions we find an identical local description in a subspace birational to $\bP^1 \times \bP^1$ with homogeneous coordinates $[e_4 :b_1] [w:t]$. Finally, by setting $s_1 =0$ and working in the open set $e_1 e_2 e_3=0$, and making the coordinate redefinitions $d_0 \rightarrow d_0/e_1, b_1 \rightarrow b_1/(e_2 e_3),e_4 \rightarrow e_4/(e_1 e_2 e_3)$, we may write
	\begin{align}
		\hat D_0|_{s_2= s_1 =0} ~:~ e_4 t + s_5 w e_4^2 + s_6 w e_4 v + d_0 b_1 v^2 w =0
	\end{align}
in a subspace birational to a Hirzebruch surface\footnote{$\F_{n}$ denotes the Hirzebruch surface $\bP(\sO \oplus \sO(n) ) \rightarrow c$, with divisor classes $f, c$ satisfying $f^2 = 0, f \cdot c = 1, c^2 = -n$.} $\F_1$ with coordinates $[e_4:v] [w:t] \cong [\lambda_0 e_4 : \lambda_0 v] [\lambda_5 w : \lambda_0 \lambda_5 t ]$. In all three cases, we see that the zero section wraps a full rational curve, i.e., the elliptic fiber splits as $F \to F' + F''$, with $F', F'' \cong \bP^1$. As discussed, since these exceptional curves $F'$ lie in $\hat D_0$, the primitive BPS particles corresponding to M2 branes wrapping $F'$ carry non-trivial KK charge given by $\hat D_{\bar 0} \cdot F' \ne 0$.

Having specified the geometric origin of the light BPS particles in the 3D spectrum with non-trivial $\au(1)_{\text{KK}}$ charges, we next turn our attention to local matter charged under $\asu(3) \oplus \asu(2) \oplus \au(1)$.

%Setting $\lambda_5 = v/t, \lambda_4  b_1 d_0 v/u, \lambda_3 = \lambda_2/e_2, \lambda_2 =1/b_1, \lambda_1 = 1/d_0, \lambda_0 = 1/v$ and defining $\tilde e_5 = t e_5/v $, we find that the generating section may be expressed as the set of points over the locus
%	\begin{align}
%		\hat D_1 ~:~  \{ \tilde e_5 - s_2 = 0\}.
%	\end{align}
%Clearly the intersection $\hat D_0 \cap \hat D_1$ occurs over the locus $s_2 = 0$.

\subsection{\texorpdfstring{$(\textbf{1},\textbf{1})_{w_{\bar 1}}$}{}}

The pure $\au(1)$ charged matter loci were found in \cite{Raghuram:2019efb} to corresponds to singular fibers of $X_0$ over the following components of the discriminant locus $\Delta = 0$:
	\begin{align}
	\begin{split}
		\text{$(\textbf{1},\textbf{1})_{1}$} ~&:~  V_{q=1} = 0 \\
			\text{$(\textbf{1},\textbf{1})_{2}$}  ~&:~  s_1 = s_2 = 0\,.
	\end{split}
	\end{align}
In the above equation
\begin{align}
	\begin{split}
		V_{q=1} &= V \backslash ( \{ s_1 = s_2 = 0\} \cup \{ b_1  = s_2 = 0\} \cup \{ d_0  = s_2 = 0\} )\\
		V &= \left\{ \begin{array}{c} - d_0  b_1 s_1^2 s_2 ( d_1 s_2^2 - 2 d_0 s_2 s_5 + 2 d_0  s_1 s_6) + d_0 ^3 b_1^2 s_1^4 \\
		+ d_2 s_2^6 - s_2^2 ( s_2 s_5 - s_1 s_6 ) ( d_1 s_2^2 - d_0 s_2 s_5 + d_0  s_1 s_6 )\\
		=- b_1  d_1 s_1 s_2^3 + 2 d_0  b_1 s_1 s_5 s_2^2 - 3 d_0 b_1  s_1^2 s_2 s_6\\
		+ 2 b_1^2 d_0^2 s_1^3 + s_2^4 s_8 - s_2^3 s_5 s_6 + s_1 s_2^2 s_6^2 \end{array} \right\}\,.
	\end{split}
	\end{align}
%	\begin{align}
%	\begin{split}
%		V_{q=1} &= V \backslash ( \{ s_1 = s_2 = 0\} \cup \{ b_1 e_2 e_3 = s_2 = 0\} \cup \{ d_0 e_1 = s_2 = 0\} )\\
%		V &= \left\{ \begin{array}{c} - d_0 e_1 b_1 e_2 e_3 s_1^2 s_2 ( d_1 s_2^2 - 2 d_0 e_1 s_2 s_5 + 2 d_0 e_1 s_1 s_6) + (d_0 e_1)^3 (b_1 e_2 e_3)^2 s_1^4 \\
%		+ d_2 s_2^6 - s_2^2 ( s_2 s_5 - s_1 s_6 ) ( d_1 s_2^2 - d_0 e_1 s_2 s_5 + d_0 e_1 s_1 s_6 )\\
%		=- b_1 e_2 e_3 d_1 s_1 s_2^3 + 2 d_0 e_1 b_1 e_2 e_3 s_1 s_5 s_2^2 - 3 d_0 e_1 b_1 e_2 e_3 s_1^2 s_2 s_6\\
%		+ 2 (b_1 e_2 e_3)^2 (d_0 e_1)^2 s_1^3 + s_2^4 s_8 - s_2^3 s_5 s_6 + s_1 s_2^2 s_6^2 \end{array} \right\}\,.
%	\end{split}
%	\end{align}
To see that the fibers in the resolution $X_5$ degenerate as expected over the total transforms of both loci described above, we work in the open set $ d_0 e_1 b_1 e_2 e_3\ne 0$. For the charge $q=1$ locus, we further restrict to the open set $s_2 \ne 0$ and solve for $d_2, s_8$ in the equation $V_{q=1} = 0$. Doing so, we find that
	\begin{align}
	\begin{split}
		&F|_{V_{q=1} = 0 } = F_{q=1}^+ + F_{q=1}^-
	\end{split}
	\end{align}
%Making the redefinitions $ u \to  u/ (e_2 e_1), w \to  w/ (e_3 e_1 )$ we may write
%	\begin{align}
%	\begin{split}
%		&F_{\{ V_{q=1} = 0\} } = F_{q=1}^+ + F_{q=1}^-
%	\end{split}
%	\end{align}
%where locally
%	\begin{align}
%	\begin{split}
%		F_{q=1}^+ ~&:~ \{ (\tilde b_1 \tilde d_0  s_1^2- s_2 s_6 s_1+ s_2^2 s_5) \tilde e_4\tilde w+s_2^2 \tilde t = 0\} \\
%		F_{q=1}^- ~&:~ \{( \tilde b_1 d_1s_2^3-\tilde b_1 \tilde d_0 s_5 s_2^2+\tilde b_1 \tilde d_0 s_1 s_6 s_2-\tilde b_1^2 \tilde d_0^2 s_1^2) \tilde e_4 \tilde e_5^2 \tilde w^2\\
%		&~~~~~(s_6 s_2^3-2 \tilde b_1 \tilde d_0 s_1 s_2^2) \tilde e_4  \tilde u\tilde e_5 \tilde w+\tilde b_1 \tilde d_0s_2^2  \tilde e_5^2 \tilde t \tilde w+s_2^4 \tilde e_4  \tilde u^2=0 \}
%	\end{split}
%	\end{align}
%in the ambient space
%	\begin{align}
%	\bP^2_{[ \tilde e_4\tilde u:(\tilde e_5 \tilde t - \tilde e_4 \tilde u s_1)/s_2:\tilde e_4\tilde e_5\tilde w]} \setminus \{ \tilde u =\tilde e_5 \tilde w  = 0\} \cup \{ \tilde w = \tilde t =0\}.
%	\end{align}
%The intersections $\hat D_0 \cap F_{q=1}^+$, $\hat D_1 \cap F_{q=1}^-$ are distinct points, whereas the intersections $\hat D_0 \cap F_{q=1}^-, \hat D_1 \cap F_{q=1}^+$ are empty.
We learn also that the zero section intersects $F_{q=1}^{+}$ in a point, while the generating section intersects $F_{q=1}^{-}$ in a point.
%	\begin{align}
%		\hat D_0 \cdot F_{q=1}^+ =1,~~~~ \hat D_1 \cdot F_{q=1}^- = 1.
%	\end{align}
Moreover, it is straightforward to verify that the intersection $F_{q=1}^+ \cap F_{q=1}^-$ consists of two distinct points, hence we see the fibers over $V_{q=1} = 0$ are type I$_2$, as anticipated, where $F_{q=1}^+$ as the affine component of the affine $\asu(2)$ Dynkin diagram describing the intersection structure of the enhancement. The enhancement of the elliptic fiber to an I$_2$ Kodaira fiber indicates the presence of charged matter and since this enhancement occurs away from the components of the discriminant locus associated to nonabelian gauge symmetry, we conclude this matter can only be charged under the $\au(1)$ of the 4D gauge algebra. The $\au(1)$ charges can be checked by computing intersection products $\hat D_{\bar 1 } \cdot F_{q=1}^{\pm{}}$.

We treat the charge $q=2$ locus in a similar manner. Setting $s_1 = s_2 =0$, \cref{eq:X5hyp} splits into two components, $p_{5,2}(s_1 =s_2 =0 ) = e_5 t = 0$. It follows that the fibers over this locus degenerate:
	\begin{align}
		F|_{s_1 = s_2 =0 } = F_{q=2}^{+} + F_{q=2}^-, ~~~~ F_{q=2}^+ = F|_{ s_1 = s_2 = t =0},~~~~F_{q=2}^-= F|_{s_1= s_2 = e_5 =0}\,.
	\end{align}
Notice that the above definitions imply that the zero section wraps the exceptional curve $F_{q=2}^-$.
%, that is (using the adjunction formula and the fact that the exceptional curve $F_{q=2}^{-} \cdot  F_{q=2}^{-}|_{\hat D_0} = -1$)
%	\begin{align}
%		\hat D_0 \cdot F_{q=2}^- = K_{\hat D_0} \cdot F_{q=2}^-|_{\hat D_0} =-1.
%	\end{align}
To study the geometry of the degenerated fibers more explicitly, we use local descriptions of the fibers in affine open sets, as we now describe.

For $F_{q=2}^{+}$,
%we set $\lambda_5 = 1/(w \lambda_0 \lambda_1 \lambda_3 \lambda_4)$, $\lambda_{i=1,3,4} = e_i$ and
Making the redefinitions $ u \to u/(e_1 e_2 e_4 ),  w \to w/(e_1 e_3 e_4 )$,
%so that
%	\begin{align}
%	\begin{split}
%		F_{q=2}^+ ~:~ &\{ \tilde b_1 d_1 e_5 v \tilde w+\tilde b_1 d_2 e_5^2 \tilde w^2+\tilde b_1 \tilde d_0 v^2+e_5 s_8 \tilde u \tilde w+s_5 \tilde u^2+s_6 \tilde u  v = 0\} \\
%		~&\subset~ \bP^2_{[\tilde u :v:\tilde we_5]} \setminus \{[0:1:0]\}.
%	\end{split}
%	\end{align}
we can describe $F_{q=2}^+$ locally as a smooth conic in $\bP^2$ with homogeneous coordinates $[ u :v: we_5]$ with the point $[0:1:0]$ removed. For $F_{q=2}^-$,
%we set $\lambda_4 = 1/(u \lambda_0 \lambda_1 \lambda_2), \lambda_{i=1,2,3} = e_i$ and define
we make the redefinitions $ w \to  e_2  e_4 uw/e_3, e_4 \to e_4/( e_1 e_2 u)$ so that
%	\begin{align}
%		F_{q=2}^- ~:~ &\{ \tilde e_4 t+(\tilde b_1 \tilde d_0 v^2 +s_5\tilde e_4^2 +s_6 \tilde e_4 v) \tilde w =0\} ~ \subset ~{\F_1}_{[\tilde e_4:v] [ \tilde w:t] }
%	\end{align}
$F_{q=2}^{-}$ can be described locally as a sum of base and fiber curves in ${\F_1}$ with homogeneous coordinates $[ e_4 :v] [w:t] \cong [  \lambda_0 e_4 :  \lambda_0 v] [\lambda_5  w: \lambda_0 \lambda_5 t]$.

The generating section $\hat D_1$ intersects $F_{q=2}^-$ in a point, but does not intersect $F_{q=2}^+$.
%	\begin{align}
%		\hat D_1 \cdot F_{q=2}^{-} = 1,~~~~ \hat D_1 \cdot F_{q=2}^{+} =0.
%	\end{align}
Moreover, the intersection $F_{q=2}^+ \cap F_{q=2}^-$ consists of two distinct points, hence the fibers over the charge $q=2$ locus are type I$_2$. Once again, the charges can be determined explicitly by computing the intersection products $\hat D_{\bar 1} \cdot F_{q=2}^{\pm{}}$.

\subsection{\texorpdfstring{$(\textbf{1},\textbf{2})_{w_{\bar 1}}$}{}}
\label{SU2matter}
\begin{figure}
	\begin{center}
	\includegraphics{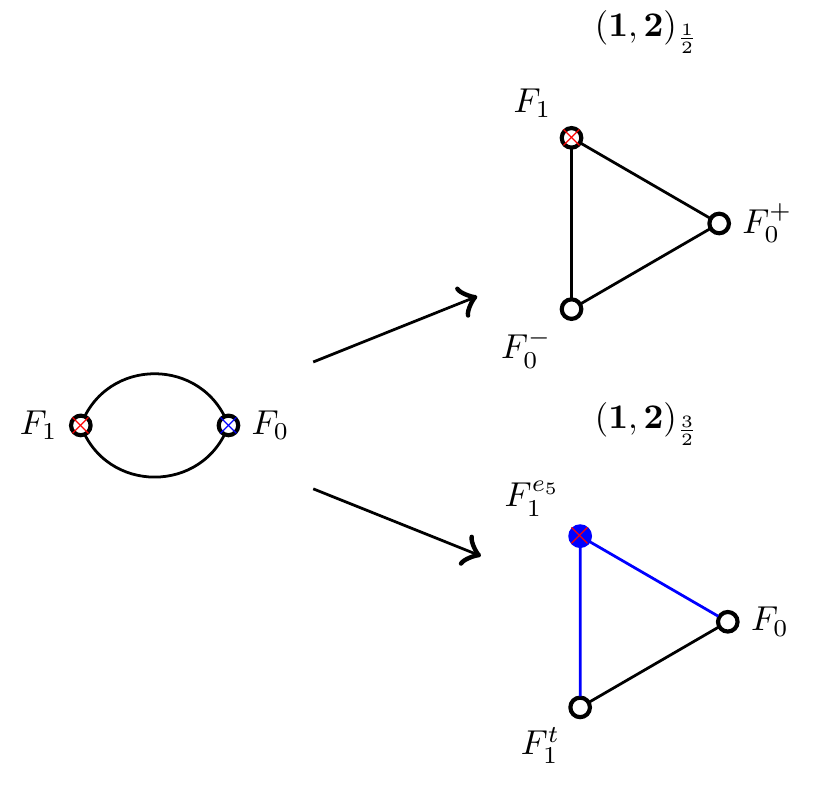}
	\end{center}
	\caption{Schematic depiction of the degeneration of the $\text{I}_2$ Kodaira fibers over the codimension-two loci corresponding to local matter transforming in the representations $(\mathbf 1,\mathbf 2)_{1/2}$ (over $d_0e_1=\Delta_{(a)}=0$) and $(\mathbf 1,\mathbf 2)_{3/2}$ (over $d_0 e_1 = s_2 =0$). Note that a blue (respectively, red) $\times$ in the center of a node corresponding to an irreducible $\bP^1$ component of the singular elliptic fiber indicates that $\hat D_0$ (respectively, $\hat D_1$) intersects the said $\bP^1$ in a point. The blue node is wrapped entirely by $\hat D_0$.}
	\label{su2matter}
\end{figure}

The residual discriminant locus intersecting the codimension-one I$_2^{}$ locus $d_0 = 0$ is
	\begin{align}
	\begin{split}
	\label{I2resid}
	f&= \left(4 b_1 d_1 s_2-s_6^2\right){}^2 + \cO(d_0)\\
	g&=\left(4 b_1 d_1 s_2-s_6^2\right){}^3 + \cO(d_0)\\
		\Delta &=- \frac{1}{16} d_0^2    b_1^3 s_2 \Delta_{(a)}\left(4 b_1 d_1 s_2-s_6^2\right){}^2 + \cO(d_0^3)\,,
	\end{split}
	\end{align}
where
	\begin{align}
	\begin{split}
		\Delta_{(a)} &= b_1^2 d_1^3 s_1^2+b_1 d_1^2 s_2 s_5^2-b_1 d_1^2  s_1 s_5 s_6-2 b_1 d_1^2 s_1 s_2 s_8\\
		&~~-2 b_1 d_2 d_1  s_2^2 s_5+3 b_1 d_2 d_1  s_1 s_2 s_6+b_1 d_2^2 s_2^3+d_1 s_2^2 s_8^2\\
		&~~+d_1 s_1 s_6^2 s_8-d_1 s_2 s_5 s_6 s_8-d_2 s_1 s_6^3+d_2 s_2 s_5 s_6^2-d_2 s_2^2 s_6 s_8\,.
	\end{split}
	\end{align}
thus according to Tate's algorithm \cite{KatzEtAlTate} we can anticipate the presence of localized matter charged under $\asu(2)$ for all codimension-two loci except for $d_0 =4 b_1 d_1 s_2-s_6^2=0$. According to the analysis of \cite{Raghuram:2019efb}, we anticipate the following representations
	\begin{align}
	\begin{split}
		(\textbf{1},\textbf{2})_{\frac{1}{2}} ~&:~ d_0 = \Delta_{(a)} =0\,, \\
		(\textbf{1},\textbf{2})_{\frac{3}{2}} ~&:~  d_0 = s_2 =0\,.
	\end{split}
	\end{align}
In this subsection we focus on the local matter away from the collision of I$_2$ and I$_3^\text{split}$ fibers; see \cref{su2matter}. Bifundamental matter, which is localized at the collision of the I$_2$ and I$_3^{\text{split}}$ fibers, is discussed in \cref{KV}.

Judging from \cref{I2resid}, we expect the I$_2^{}$ fiber to enhance to I$_3^\text{split}$ over $\Delta_{(a)} = 0$. Unfortunately, an explicit algebraic description of the degeneration of the elliptic fibers over $\Delta_{(a)}=0$ is difficult to obtain so we simply verify that the discriminant $\Delta_{F_0}$ of the cubic polynomial defining $F_0$ in \cref{eq:su2comp} vanishes on this locus. In particular, we find
	\begin{align}
		\Delta_{F_0} = (\cdots) \Delta_{(a)}\,.
	\end{align}
In the above equation, $ (\cdots)$ denotes factors of $b_1,e_2,e_3$ responsible for the degeneration leading to bifundamental matter; we analyze this degeneration in detail in \cref{KV}. Schematically, we write
	\begin{equation}
	\label{eq:Deltaa}
		F_0 |_{\Delta_{(a)} = 0 } = F_0^+ + F_0^-\,.
	\end{equation}
Unfortunately, without an explicit algebraic description, it is not precisely clear how $\hat D_0, \hat D_1$ intersect the irreducible components $F_0^{\pm{}}$ of the elliptic fiber over this locus. However, given that the BPS particles corresponding to M2 branes wrapping $F_0^{\pm{}}$ have much larger KK central charges than Coulomb branch central charges, we can further anticipate that $\hat D_0$ only intersects either $F_0^+$ or $F_0^-$ in a point, rather wrapping the entire curve.

%We can anticipate that $F_1$ degenerates over the locus $\{s_2=0\}$ by computing the discriminant of the polynomial in \cref{F1}:
%	\begin{align}
%	\label{F0codim2}
%		 \Delta_{F_1} =  (b_1 e_2 e_3)^2 s_2.
%	\end{align}
Turning next to $ s_2 = 0$ and keeping in mind $p_{5,2}(e_1=s_2=0) = t e_5 = 0$, we find
	\begin{align}
		F_1|_{s_2 =0} = F_{1}^t + F_{1}^{e_5}\,.
	\end{align}
%In order to see the above degeneration, we observe that the polynomial $p_{5,2}$ appearing in \cref{eq:X5hyp} factors:
%	\begin{align}
%		p_{5,2}(e_1=s_2=0) = t e_5 = 0
%	\end{align}
The above equation describes a pair of curves intersecting transversally. A local algebraic description of $F_{1}^{t}$ can be obtained by restricting the description of $F_1$ (see \cref{eq:su2comp}) to the locus $s_2=0$. Making appropriate coordinate redefinitions, $F_{1}^{e_5}$ can be described locally as a sum of a base and fiber curve in subspace birational to $\F_1$ with coordinates
%	\begin{align}
%		F_{1}^-  ~:~ \{ \tilde u^2 \tilde t + ((b_1e_2 e_3)d_0 + s_6 \tilde u) \tilde w = 0\} ~~ \subset ~~  {\F_1}_{[\tilde u: d_0][\tilde w:\tilde t]}
%	\end{align}
$[u:d_0] [w:t] \cong [\lambda_1 u: \lambda_1 d_0][\lambda_1 \lambda_5 w:\lambda_5 t]$. A straightforward computation shows that the pairwise intersections $F_{0} \cap F_{1}^{t}, F_0 \cap F_{1}^{e_5}, F_{1}^t \cap F_{1}^{e_5}$ are distinct points and hence the degeneration over $s_2 =0$ enhances the I$_2^{}$ fiber to an I$_3^\text{split}$ fiber, indicating matter transforming in the $\textbf{2}$ of $\asu(2)$

Note that $\hat D_0$ wraps the entire curve $F_1^{e_5}$, and both $\hat D_0 , \hat D_1$ intersect $F_1^t$ in distinct points.
%	\begin{align}
%		\hat D_0 \cdot F_1^{-} = -1,~~~~ \hat D_0 \cdot F_1^+ = 1,~~~~\hat D_1 \cdot F_1^{+} = 1.
%	\end{align}
%Using the fact that $F_1$ is a fiber in $\hat D_3$, we infer $\hat D_3 \cdot F_1 = -2$ and hence $\hat D_3 \cdot F_1^{+} = -1$. This in turn implies
%	\begin{align}
%		\sigma_1^I \hat D_i \cdot F_1^{+} &= ( - \hat D_0 + \frac{1}{2} \hat D_3 ) \cdot F_1^- =- \frac{3}{2}
%	\end{align}
%as claimed.

\subsection{\texorpdfstring{$(\textbf{3},\textbf{1})_{w_{\bar 1}}$}{}}
\begin{figure}
	\begin{center}
	\includegraphics{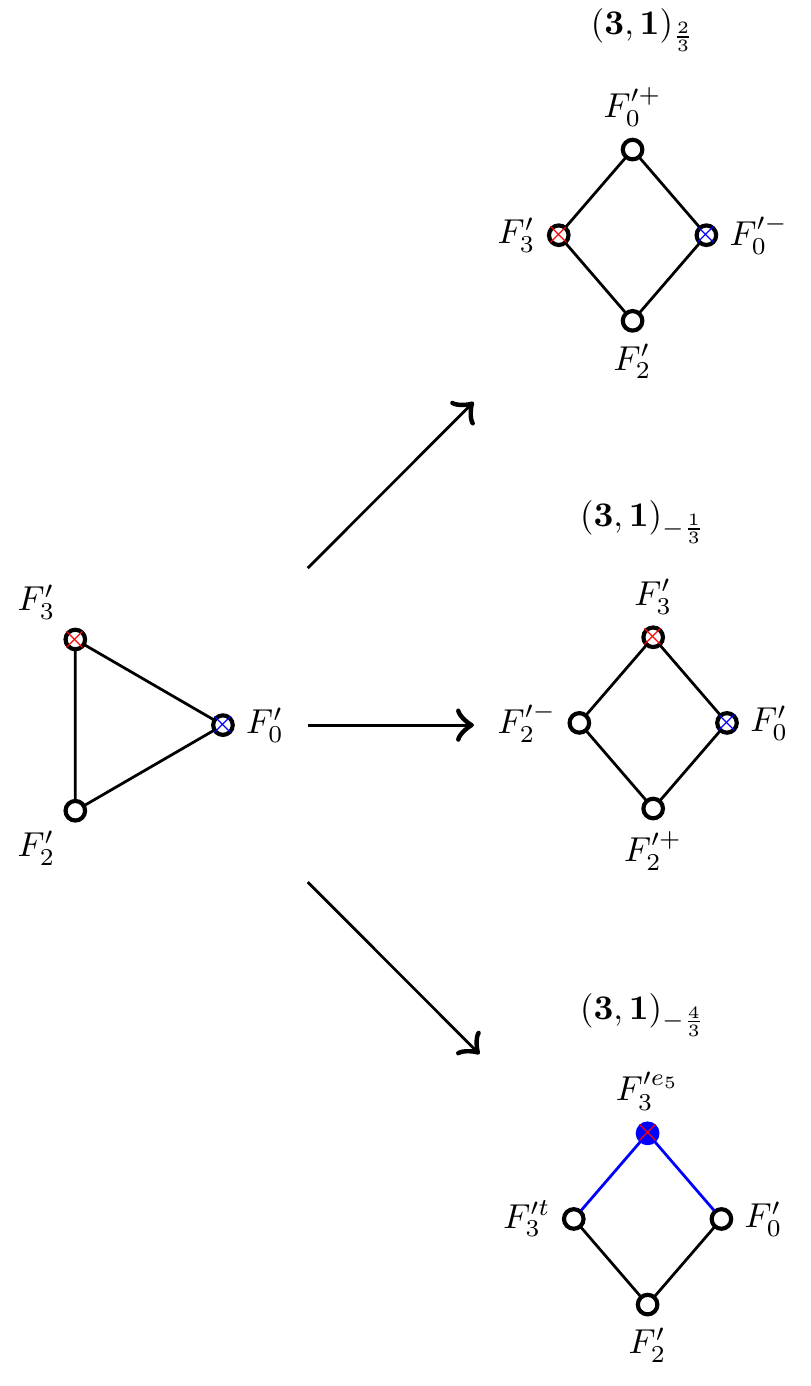}
	\end{center}
\caption{Schematic depiction of the degeneration of the $\text{I}_3^{\text{split}}$ Kodaira fibers over the codimension-two loci corresponding to local matter transforming in the representations $(\mathbf 3,\mathbf 1)_{2/3}$ (over $b_1 e_2 e_3=s_8 s_2^2 - s_5 s_2 s_6 + s_1 s_6^2 =0$), $(\mathbf 3,\mathbf 1)_{-1/3}$ (over $b_1 e_2 e_3 = d_2 s_6^2 - d_1 s_6 s_8 + d_0 e_1 s_8^2 =0$), and $(\mathbf 3,\mathbf 1)_{-4/3}$ (over $b_1 e_2 e_3 = s_2 =0$). Note that a blue (respectively, red) $\times$ in the center of a node corresponding to an irreducible $\bP^1$ component of the singular elliptic fiber indicates that $\hat D_0$ (respectively, $\hat D_1$) intersects the said $\bP^1$ in a point. The blue node in the bottom-most graph is wrapped entirely by $\hat D_0$.}
\label{su3matter}
\end{figure}

The residual discriminant intersecting the I$_3^\text{split}$ locus $b_1  = 0$ is
	\begin{align}
		\begin{split}
			f&= - \frac{s_6^4}{48} + \cO(b_1^2)\,, \\
			g&= \frac{s_6^6}{864} + \cO(b_1^4)\,, \\
			\Delta &= - \frac{1}{16} b_1^3 d_0^2 s_2 s_6^3 (s_8 s_2^2+s_1 s_6^2-s_2 s_5 s_6)(d_0 e_1 s_8^2+d_2 s_6^2-d_1 s_8 s_6)\\
			 &~~~+ \cO(b_1^4)\,.
		\end{split}
	\end{align}
Following Tate's algorithm, we expect to see localized matter charged under $\asu(3)$ at all codimension-two components of the discriminant locus except for $ b_1 = s_6 =0$.\footnote{The locus $b_1 = s_6 = 0$ is an IV locus, and does not support any charged matter \cite{Raghuram:2019efb}.} The analysis of \cite{Raghuram:2019efb} anticipates the following $\asu(3)$ charged matter representations (see \cref{su3matter}):
	\begin{align}
	\begin{split}
		(\textbf{3}, \textbf{1})_{ \frac{2}{3} } ~&:~  b_1= s_8 s_2^2 - s_5 s_2 s_6 + s_1 s_6^2= 0 \\
		(\textbf{3}, \textbf{1})_{- \frac{1}{3} } ~&:~  b_1 = d_2 s_6^2 - d_1 s_6 s_8 + d_0e_1 s_8^2 = 0  \\
		(\textbf{3}, \textbf{1})_{- \frac{4}{3} } ~&:~  b_1 = s_2 = 0 \,.
	\end{split}
	\end{align}
%To study the behavior of $F_0'$, we first note that the determinant of the polynomial in \cref{F0prime} is
%	\begin{align}
%		\Delta_{F_0'} &=-2 \left(s_8 s_2^2+s_1 s_6^2-s_2 s_5 s_6\right).
%	\end{align}
%Thus we see
Over the locus $ s_8 s_2^2 - s_5 s_2 s_6 + s_1 s_6^2 = 0$, we find that the fibers in the resolution $X_5$ behave as follows:
	\begin{align}
	\label{F0pdegen}
		F_{0}'|_{s_8 s_2^2+s_1 s_6^2-s_2 s_5 s_6 = 0} = F_0'^{+} + F_0'^{-}\,.
	\end{align}
%While the description in \cref{F0prime} is useful for analyzing the local degeneration, in order to study the full intersection structure of the I$_3^\text{split}$ fibers we need to allow the variable $e_3$ to vanish.
%Setting $\lambda_{i=4,5}=e_i, \lambda_2 = 1/ (e_4 u \lambda_0 \lambda_1)$ and
Substituting $w \to  w/(e_4 e_5), e_2 \to e_2/(e_4 u)$ we find in the open set $s_2 \ne 0$ that $F_0'^{\pm{}}$ can be given a local algebraic description
%	\begin{align}
%	\begin{split}
%		F_0'^+ ~&:~ \{ s_2\tilde e_2 +s_6 \tilde w=0 \} ~\subset~ \bP^2_{[e_1 e_3 \tilde e_2 :v:e_1 e_3 \tilde w]}\\
%		F_0'^- ~&:~ \{ e_1 e_3 s_5 s_2 \tilde w-e_1 e_3 s_1 s_6 \tilde w+e_1 e_3 s_1 s_2\tilde e_2 +s_2^2 v= 0\} ~ \subset ~ \bP^2_{[e_1 e_3 \tilde e_2 :v:e_1 e_3 \tilde w]}
%	\end{split}
%	\end{align}
as a pair of lines in $\bP^2$ with coordinates $[e_1 e_3 e_2 :v:e_1 e_3 w]$ and the point $[1:0:0]$ removed. Given this description, it is straightforward to verify that the pairwise intersections $F_{0}'^{-} \cap F_2', F_0'^{+} \cap F_3', F_{0}'^+ \cap F_{0}'^-$ are distinct points, hence the fiber enhances to I$_4^\text{split}$ indicating matter in the $\textbf{3}$. Note that $\hat D_0$ intersects $ {F'_{0}}^-$ in a point along $w=0$, whereas $\hat D_0$ does not intersect $ {F'_{0}}^+$.
%	\begin{align}
%		\hat D_0 \cdot {F'_0}^- = 1,~~~~ \hat D_0 \cdot {F'_0}^+ = 0.
%	\end{align}
%Using the fact that
%%$( \hat D_3 ,\hat D_4 , \hat D_5) \cdot {F'_0}^+ =(1,0,1),
%$(\hat D_4 , \hat D_5) \cdot {F'_0}^- =  (1,0) $ we find the $\U(1)$ charge of ${F'_0}^-$ is given by
%	\begin{align}
%		\sigma_1^I \hat D_I \cdot {F'_0}^- &=(-\hat D_0 + \frac{1}{3} \hat D_4 ) \cdot {F'_{0}}^-=- \frac{2}{3}.
%	\end{align}

Similarly, over the locus $ d_0 e_1 s_8^2+d_2 s_6^2-d_1 s_8 s_6 = 0$ the curve $F_2'$ degenerates as follows:
	\begin{align}
		F_2'|_{ d_0 e_1 s_8^2+d_2 s_6^2-d_1 s_8 s_6 = 0} = F_2'^+ + F_2'^-\,.
	\end{align}
%Setting $\lambda_{i=1,4,5} = e_i, \lambda_3 = 1/(e_4 e_5 w \lambda_0 \lambda_1)$ and
Making the coordinate substitutions $e_3 \to  e_3/ (e_4 e_5 w),  u \to u/e_4$ we find that in the open set $s_6 s_8 \ne 0$ that $F_2'^{\pm{}}$ can be described locally as
%	\begin{align}
%	\begin{split}
%		F_2'^+~&:~ \{ s_8 e_1 \tilde e_3  +s_6 v =0 \}~\subset ~ {\F_1}_{[v:e_1 \tilde e_3] [\tilde u:b_1]}\\
%		F_2'^- ~&:~ \{ b_1 d_0 s_8 v+\tilde e_3 b_1 d_2 s_6+s_8 s_6 \tilde u = 0\}~\subset ~ {\F_1}_{[v:e_1 \tilde e_3] [\tilde u:b_1]},
%	\end{split}
%	\end{align}
as a pair of lines in $\F_1$ (the homogeneous coordinates for this $\F_1$ are described in \cref{sec:SUgauge}). The pairwise intersections $F_2'^+ \cap F_0', F_2'^- \cap F_3', F_2'^+ \cap F_2'^-$ are distinct points, and thus we find an enhancement to an I$_4^\text{split}$ fiber again indicating matter in the $\textbf{3}$ of $\SU(3)$.
%Using the fact that $(\hat D_4, \hat D_5) \cdot F_2'^+= (-1,0)$, the $\U(1)$ charge is given by
%	\begin{align}
%		\sigma_1^I \hat D_I \cdot F_2'^+ =  \frac{1}{3} \hat D_4  \cdot F_2'^+ = - \frac{1}{3}.
%	\end{align}

Finally we come to the locus $ s_2 =0$. The component $F_3'$ degenerates as follows:
	\begin{align}
	\label{eqn:F3prime}
		F_3'|_{s_2 =0 } =  F_3'^t + F_3'^{e_5}\,.
	\end{align}
%Setting $\lambda_1 = 1/d_0, \lambda_0 = 1/v$ and d
Making coordinate redefinitions $t \to vt, u  \to d_0 v u,  w \to d_0 v w$, we find that the irreducible components of the above degeneration can be described by
	\begin{align}
	\begin{split}
		F'^{t}_3 &= \{ t =  b_1  w+e_2 e_4  t  u^2+e_4 s_6  u  w =0 \}~\subset~ \bP^1_{[e_4  u: b_1]} \times \bP^1_{[ w: t]}\\
		 F'^{e_5}_3 &= \{e_5 =b_1  w+e_2 e_4  t  u^2+e_4 s_6  u  w  = 0\}~\subset ~ {\F_1}_{[e_4  u: b_1][ e_2 :e_4 e_5  w]}
	\end{split}
	\end{align}
where $[e_4  u: b_1] [ e_2:e_4 e_5  w] \cong [\lambda_2 e_4  u:\lambda_2 b_1] [\lambda_3 \lambda_2^{-1}e_2 : \lambda_3 e_4 e_5  w ]$. The pairwise intersections $F_3'^{t} \cap F_2', F_3'^{e_5} \cap F_0', F_3'^{t} \cap F_3'^{e_5}, F_0' \cap F_2'$ are all distinct points, hence we again find an enhancement of type I$_4^\text{split}$.

Note that $\hat D_0$ wraps $F_3'^{e_5}$ and $\hat D_1$ intersects $F_3'^{e_5}$ in a point $\hat D_0 \cap \hat D_1$ away from the intersection $F_3'^{t} \cap F_3'^{e_5}$.
%	\begin{align}
%		\hat D_0 \cdot F_3'^- = -1,~~~~ \hat D_1 \cdot F_3'^- = 1.
%	\end{align}

\subsection{\texorpdfstring{$(\textbf{3},\textbf{2})_{w_{\bar 1}}$}{}}
\label{KV}
In this section we describe the collision of I$_2^{}$, I$_3^\text{split}$ fibers over the codimension-two locus $b_1  = d_0 = 0$:
	\begin{align}
	\label{eqn:32fibers}
		F|_{  b_1 e_2 e_3=d_0 e_1 =0 } = F_{00} + F_{02} + F_{10} + F_{12} + F_{13}\,,
	\end{align}
where $F_{ab}$ denotes the curve obtained by setting $e_a = e_b = 0$ for $a,b \ne 0$, or $d_0,b_1 =0$ for (resp.) $a,b=0$.

\begin{figure}
	\begin{center}
	\includegraphics{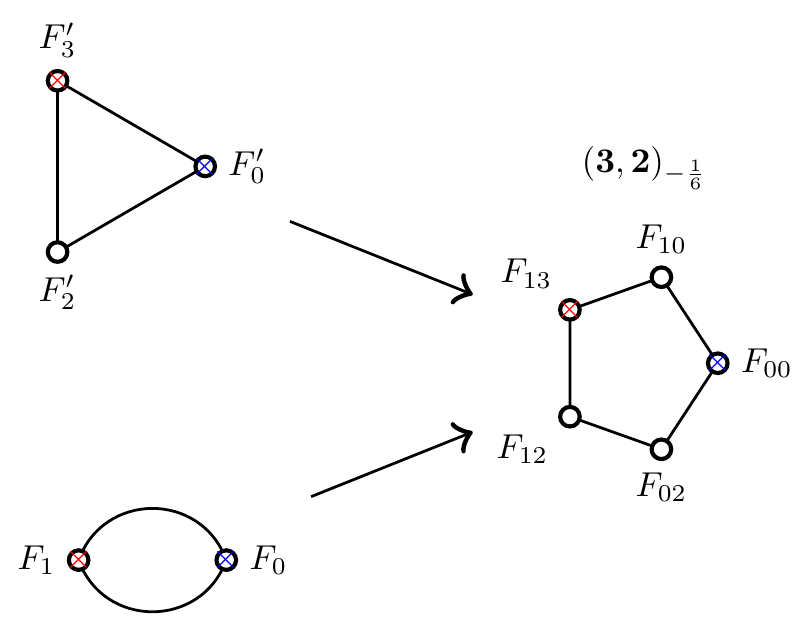}
	\end{center}
	\caption{Schematic depiction of the collision of $\text{I}_2$ and $\text{I}_3^{\text{split}}$ Kodaira fibers over the codimension-two locus corresponding to local matter transforming in the representation $(\mathbf 3,\mathbf 2)_{-1/6}$ (over $d_0 e_1 = b_1 e_2 e_3=0$). Note that a blue (respectively, red) $\times$ in the center of a node corresponding to an irreducible $\bP^1$ component of the singular elliptic fiber indicates that $\hat D_0$ (respectively, $\hat D_1$) intersects the said $\bP^1$ in a point.}
	\label{su3su2matter}
\end{figure}

The analysis of \cite{Raghuram:2019efb} indicates that $\asu(3) \oplus \asu(2)$ charged matter is localized at the collision of the $\text{I}_2, \text{I}_3^{\text{split}}$ singularities (see \cref{su3su2matter}):
	\begin{align}
		( \textbf{3}, \textbf{2} )_{\frac{1}{6}} ~:~  d_0  = b_1 = 0\,.
	\end{align}
Working in the affine open set $\lambda_5 = e_5 \ne 0$, we can give the above irreducible components local descriptions as follows. For $F_{00}$,
%set $\lambda_{i=3,4} = e_i, \lambda_2 = 1/(\lambda_0 \lambda_1 e_4 u)$ and
making the redefinitions $ w \to w/(e_3 e_4 e_5 ),  e_2 \to  e_2/(e_4 u)$ leads to the local description of $F_{00}$ as a smooth conic in $\bP^2$ with coordinates $[e_1 e_2 :v:e_1 w]$ with the point $[0:1:0]$ removed.
% so that
%	\begin{align}
%	\begin{split}
%		&F_{00} ~:~\{ e_1 s_8 \tilde w^2+  s_5e_1 \tilde e_2 \tilde w+e_1 s_1 \tilde e_2^2 +s_6 v \tilde w+ s_2 \tilde e_2 v =0 \}\\
%		 ~&~~~~~~~~~\subset ~ \bP^2_{[e_1 \tilde e_2: v: e_1 \tilde w]} \setminus \{ [ 0:1:0]\}.
%	\end{split}
%	\end{align}
For $F_{02}$,
%set $\lambda_{i=3,4} = e_i, \lambda_1 = 1/ (e_3 e_4 e_5 w \lambda_0)$ and
redefining $ u \to e_3 e_5 wu,  e_2 \to e_2/ e_3$ leads to a description as curve of bi-degree $(1,1)$ in $\bP^1 \times \bP^1$ with coordinates $[v:e_1 w][u:b_1]$.
%	\begin{align}
%		F_{02}~:~\{ d_1 b_1  v+d_2 b_1 e_1 \tilde w+ s_8 \tilde u e_1 \tilde w+s_6 \tilde u v = 0\} ~ \subset~ \bP^1_{[ v:e_1 \tilde w]} \times \bP^1_{[\tilde u:b_1]}.
%	\end{align}
For $F_{10}$,
%set $\lambda_4 = e_4, \lambda_2 = 1/(\lambda_0 \lambda_1 e_4 u), \lambda_0 = 1/ v$ and
redefining $ w \to v w/(e_4 e_5),  e_2 \to v e_2/e_4 u $ allows us to describe $F_{10}$ locally as a line in $\bP^2$ with homogeneous coordinates $[e_3 e_2:e_3 w:d_0]$.
%	\begin{align}
%		F_{10} ~:~\{ s_2 \tilde e_2 + s_6 \tilde w = 0 \} ~ \subset ~ \bP^2_{[ e_3 \tilde e_2:e_3 \tilde w:d_0]}.
%	\end{align}
For $F_{12}$,
%set $\lambda_4 = e_4, \lambda_3 = 1/ (\lambda_0 \lambda_1 e_4 e_5 w)$ and
making the redefinitions $ u \to vu/e_4,  e_3 \to ve_3/ (e_4e_5w )$, we find that $F_{12}$ can described as a curve of bi-degree $(1,1)$ in $\F_1$ with homogeneous coordinates $[e_3:d_0][u:b_1]$.
%	\begin{align}
%		F_{12} ~:~ \{ (d_0 + d_1 \tilde e_3) b_1 + s_6 \tilde u = 0\} ~\subset~ {\F_1}_{[ \tilde e_3:d_0] [ \tilde u: b_1]}.
%	\end{align}
Finally, for $F_{13}$, making the redefinitions $e_4 \to d_0 v e_4/u, w \to u w/e_5$ and rescaling the coordinates appropriately, we find that $F_{13}$ can be described by the equation
	\begin{align}
		F_{13} = \{s_2  e_2 e_4 + b_1 w + s_6 e_4 w = 0\} ~\subset ~ \F_{1}
	\end{align}
where the $\F_1$ in the above equation has homogeneous coordinates $[e_4:b_1] [e_2 :e_4 w] \cong [\lambda_2 e_4: \lambda_2 b_1] [ \lambda_2^{-1} \lambda_3 e_2 : \lambda_3 e_4 w]$. %set $\lambda_4 = 1/(\lambda_2 u), \lambda_2 = v d_0/(e_4 u), \lambda_0 = 1/v$ and
%we define $ w \to  d_0 v w/(e_4 e_5), b_1 \to  e_4 u b_1/(d_0 v), e_2 \to d_0 v e_2 /( e_4 u) $ so that $F_{13}$ is given by the local description
%	\begin{align}
%		F_{13} ~:~ \{ s_2  e_2 +  (b_1  + s_6)  w = 0\} ~\subset ~ \C_{( b_1)} \times \bP^1_{[ e_2 : w]}.
%	\end{align}
It is straightforward to verify that the only intersections $F_{00} \cap F_{02}, F_{00} \cap F_{10}, F_{02} \cap F_{12}, F_{10} \cap F_{13}, F_{12} \cap F_{13}$ are distinct points, hence we find that the elliptic fiber is type I$_{5}^\text{split}$.

Note that $\hat D_0$ intersects $F_{00}$ in a point, while $\hat D_1$ intersects $F_{13}$ in a point.
%	\begin{align}
%		\hat D_0 \cdot F_{00} = 1,~~~~\hat D_1 \cdot F_{13} = 1.
%	\end{align}

\section{Codimension three: Yukawa interactions}
\label{sec:Yukawa}

\begin{figure}
	\begin{center}
	\includegraphics{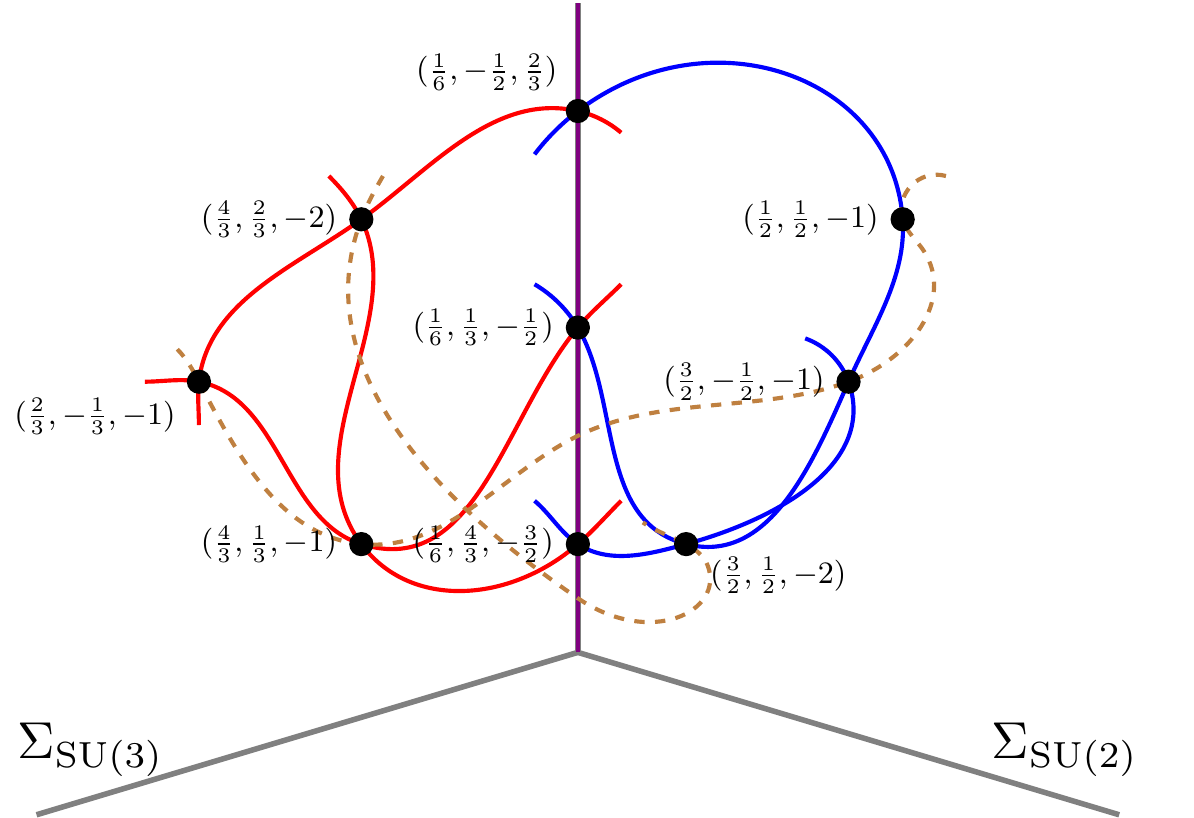}
	\end{center}
	\caption{Cartoon depiction of the various junctions of triples of matter curves $C_{\mathsf R}, C_{\mathsf R'}, C_{\mathsf R''} \in B$  associated to candidate Yukawa interactions appearing in the low-energy 4D $\cN=1$ action. Each codimension-three point in $B$ in the above figure (represented by a black dot) is labeled by the triple of $\au(1)$ charges $w_{\bar 1},w'_{\bar 1},w''_{\bar 1}$ of the three representations $\mathsf r, \mathsf r', \mathsf r''$ (see \cref{rtriple}) contracted into a gauge singlet term in the low-energy Lagrangian; note that each triple labels one of a pair of possible complex conjugate interaction terms. The red, blue, and purple curves correspond to matter curves associated to non-trivial $\asu(3), \asu(2)$, and $\asu(3) \oplus \asu(2)$ representations, respectively. By contrast, the dashed lines denote matter curves corresponding to pure $\au(1)$ charges. Note also that the $(\mathbf 1,\mathbf 1)_{-1}$ matter curve intersects the $(\mathbf 1,\mathbf 2)_{1/2}$ in a tangent point.}
	\label{spaghetti}
\end{figure}

In this section, we study the degenerations of the elliptic fibers
over codimension-three loci in $B$ that are characterized by triple intersections of matter curves,
	\begin{align}
	\label{Yuktrip}
		C_{\mathsf R_1} \cap C_{\mathsf R_2} \cap C_{\mathsf R_3} \subset B.
	\end{align}
Since the orders of vanishing of the sections $f,g,\Delta$ can further increase at these codimension-three loci, the singularity type of the elliptic fibers can enhance; we refer to this behavior as a degeneration of the elliptic fibers because analogous to the degenerations of the elliptic fibers over codimension-one or codimension-two components of the discriminant locus, these singularity enhancements also manifest themselves in the resolved geometry as the splitting of one or more of the components of the elliptic fibers into multiple irreducible $\mathbb P^1$ components. However, there is an important distinction to be made here, in that the elliptic fiber degenerations over loci of the form \cref{Yuktrip} do not produce additional generators of the cone of effective curves. This is because the degenerations are of the schematic form
	\begin{align}
	\label{codim3degen}
		F_{\mathsf R_1}  = F'_{\mathsf R_2} + F''_{\mathsf R_3}
	\end{align}
at a given point in the locus \cref{Yuktrip}. In the above equation, $F_{\mathsf R_i}$ is an irreducible component of the elliptic fibers (and hence a generator of the cone of effective curves) over a \emph{generic} point the codimension-two locus $C_{\mathsf R_i}$, and thus we see that the classes of any apparently ``new'' curves that appear as a result of these degenerations turn out to be linear combinations of existing curve classes.

\cref{codim3degen} implies that at each degeneration point in the
resolved geometry, there resides a localized interaction in the
M-theory effective action that mediates the splitting of a BPS state
corresponding to an M2-brane wrapping $F_{\mathsf R_1}$ into a pair of BPS states corresponding to M2-branes wrapping $F'_{\mathsf R_2}$ and $F''_{\mathsf R_3}$ \cite{Marsano_2011}. The representation-theoretic version of this statement that characterizes the F-theory limit is that degenerations of the form \cref{codim3degen} indicate the existence of maps
	\begin{align}
		\mathsf{R}_1 \times \mathsf{R}_2 \times \mathsf{R}_3 \rightarrow 1,
	\end{align}
which produce gauge singlets from the weights of the matter representations in the low-energy effective action. In other words, these codimension-three singularities correspond to Yukawa
interactions appearing in the low-energy effective 4D action
\cite{Donagi:2008ca,BeasleyHeckmanVafaI,BeasleyHeckmanVafaII}
(see also Section 6 of \cite{WeigandTASI} and references therein). An
important check that such interactions indeed can appear in the
low-energy effective action is to identify an explicit elliptic fiber degeneration in the resolution of the singular F-theory background.\footnote{
Note that while there are also Yukawa type couplings that include (non-chiral)
adjoint
representations of the nonabelian factors, associated with
non-localized  matter fields in the bulk of the 7-branes, we focus here
on localized Yukawa couplings associated with nonabelian fundamental
fields and the charge combinations of local matter described in the
previous section.}

In the case of the $\SM$ model, we anticipate the existence of Yukawa interactions corresponding to the following gauge singlets:
\begin{align}
	\label{rtriple}
		(\mathsf r_3, \mathsf r_2)_{w_{\bar 1}} \times (\mathsf r'_3, \mathsf r'_2)_{w_{\bar 1}'} \times (\mathsf r''_3, \mathsf r''_2)_{w_{\bar 1}'''}\,.
	\end{align}
To
ensure that the couplings are gauge singlets separately for each gauge
factor $\asu(3), \asu(2), \au(1)$ we require that the nonabelian
irreps are either singlets or pairwise complex conjugates and that the
$\au(1)$ charges sum to zero.\footnote{As the one possible exception
  to this characterization of the Yukawa interactions in the $\SM$
  model, we point out that there could also exist a Yukawa interaction
  of the form $(\textbf{3},\textbf{1})_{\frac{2}{3}} \times
  (\textbf{3},\textbf{1})_{\frac{2}{3}} \times
  (\textbf{3},\textbf{1})_{-\frac{4}{3}}$ (along with its
  conjugate). This interaction would have to be localized at the
  double intersection of the $(\textbf{3},\textbf{1})_{-\frac{4}{3}}$
  locus $b_1 e_2 e_3 = s_2 =0$ with the
  $(\textbf{3},\textbf{1})_{\frac{2}{3}}$ locus $b_1 e_2 e_3 = s_8
  s_2^2 - s_5 s_2 s_6 + s_1 s_6^2 =0$. However, the only possible
  double intersection would occur at tangent point $s_6^2 =0$, which
  lies along the IV locus $b_1e_2 e_3 = s_6 = 0$, which does not
  support any charged matter, hence there is no additional Yukawa
  interaction. \label{excepYuk}} In more detail, we compute the full list of possible
Yukawa interactions by identifying all solutions to the equation
$w_{\bar 1} + w'_{\bar 1} + w''_{\bar 1} = 0$, where $w_{\bar 1},
w'_{\bar 1}, w''_{\bar 1}$ are a subset of $\au(1)$ charges in the
low-energy (anti-)chiral spectrum, and then excluding all candidate
solutions for which the nonabelian factors cannot be combined into a
gauge singlet. With the exception of the possible gauge singlet described in \cref{excepYuk}, it turns out that all possible Yukawa interactions consistent with the matter representations of the $\SM$ model
arise as elliptic fiber singularity enhancements over codimension-three
loci in $B$,
assuming generic characteristic data. In the following, we discuss one
of a pair of complex conjugate Yukawa interactions associated to each
codimension-three singularity.

\subsection{\texorpdfstring{$(\textbf{3}^*,\textbf{2})_{w_{\bar 1}} \times (\textbf{3},\textbf{1})_{w_{\bar 1}'} \times (\textbf{1},\textbf{2})_{w_{\bar 1}''}$}{}}

First we consider the case $(w_{\bar 1}, w_{\bar 1}', w_{\bar 1}'') = (-\frac{1}{6},-\frac{4}{3},\frac{3}{2})$. Using the description of the fiber degeneration \cref{eqn:32fibers} associated with local matter transforming in $(\mathbf 3, \mathbf 2)_{\frac{1}{6}}$, over the locus $s_2=0$ we observe an additional degeneration $F'_{13} \to F'^t_{13} + F'^{e_5}_{13}$ inherited from the $(\mathbf 3, \mathbf 1)_{-\frac{4}{3}}$ degeneration $F'_3 \to F'^t_3 + F'^{e_5}_3$ described in \cref{eqn:F3prime}, so that
	\begin{align}
		F|_{b_1 e_2 e_3 = d_0 e_1 = s_2 =0} = F_{00} + F_{10}^{e_5} + F_{13}^{e_5} + F_{13}^t + F_{12}^t + F_{02}^t\,.
	\end{align}
The nontrivial intersections are the distinct points $F_{00} \cap
F_{10}^{e_5}, F_{10}^{e_5} \cap F_{13}^{e_5}, F_{13}^{e_5} \cap
F_{13}^t, F_{13}^t \cap F_{12}^t, F_{12}^t \cap F_{02}^t, F_{02}^t
\cap F_{00}$, thus we observe an enhancement $\text{I}_5^{\text{split}
}\rightarrow \text{I}_6^{\text{split}}$. Note that $\hat D_0$ wraps
$F_{10}^{e_5}$ and $F_{13}^{e_5}$, while $\hat D_1$ intersects
$F_{13}^{e_5}$ in a point.

The second case is $(w_{\bar 1}, w_{\bar 1}', w_{\bar 1}'') = (-\frac{1}{6},\frac{2}{3},-\frac{1}{2})$. In the open set $e_5 \ne 0$ we observe the further degeneration $F_{00} \rightarrow F_{00}^{+} + F_{00}^-$, that is
	\begin{align}
		F|_{b_1 e_2 e_3 = d_0 e_1 = s_8 s_2^2 - s_2 s_5 s_6 + s_1 s_6^2 =0} = F_{00}^{+} + F_{00}^- + F_{02} + F_{10} + F_{12} + F_{13}\,.
	\end{align}
Setting $s_8 = (s_2 s_5 s_6-s_1 s_6^2)/s_2^2$, and making the coordinate redefinitions $w \to w/(e_3e_4 e_5)$, the new rational curves can be described as
	\begin{align}
	\begin{split}
		F_{00}^{+} ~&=~ \{s_2 e_2  +  s_6  w = 0\}\,,\\
		F_{00}^- ~ &=~ \{s_1 s_2 e_1 e_2 + s_2^2 v + s_2 s_5 e_1 w - s_1 s_6 e_1 w=0\}\,.
	\end{split}
	\end{align}
The new intersection points are $F_{00}^+ \cap F_{00}^-, F_{00}^{-} \cap F_{02}, F_{00}^+ \cap F_{10}$, with all other intersection points unchanged and thus we find another $\text{I}_6^{\text{split}}$ enhancement. Here, $\hat D_0$ intersects $F_{00}^-$ in a point.

The final case is $(w_{\bar 1}, w_{\bar 1}', w_{\bar 1}'') = (-\frac{1}{6},-\frac{1}{3},\frac{1}{2})$. We observe the further degeneration $F_{02} \rightarrow F_{02}^+ + F_{02}^-$, i.e.,
	\begin{align}
		F|_{b_1 e_2 e_3 = d_0 e_1 = d_2 s_6^2 - d_1 s_6 s_8 + d_0 e_1 s_8^2 = 0} = F_{00} + F_{02}^+ + F_{02}^- + F_{10} + F_{12} + F_{13}\,,
	\end{align}
where setting $d_2 = d_1 s_8/s_6$ and making the coordinate redefinitions $u \to e_3 e_5 w u, w \to w / (e_1 e_3 e_4 e_5)$, we may write
	\begin{align}
	\begin{split}
		F_{02}^+ ~&=~ \{d_1 b_1 + s_6 u =0\}\,,\\
		F_{02}^- ~ &=~ \{s_6 v + s_8 w =0\}\,.
	\end{split}
	\end{align}
The new intersection points are $F_{02}^+ \cap F_{02}^-, F_{02}^+ \cap F_{12}, F_{02}^- \cap F_{00}$ with all other intersection points remaining the same, hence we again find an $\text{I}_6^{\text{split}}$ enhancement.

\begin{figure}
		\begin{center}
\includegraphics{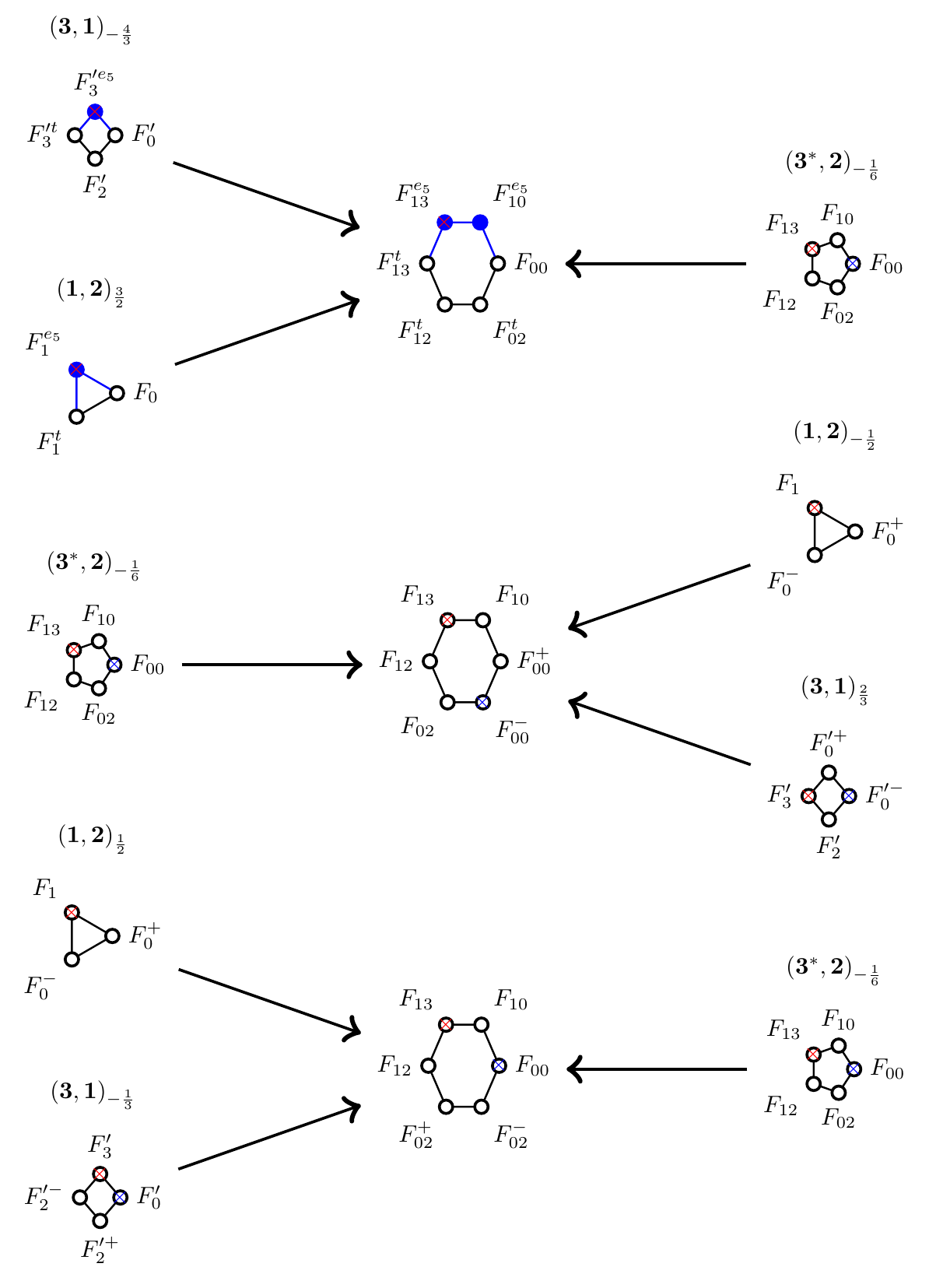}
		\end{center}
		\caption{Singular fibers associated to Yukawa interactions of the form $(\textbf{3}^*,\textbf{2})_{w_{\bar 1}} \times (\textbf{3},\textbf{1})_{w_{\bar 1}'} \times (\textbf{1},\textbf{2})_{w_{\bar 1}''}$.}
\label{321Yukawa}
	\end{figure}

See \cref{321Yukawa} for a schematic depiction of the singular fibers over these special points in $B$.

\subsection{\texorpdfstring{$(\textbf{3}^*,\textbf{1})_{w_{\bar 1}} \times (\textbf{3},\textbf{1})_{w_{\bar 1}'} \times (\textbf{1},\textbf{1})_{w_{\bar 1}''}$}{}}

First, consider the case $(w_{\bar 1 }, w_{\bar 1}', w_{\bar 1}'') = (\frac{4}{3},\frac{2}{3},-2)$. We use the description of the $(\textbf{3},\textbf{1})_{-\frac{4}{3}}$ degeneration given in \cref{eqn:F3prime}, in which we already see the degeneration $F'_3 \rightarrow F'^{t}_3 + F'^{e_5}_3$. Over the locus $s_1 =0$ an additional exceptional curve $F'^{e_5}_0$ appears due to the intersection of the zero section $\hat D_0$ with the affine component $F'_0$ of the singular elliptic fiber, leading to the degeneration
	\begin{align}
		F|_{b_1 e_2 e_3 =  s_2 =s_1=0} = F'^{t}_0 + F'^{e_5}_0 + F'_2 + F'^{t}_3 + F'^{e_5}_3\,.
	\end{align}
The new intersection points are $F'^{e_5}_0 \cap F'^{t}_0, F'^{e_5}_0 \cap F'^{e_5}_3, F_0'^{t} \cap F_2'$, hence we find an enhancement $\text{I}_4^{\text{split}} \rightarrow \text{I}_5^{\text{split}}$. Notice that $\hat D_0$ now wraps both $F'^{e_5}_3, F'^{e_5}_0$, while $\hat D_1$ still intersects $F'^{e_5}_3$ in a point.

Next, consider the case $(w_{\bar 1 }, w_{\bar 1}', w_{\bar 1}'') = (-\frac{2}{3},-\frac{1}{3},1)$. We work with the description of the $(\textbf{3},\textbf{1})_{\frac{2}{3}}$ degeneration in \cref{F0pdegen}, in which $F'_0 \rightarrow F'^{+}_0 + F'^-_0$. Here, consistent with the local matter representation $(\mathbf 3 , \mathbf 1)_{-\frac{1}{3}}$, for which we see the degeneration $F'_2 \rightarrow F'^{+}_2 + F'^{-}_2$, we find that the elliptic fiber decomposes as
	\begin{align}
		F|_{b_1e_2 e_3 = d_2 s_6^2 - d_1 s_6 s_8 + d_0 e_1 s_8^2 = s_8 s_2^2 - s_2 s_5 s_6 + s_1 s_6^2 =0} = F'^{+}_0 + F'^{-}_0  + F'^{+}_2 + F'^{-}_2+ F'_3\,.
	\end{align}
Solving simultaneously for $s_8, d_1$ and making the coordinate redefinitions $w \to w/ (e_1 e_3 e_4 e_5)$, $t \to s_2 t/ e_5$, and $u \to u / e_4$, we may write
	\begin{equation}
		\begin{split}
			F'^{+}_2 ~&= ~\{ s_6 t + s_8 w=0 \}\,, \\
			F'^{-}_2~&= ~\{  b_1 d_0 e_1 s_6 t + e_1 s_6^2 u + b_1 d_1 s_6 w- b_1 d_0 e_1 s_8 w = 0\}\,.
		\end{split}
	\end{equation}
The new intersections points are $F'^{+}_2 \cap F'^{-}_2, F'^{+}_2 \cap F'^{-}_0, F'^{-}_2 \cap F_3'$, hence we find another enhancement of the form $\text{I}_4^{\text{split}} \rightarrow \text{I}_5^{\text{split}}$.

Finally, consider the case $( w_{\bar 1} , w_{\bar 1}' ,  w_{\bar 1}'') = ( \frac{4}{3}, - \frac{1}{3}, -1)$. In this case we again work with the $(\mathbf 3, \mathbf 1)_{-\frac{4}{3}}$ description given in \cref{eqn:F3prime}. Consistent with the $(\mathbf 3, \mathbf 1)_{\frac{1}{3}}$ degeneration $F'_2 \to F'^+_2 + F'^-_2$, we find
	\begin{equation}
		F|_{b_1 e_2 e_3 = s_2 =d_2 s_6^2 - d_1 s_6 s_8 + d_0 e_1 s_8^2 = 0} = F'_0 + F'^+_2 + F'^-_2 + F'^t_3 + F'^{e_5}_3\,.
	\end{equation}
Using the projective scaling symmetry of the ambient space and making the coordinate redefinitions $u \to u/(e_1 e_4), e_3 \to e_3/(e_1 e_4 e_5 w), d_0 \to d_0/e_1$, the new irreducible curves can be expressed as
	\begin{equation}
		\begin{split}
			F'^+_2 ~&=~\{ s_6 v + s_8 e_3  = 0\}\\
			F'^-_2 ~&=~\{ s_6^2 u + s_6 d_0 b_1 v + s_6 d_1 b_1 e_3 - s_8 b_1 d_0 e_3 = 0\}
		\end{split}
	\end{equation}
The new points of intersection are $F'^+_2 \cap F'^-_2, F'^+_2 \cap F'_0, F'^-_2 \cap F'^t_3$ and therefore we see an enhancement of the form $\text{I}_4^{\text{split}} \rightarrow \text{I}_5^{\text{split}}$.

\begin{figure}
		\begin{center}
\includegraphics{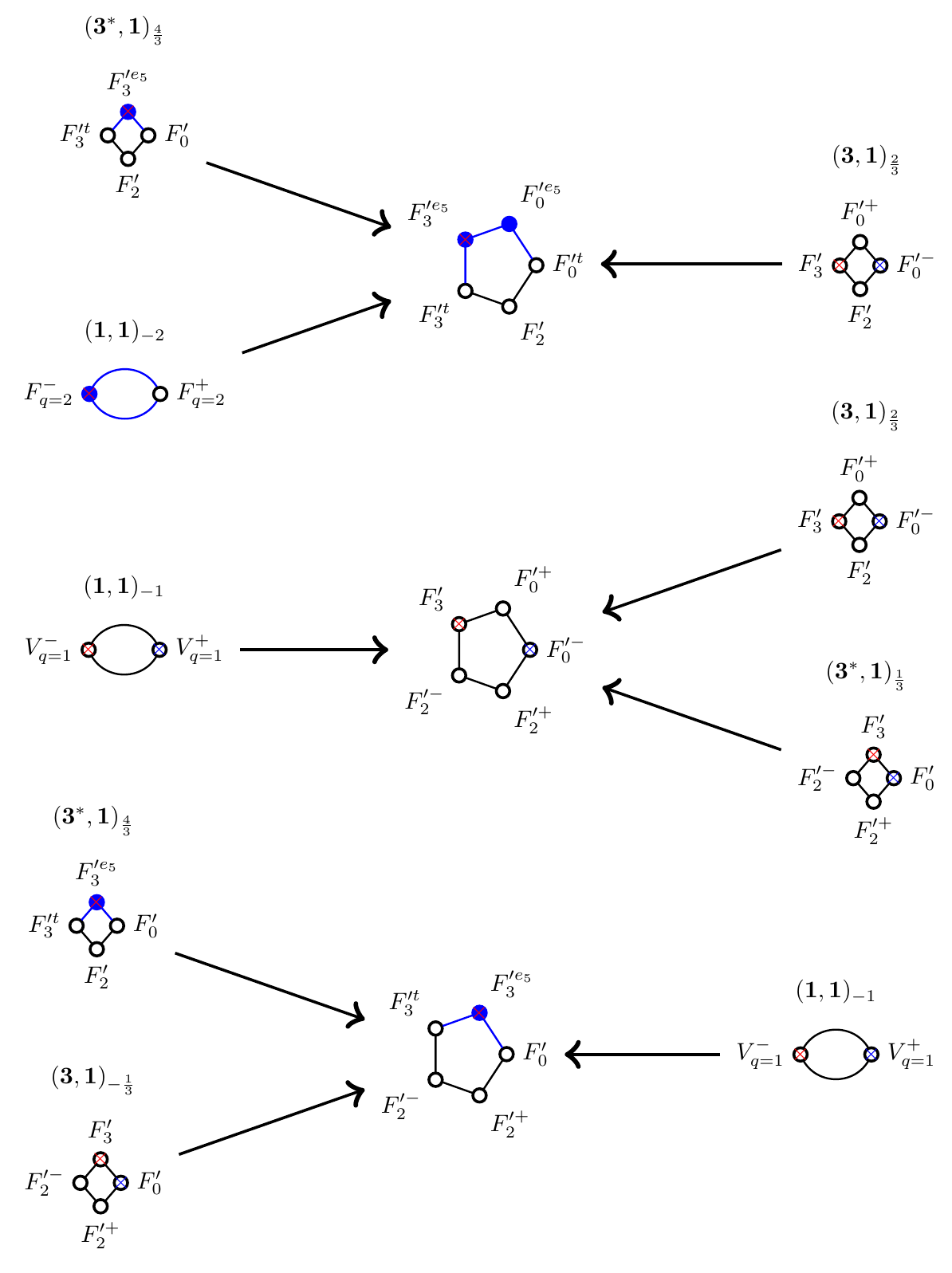}
		\end{center}
		\caption{Singular fibers associated to Yukawa interactions of the form $(\textbf{3}^*,\textbf{1})_{w_{\bar 1}} \times (\textbf{3},\textbf{1})_{w_{\bar 1}'} \times (\textbf{1},\textbf{1})_{w_{\bar 1}''}$.}
		\label{331Yukawa}
	\end{figure}

See \cref{331Yukawa} for a schematic depiction of the singular fibers over these special points in $B$.

\subsection{\texorpdfstring{$(\textbf{1},\textbf{2})_{w_{\bar 1}} \times (\textbf{1},\textbf{2})_{w_{\bar 1}'} \times (\textbf{1},\textbf{1})_{w_{\bar 1}''}$}{}}

The first case is $( w_{\bar 1}, w'_{\bar 1} , w''_{\bar 1}) = ( \frac{1}{2}, -\frac{3}{2}, 1)$. In this case, we borrow the description of the fiber degeneration $F_1 \to F_1^t + F_1^{e_5}$ that occurs over the codimension-two locus associated to local matter transforming in the representation $(\mathbf 1, \mathbf 2)_{\frac{3}{2}}$. In codimension three, we see the degeneration
	\begin{equation}
		F|_{d_0 e_1 = s_2 = \Delta_{(a)}/s_1= 0} = F_0^+ + F_0^- + F_1^{t} + F_1^{e_5}
	\end{equation}
where now, in contrast to the discussion in \cref{SU2matter}, the polynomial $\Delta_{(a)}$ factors over the locus $s_2 =0$:
	\begin{equation}
		\frac{\Delta_{(a)}}{s_1}  =  s_1 d_1^3 b_1^2  e_2^2 e_3^2 - s_6 s_5 d_1^2 b_1 e_2 e_3 - s_6^3 d_2 + s_6^2 s_8 d_1 + \cO(s_2)\,.
	\end{equation}
The above factorization permits us to straightforwardly solve for an explicit algebraic description of the curves $F_0^{\pm{}}$. Solving $\Delta_{(a)}/s_1 =s_2 = 0$ for $s_8$ and eliminating the unit $e_2$ in the affine open set $s_1 e_1 e_3 e_4 u \ne 0$, we may use the following local descriptions
	\begin{align}
		\begin{split}
			F_0^+ ~&=~ \{ b_1 d_1 e_3 e_5 w+s_6 u = 0 \}\,,\\
			F_0^- ~&=~ \{ -b_1 d_1^2 e_3 e_5^2 t^2 w+d_1 e_5 s_6 t^2 u+d_1 e_1 e_3 e_4 e_5 s_5 s_6 t u w\\
			&~~~~~+d_1 e_1 e_3 e_4 s_1 s_6^2 u v w+d_2 e_1^2 e_3^2 e_4^2 e_5 s_1 s_6^2 u w^2=0\}\,.
		\end{split}
	\end{align}
The new points of intersection are $F_0^+ \cap F_0^-, F_0^+ \cap F_1^t, F_0^- \cap F_1^{e_5}$, hence we see an enhancement $\text{I}_3^{\text{split}} \to \text{I}_{4}^{\text{split}}$.

The next case is $(w_{\bar 1} , w'_{\bar 1} ,w''_{\bar 1} ) = (\frac{1}{2}, \frac{3}{2} ,-2)$. This Yukawa interaction can be analyzed in a manner analogous to that of the previous case, with the key distinction being that we now solve $\Delta_{(a)}|_{s_2=0} = 0$ by setting $s_1 =0$. Here, it is illuminating to denote both degenerations by $F_0 \to F_0^t + F_0^{e_5}, F_1 \to F_1^{t} + F_1^{e_5}$:
	\begin{equation}
		F|_{d_0 e_1 = s_2 = s_1 = 0} = F_0^t + F_0^{e_5} + F_{1}^t + F_1^{e_5}\,,
	\end{equation}
where the  points of intersection are $F_0^t \cap F_0^{e_5}, F_1^t \cap F_1^{e_5} F_0^t \cap F_1^t, F_0^{e_5} \cap F_1^{e_5}$. We thus see an enhancement of the form $\text{I}_{3}^{\text{split}} \to \text{I}_{4}^{\text{split}}$. Notice that $\hat D_0$ now wraps both $F_0^{e_5}$ and $F_1^{e_5}$, while $\hat D_1$ continues to intersect $F_1^t$ in a point.

The final case is $(w_{\bar 1} , w'_{\bar 1} ,w''_{\bar 1} ) = (\frac{1}{2}, \frac{1}{2} ,-1)$. The restriction of the locus $\Delta_{(a)} = V_{q=1} = 0$ to $d_0 e_1 =0$ can be expressed as
	\begin{equation}
		\{ \Delta_{(a)}=V_{q=1}=0\}|_{d_0 e_1 =0} = \{ d_2 s_2^2-d_1 s_5 s_2+d_1 s_1 s_6 = b_1 d_1 e_2 e_3 s_1 s_2-s_8 s_2^2+s_5 s_6 s_2-s_1 s_6^2 = 0\}\,.
	\end{equation}
Note that the intersection of $V_{q=1} = 0$ and $\Delta_{(a)} =0$ occurs at a tangent point as depicted in \cref{spaghetti}, i.e., if we write $V_{q=1}: = \{ V_{q=1,1} =V_{q=1,2}  =0\}$ then
	\begin{equation}
	\label{tangent}
		\Delta_{(a)}|_{V_{q=1,1} = 0} = V_{q=1,2}^2,~~~~ \Delta_{(a)}|_{V_{q=1,2} = 0} = V_{q=1,1}^2\,.
	\end{equation}
We solve the above equations by eliminating $d_2, s_8$. Over this codimension-three locus, using the degeneration $F_0 \to F_0^+ + F_0^-$ inherited from the codimension-two enhancement associated with local matter in the representation $(\mathbf 1 ,\mathbf 2)_{\frac{1}{2}}$ described in \cref{eq:Deltaa}, we see the further degeneration $F_0^+ + F_0^- \to F_0^{++} + F_0^{--} + F_0$:
	\begin{equation}
		F|_{d_0 e_1 = \Delta_{(a)} = V_{q=1}=0} = F_0^{++} + F_0^{--} + F_0^0 + F_1\,,
	\end{equation}
where the points of intersection of the irreducible components are $F_0^{++} \cap F_0^0 , F_0^{--} \cap F_0^0$, signaling an enhancement $\text{I}_{3}^{\text{split}} \to \text{I}_{4}^{\text{split}}$.

\begin{figure}
		\begin{center}
\includegraphics{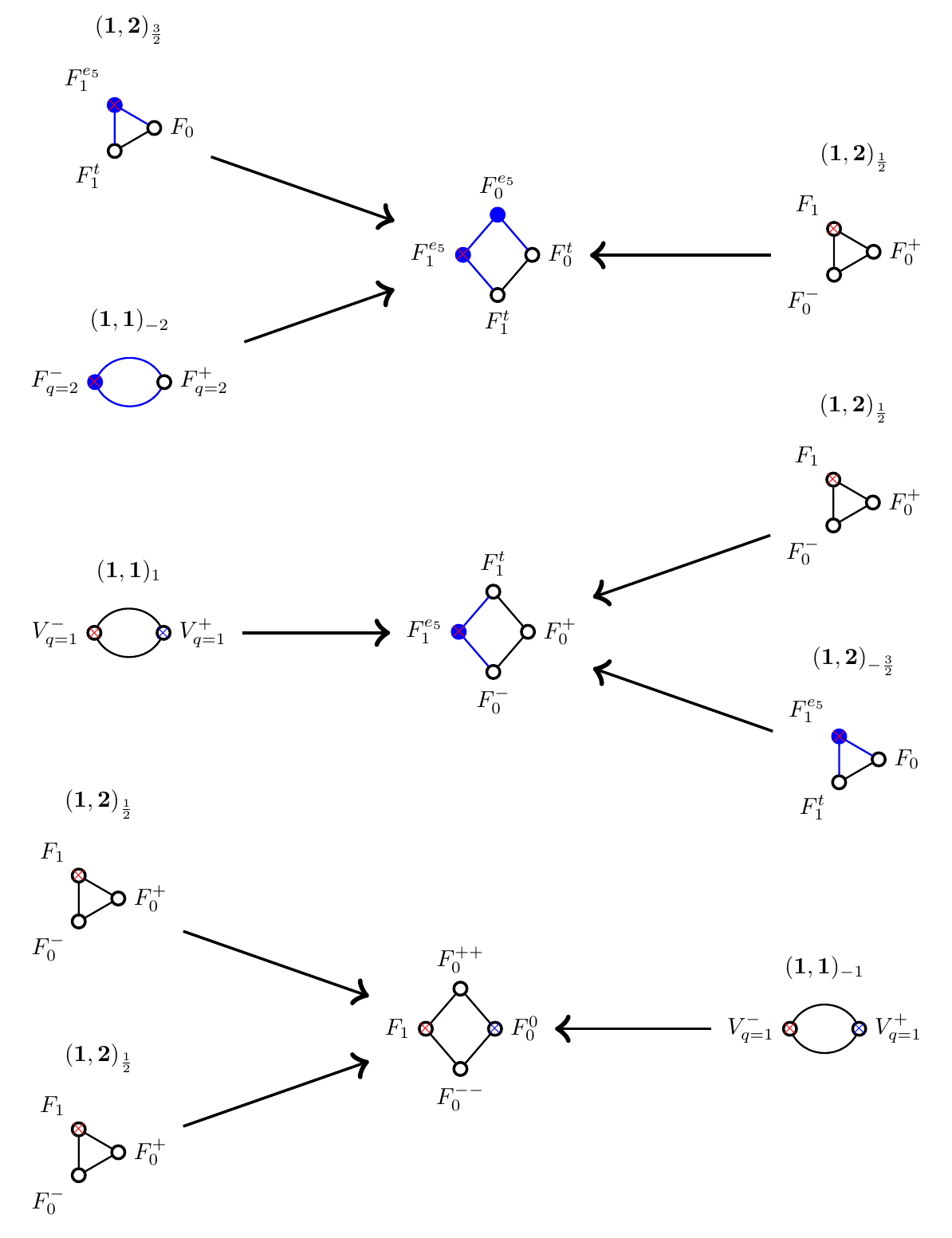}
		\end{center}
		\caption{Singular fibers associated to Yukawa interactions of the form $(\textbf{1},\textbf{2})_{w_{\bar 1}} \times (\textbf{1},\textbf{2})_{w_{\bar 1}'} \times (\textbf{1},\textbf{1})_{w_{\bar 1}''}$.}
		\label{221Yukawa}
	\end{figure}

See \cref{221Yukawa} for a schematic depiction of the singular fibers over these special points in $B$.

\section{Intersection theory, 3D Chern--Simons couplings, and chiral matter multiplicities}
\label{sec:intersection}

In the preceding sections, we have studied the geometry of the resolution
$X_5$.  This analysis confirms that the singularities of $X_0$ are
consistent with the expected kinematics of the low-energy 4D
theory \cite{Raghuram:2019efb}, and provides a geometric grounding for further investigation
of the structure of the matter spectrum and Yukawa interactions in the
theory.
In this section, we shift our attention to computing the
chiral excesses introduced to the 4D matter spectrum by switching on a
non-trivial flux background.
 Since we
closely follow the approach described in \cite{Jefferson:2021bid}, we omit here many technical details
of the methodology involved and mostly focus on describing the results of using these methods to analyze the $\SM$ model;
the interested reader should refer to \cite{Jefferson:2021bid} for
details of the approach that are not spelled out here explicitly, and
to \cite{WeigandTASI} and further references in these two papers for
further background and the earlier literature on
chiral matter analyses in F-theory.

\subsection{Quadruple intersection numbers and the reduced
  intersection matrix}
\label{quadint}

The analysis of \cite{Jefferson:2021bid} provides a framework for
understanding chiral multiplicities from fluxes in terms of a reduced
intersection pairing matrix on the vertical part of the middle cohomology of the resolution $X_5 \rightarrow X_0$. This
reduced intersection matrix can be computed from the quadruple
intersection numbers of divisors in $X_5$. One of the central results
of \cite{Jefferson:2021bid} was the observation that the intersection
pairing matrix appears to be independent of resolution, despite the
fact that the quadruple intersection numbers are
resolution-dependent. If this result indeed always holds, it suggests that the reduced intersection pairing matrix is the underlying resolution-independent mathematical structure that encodes information about the chiral matter spectrum in 4D F-theory compactifications.
In the following discussion, we adopt this viewpoint and turn our attention towards computing the reduced intersection matrix associated to $X_5$ as a means to recover the chiral spectrum of the $\SM$ model.

The resolution $X_5$ has a
basis of divisors
	\begin{align}
		\hat D_{I = 0,1,\alpha,i_2,i_3}\,,
	\end{align}
	which includes a
rational zero section $\hat D_0$ and rational generating section $\hat
D_1$ associated to the $\au(1)$ gauge factor,
	\begin{align}
	\label{eq:ratsections}
		\hat D_0 = \{ e_5 = 0 \} \cap X_5,~~~~ \hat D_1 = \{ e_4 =0 \} \cap X_5\,,
	\end{align}
as well as an $\asu(2)$ Cartan divisor $\hat D_{i_2} = \hat D_3$, and $\asu(3)$ Cartan divisors $\hat D_{i_3} = \hat D_4, \hat D_5$ appearing as irreducible components in the pullback of the gauge divisors $\Sigma_{s}$ from $B$ to $X_5$:
	\begin{align}
	\begin{split}
		\hat \Sigma_2 &=\Sigma_2^{\alpha} \hat D_\alpha =  (\hat \Sigma_2 - \hat D_3) + \hat D_3 = \{ d_0 e_1 = 0 \} \cap X_5\,,\\
		 \hat \Sigma_3 &=\Sigma_3^{\alpha} \hat D_\alpha= (\hat \Sigma_3 - \hat D_4 - \hat D_5) + \hat D_4 + \hat D_5 = \{ b_1 e_2 e_3 =0 \} \cap X_5\,.
	\end{split}
	\end{align}
The $\au(1)_{\text{KK}}$ and $\au(1)$ divisors (i.e., the images of $\hat D_0, \hat D_1$ under the Shioda map \cite{MorrisonParkU1}) are given by
	\begin{align}
	\label{U1div}
		\hat D_{\bar 0} &= \sigma_{\bar 0}^{I} \hat D_I = \hat D_0 - \frac{1}{2} \hat K\,,\\
		\hat D_{\bar 1} &= \sigma_{\bar 1}^{I} \hat D_{I} = \hat D_1 - \hat D_0 + \hat K - \hat Y  + \frac{1}{2} \hat D_3 + \frac{1}{3} ( \hat D_4 + 2 \hat D_5)\,,
	\end{align}
where $\hat K = K^\alpha \hat D_\alpha, \hat Y = Y^\alpha \hat D_\alpha$. Note that our convention for ordering the indices $i_s$ of the Cartan divisors $\hat D_{i_s}$ matches the following presentation of the $\asu(2) \oplus \asu(3)$ Cartan matrix:
	\begin{align}
		[[\kappa_{i_s j_t}]] &= \begin{pmatrix} 2 & 0 &0 \\0 &2 & -1\\ 0 &-1 & 2 \end{pmatrix} \,.
	\end{align}

Now that we have identified a suitable basis of divisors, our next task is to compute their quadruple intersection numbers. We use the adaptation of the pushfoward technology of \cite{Esole:2017kyr} to compute the pushforwards $W_{IJKL}$ of the quadruple intersection numbers
	\begin{align}
		\hat D_I \cdot \hat D_J \cdot \hat D_K \cdot \hat D_L
	\end{align}
(here, $\cdot$ denotes the intersection product in the Chow ring of $X_5$) to the Chow ring of $B$. In more detail, we first represent the pullbacks of the divisor classes $\hat D_{ I}$ as linear combinations of the divisors $\bm{D}_\alpha ,\bm{H} , \bm{E}_{i} = [e_i ]$ restricted to the class $\bm{X}_5$ of the hypersurface $X_5 \in Y_5$:
	\begin{align}
		\hat{\bm{D}}_I \cdot \bm{X}_5\,.
	\end{align}
In the above equation, the class $ \bm{X}_5$ is given by
	\begin{align}
		\bm{X}_{5} := [ p_{5,i} ] = 3 \bm{H} - \bm{K} - \boldsymbol \Sigma_2 +2 \boldsymbol \Sigma_3 - 2 \bm{Y} - 2 \bm{E}_1 - \bm{E}_2 - \bm{E}_3 - \bm{E}_4 - \bm{E}_5
	\end{align}
and
	\begin{align}
	\begin{split}
	\label{321Cartandiv}
		 \hat{\bm{D}}_0&= \bm{E}_5 ,~~~~ \hat{\bm{D}}_1 =  \bm{E}_4,~~~~\hat{\bm{D}}_3 = \bm{E}_1 ,~~~~ \hat{\bm{D}}_4 = \bm{E}_2 - \bm{E}_3 ,~~~~ \hat{\bm{D}}_5 = \bm{E}_3
	\end{split}
	\end{align}
	so that in terms of the projection
		\begin{align}
			\pi\colon Y_5 \to B\,,
		\end{align}
the pushforwards of the quadruple intersection numbers can be expressed as
	\begin{align}
	\label{eqn:ChowXtoY}
		W_{IJKL} = \pi_{*}( \hat{\bm{D}}_I \cdot \hat{\bm{D}}_J \cdot \hat{\bm{D}}_K \cdot \hat{\bm{D}}_L \cdot \bm{X}_5)\,.
	\end{align}
The above formulas allow us to represent all intersection products as products of divisor classes in the Chow ring of $Y_5$. We then use the fact that the projection map $\pi$ can be represented as a composition of maps,
	\begin{equation}
		\pi =\varpi \circ f_1 \circ \cdots \circ f_5\,,
	\end{equation}
where $f_i$ are the blowups appearing in \cref{eq:321res} and the $\varpi$ is the canonical projection $Y_0 \rightarrow B$ in \cref{canonproj}, to represent the pushforward map $\pi_*$ acting on elements of the Chow ring of $Y_5$ as the composition
	\begin{equation}
		\pi_* = \varpi_* \circ f_{1*} \circ \cdots \circ f_{5*}\,.
	\end{equation}
Since the action of $\varpi_*, f_{i*}$ on any formal power series in the appropriate Chow ring is known explicitly, this decomposition enables us to compute the pushforward of any quadruple intersection number---see Appendix E of \cite{Jefferson:2021bid} for more details.

For convenience, we follow the approach of \cite{genfun} and arrange the intersection numbers into the generating function
	\begin{equation}
	\label{generator}
	 e^{\hat J}= \exp( \varphi^I \hat{\boldsymbol{D}}_I )\cdot \boldsymbol X_5\,.
	\end{equation}
The real parameters $\varphi^I$ in the above expression, which may be regarded as real numbers parametrizing a choice of K\"ahler form
	\begin{align}
			\hat J = \varphi^I \hat D_I\,,
	\end{align}
provide a useful parameterization for the generating function. The pushforward of \cref{generator} is
	\begin{equation}
		Z_\varphi= \pi_*(e^J)\,.
	\end{equation}
The intersection numbers can be extracted from $Z_\varphi$ by computing derivatives:
\begin{equation}
		W_{IJKL} = \left. \frac{\partial^4}{\partial \varphi^I \partial \varphi^J \partial \varphi^K \partial \varphi^L } Z_\varphi \right|_{\varphi =0 }\,.
	\end{equation}
The pushforward of the generating function $e^J$
under the map $\pi_*$ encodes all of the intersection numbers of
$X_5$:
	\begin{align}
	\begin{split}
		Z_\varphi
		&=\exp( \varphi^\alpha D_\alpha ) \cdot \left( \frac{\cZ_1}{K}+\frac{\cZ_4+\cZ_6}{(K+Y)\cdot \left(K+\Sigma _3+Y\right)\cdot \left(2 K+\Sigma _3+Y\right)\cdot \left(2 K+\Sigma _2+\Sigma _3+Y\right)}\right. \\
		&-\frac{\cZ_2+\cZ_3}{\left(K+\Sigma _3\right) \cdot \left(K+\Sigma _2+\Sigma _3\right)\cdot (K+ Y)} +\frac{\cZ_5}{\left(K+\Sigma _3+Y\right) \cdot \left(2 K+\Sigma _2+\Sigma _3+Y\right)}\\
		&\left.+\frac{\cZ_7}{K \cdot \left(K+\Sigma _3\right)\cdot (K+Y) \cdot\left(2 K+\Sigma _3+Y\right)}\right)
	\end{split}
	\end{align}
where the functions $\cZ_{1},\dots,\cZ_{7}$ are given by
	\begin{align}
		\begin{split}
			\cZ_1&= e^{ \varphi^1 K+ \varphi^3 \Sigma _2+ \varphi^5 \Sigma _3 + \varphi^0 Y}\,,\\
			\cZ_2&=\Sigma _2 \cdot Y  \cdot e^{\varphi^3 \left(K+\Sigma _2+\Sigma _3\right)+\varphi^0 (K+Y)}\,,\\
			\cZ_3&=\left(-K-\Sigma _2-\Sigma _3+Y\right) \cdot  (K+ \Sigma _3) \cdot e^{\varphi^0 (K+Y)}\,, \\
			\cZ_4&=- \left(3 K+\Sigma _2+\Sigma _3+Y\right) \cdot  (2 K+Y+ \Sigma _3) \cdot \Sigma _3 \cdot  e^{\varphi^4 (-K-Y)}\,, \\
			\cZ_5&=- \left(4 K+\Sigma _2+2 \Sigma _3+2 Y\right) \cdot e^{ \varphi^4 \Sigma _3 }\,, \\
			\cZ_6&=-   \left(K \cdot \Sigma _3 \cdot e^{\varphi^4 (-K-Y)}\right. \\
			&\left. + \left(2 K+\Sigma _3+Y\right) \cdot (K+Y)\cdot e^{\varphi^4\Sigma _3 } \right)\cdot \Sigma _2 \cdot e^{\varphi^3 \left(2 K+\Sigma _2+\Sigma _3+Y\right)}\,,\\
			\cZ_7 &=- \left(Y \cdot \left(2 K+\Sigma _3+Y\right)\cdot  e^{\left(\varphi^4-\varphi^5\right) \left(-K-\Sigma _3\right)+\varphi^0 (K+Y)}\right.\\
			&\left. +\left(K+\Sigma _3\right)\cdot K  \cdot e^{\left(\varphi^5-\varphi^4\right) \left(2 K+\Sigma _3+Y\right)}\right) \cdot \Sigma _3 \cdot  e^{\varphi^4 \left(K+\Sigma _3\right)+ \varphi^3 \Sigma _2} \,.
		\end{split}
	\end{align}
Following the analysis of \cite{Jefferson:2021bid}, we  organize the pushforwards of the quadruple intersection numbers into a matrix of intersection pairings:
	\begin{equation}
	\label{intMdef}
		M_{(IJ)(KL)} = W_{IJKL} = S_{IJ} \cdot S_{KL},~~~~ S_{IJ} := \hat D_I \cdot \hat D_J \in \Lambda_S\,,
	\end{equation}
where we view the matrix $M_{(IJ)(KL)}$ as acting on an integer
lattice $\Lambda_S$ spanned by the elements $S_{IJ}$. The components
of $M_{(IJ)(KL)}$ are displayed in \cref{fig:321intnum}. Although the
matrix $M_{(IJ)(KL)}$ may appear at face value to simply be a trivial
repackaging of the intersection numbers,
as discussed in
\cite{Jefferson:2021bid} we expect the nondegenerate part of this
matrix to be independent of resolution (up to a choice of basis),
despite the fact that complete set of intersection numbers does not share this feature.
 The matrix $M_{(IJ)(KL)}$ has
a non-trivial null space
	\begin{align}
		\{  \nu^{IJ} \in \Lambda_S | M_{(KL)(IJ)} \nu^{IJ} =0 \text{ for all } (KL)\}\,,
	\end{align}
which is spanned by (at least) the following vectors:
	\begin{align}
	\label{nullspace}
		\left\{
\begin{array}{c}
 Y^\alpha S_{1\alpha }-S_{01} \\
 S_{04} \\
 S_{14} \\
 \Sigma _2^\alpha S_{1\alpha }-S_{13} \\
 K^\alpha S_{1\alpha }-S_{11} \\
 \Sigma _3^\alpha S_{1\alpha }-S_{15} \\
 (K+Y)^\alpha S_{0\alpha }-Y^\alpha S_{1\alpha }-S_{00} \\
  \left(K \Sigma _3+\Sigma _3 Y\right)^\alpha S_{2\alpha }+\left(-K+\Sigma _3-Y\right)^\alpha S_{\alpha 4} -\Sigma _3^\alpha S_{0\alpha }-\Sigma _3^\alpha S_{1\alpha }-S_{44} \\
 \left(2 K+\Sigma _2+\Sigma _3+Y\right)^\alpha S_{\alpha 3}-2 \Sigma _2^\alpha S_{1\alpha }-S_{03}-S_{33}-S_{35} \\
  \left(2 K+\Sigma _2+\Sigma _3+Y\right)^\alpha S_{\alpha 3}-2 \Sigma _2^\alpha S_{1\alpha }-\Sigma _2^\alpha S_{\alpha 5}-S_{03}-S_{33} \\
 (-K-Y)^\alpha S_{\alpha 5}+\Sigma _3^\alpha S_{1\alpha }+S_{05}-S_{45} \\
  \left(2 K+\Sigma _3+Y\right)^\alpha S_{\alpha 5}-2 \Sigma _3^\alpha S_{1\alpha }-S_{05}-S_{55} \\
\end{array}
\right\}\,.
	\end{align}
The quotient of the lattice $\Lambda_S$ by the complete set of null vectors is
isomorphic to the nondegenerate lattice
$H_{2,2}^{\text{vert}}(X_5,\Z)$ defined by the pairing
$M_{\text{red}}$.  While in simple cases (such as the base $\bP^3$),
the set \cref{nullspace} is a complete basis of the nullspace of $M$, there may
be further null vectors, associated, for example, with
fluxes
$\phi^{\alpha 3}D_\alpha = D$, with $D \cdot \Sigma_2 = 0$.  We write
a general abstract form of $\mr$ in \cref{Mred} and \cref{t:mr-alternate}, in two
different choices of bases, with the understanding that some
further null vectors may need to be removed to define the reduced basis for some bases, as discussed above.
While we have computed this from a particular resolution, as in the
cases studied in \cite{Jefferson:2021bid} we expect this form of $\mr$
to be resolution-independent.
We discuss the relationship between
M-theory flux backgrounds and the middle cohomology subgroup
$H_{2,2}^{\text{vert}}(X_5,\Z)$ in more detail in the following
subsection.

\subsection{Fluxes preserving 4D \texorpdfstring{$\SM$}{} gauge symmetry}
\label{ftheoryfluxes}

\subsubsection{Fluxes through vertical surfaces}

As described in \cref{quadint}, our primary purpose in computing the quadruple intersection numbers is to determine an explicit parametrization for the lattice of  M-theory fluxes through vertical surfaces
\begin{equation}
	\label{3Dmatch}
	\Theta_{I J} %{\hat I \hat J}
        = \int_{S_{I J}} %{\hat I \hat J}}
          G_4\,.
	\end{equation}
In the above equation, $G_4=\mathrm d C_3$ is the
field strength of the M-theory 3-form $C_3$ and is expected to satisfy
the shifted quantization condition $G_4 - c_2/2 \in H^{4}(X_5,\Z)$ in consistent M-theory vacua \cite{Witten:1996md}, and $S_{IJ} := \hat D_I \cdot \hat
D_J$ (see \cref{intMdef}) are so-called ``vertical'' 4-cycles whose
homology classes generate the subgroup
$H_{2,2}^{\text{vert}}(X_5)$.
%\footnote{Note that the hatted
%  indices $\hat I$ appearing in \cref{3Dmatch} are a restriction of
%  the usual indices $I$ to a particular subset of values corresponding
%  to a choice of basis for the sublattice of symmetry-preserving
%  fluxes; see the discussion below \cref{symconst} and in particular
%  \cref{hat} for a more detailed explanation.}

While $G_4$ could in principle be any element of $H^4(X_5,\Z)$,
because the middle cohomology of smooth elliptic CY fourfolds $X$
admits an orthogonal decomposition \cite{Greene:1993vm,Braun:2014xka}
	\begin{equation}
\label{eq:orthogonal-decomposition}
		H^4(X) = H^{2,2}_{\text{vert}}(X) \oplus
                H^4_{\text{hor}}(X) \oplus H^4_{\text{rem}} (X)\,,
	\end{equation}
for the purpose of computing \cref{3Dmatch} we can restrict our
attention to vertical flux backgrounds, i.e., $G_4 \in
H^{2,2}_{\text{vert}}(X_5,\Z/2)$, possibly at the expense of
being able to determine the precise quantization of the chiral indices
(further discussion of the quantization issue appears in \cref{sec:quantization}). We follow this approach, and for convenience we parametrize the vertical part of $G_4$ by its Poincar\'e dual,
\begin{equation}
	\text{PD}(G_4^{\text{vert}}) =: \phi =  \phi^{IJ} S_{IJ}\,,
\end{equation}
where we view $\phi^{ K L} \in \Lambda_S$ as an integer\footnote{The
  flux background $\phi$ could also be a half-integer vector. This is
  required, for example, when $c_2$ is not an even class as a
  consequence of the shifted quantization $G_4 - c_2/2 \in
  H^4(X_5,\Z)$ as first pointed out in \cite{Witten:1996md};
  see \cref{c2even} for further discussion.} vector living in the
integral lattice $\Lambda_S$ of vertical flux backgrounds spanned by
the 4-cycles $S_{IJ}$. This convenient choice of parametrization leads
to the following simple linear algebraic expression for the space of
vertical fluxes:
	\begin{align}
%	  \Theta_{\hat I \hat J} = S_{\hat I \hat J} \cdot S_{\hat K \hat L} \phi^{\hat K \hat L}%|_{ \Lambda_C}
%          = M_{(\hat I \hat J)(\hat K \hat L)} \phi^{\hat K \hat L}%|_{\Lambda_C}
	  \Theta_{ I  J} = S_{ I  J} \cdot S_{ K  L} \phi^{ K  L}%|_{ \Lambda_C}
          = M_{( I  J)( K  L)} \phi^{ K  L}%|_{\Lambda_C}
          \,.
          \label{eq:tmp}
	\end{align}

\subsubsection{Symmetry constraints}

We are specifically interested in M-theory fluxes
that lift to 4D F-theory fluxes preserving local Lorentz and $\SM$
gauge symmetry.
In the above equation \labelcref{eq:tmp},
these symmetry constraints can be imposed by restricting
$\phi^{ K  L}$ to live in the sublattice
$\Lambda_C \subset \Lambda_S$ of vertical flux backgrounds that
preserve 4D Poincar\'e and $\SM$ symmetry, i.e., flux backgrounds
whose corresponding fluxes satisfy \cite{Dasgupta:1999ss,Donagi:2008ca}
\begin{equation}
	\label{symconst}
		\Theta_{I \alpha} = 0\,.
	\end{equation}
We defer explicitly solving the constraints $\Theta_{I\alpha} =0$ to
\cref{sec:321-spectrum,altsym}, where we give two complementary
approaches to analyzing the lattice of constrained fluxes. For now, we
focus on computing the constraints imposed on the %chiral spectrum
vertical fluxes $\Theta_{I J}$
by
the geometry of the $\SM$ model defined over an arbitrary base, when
these symmetry conditions are imposed. These
constraints on the vertical fluxes will reveal whether or not the full set of chiral matter
combinations compatible with 4D anomaly cancellation can be realized
in a generic F-theory compactification. In \cite{Jefferson:2021bid},
it was observed %that the full set of linear constraints on the
that the homology relations associated with the nullspace of the
matrix $M_{(IJ)(KL)}$
correspond to linear constraints on the vertical fluxes $\Theta_{I
  J}$, due to the symmetry of $M$.
Thus,
% we simply need to determine the image of
%from the null
%space described in \cref{nullspace} %in the constrained sublattice
%$\Lambda_C$. Observe that since $M_{(IJ)(KL)}$ is a symmetric matrix,
an arbitrary null vector $\nu = \nu^{IJ} S_{IJ}$ in
the span of the vectors described in \cref{nullspace} implies the existence of
a null relation among fluxes,
	\begin{align}
		\nu^{IJ} \Theta_{IJ} = \nu^{IJ} M_{(IJ)(KL)} \phi^{KL} = 0\,.
	\end{align}
This implies that there exists a set of equivalences among vertical
fluxes given by \cref{nullspace}, in which the basis elements $S_{IJ}$
are replaced by fluxes $\Theta_{IJ}$. For example, $\Theta_{04} = 0$
and $Y^{\alpha} \Theta_{1\alpha} - \Theta_{01} = 0$.
Restricting attention to fluxes obeying the symmetry constraints
%The image of these relations in the constrained sublattice
%$\Lambda_{C}$ is then just given by setting
$\Theta_{I\alpha} =0$, this leads to the following relations among vertical F-theory fluxes:
	\begin{align}
	\begin{split}
	\label{geobasisnullfluxes}
0&=\Theta_{01} =\Theta_{04} =\Theta_{14} =\Theta_{13} = \Theta_{11} =\Theta_{15} = \Theta_{00} \\
 0&= \Theta_{44} \\
 0&=\Theta_{03}+\Theta_{33}+\Theta_{35} \\
0&=  \Theta_{03}+\Theta_{33} \\
 0&=\Theta_{05}-\Theta_{45} \\
 0&=\Theta_{05}+\Theta_{55}\,.
\end{split}
	\end{align}
In \cref{chiralind}, after equating the F-theory fluxes $\Theta_{ I
  J}$ to 3D Chern--Simons couplings,
we translate the above linear relations into relations among chiral
indices and compare the resulting relations to the 4D anomaly
cancellation constraints to see whether or not F-theory geometry is
more restrictive then the physical anomaly cancellation
conditions.

\subsubsection{Integrality of the second Chern class}
\label{c2even}

The shifted quantization condition
$G_4 - c_2/2 \in H^4(X_5,\Z)$ implies that when $c_2/2$ is not
an integer class, $G_4$ must also be a half-integer
class.
This in turn implies that $\phi = \text{PD}(G_4)$ cannot take integer values.
Note, however, that $\phi$ can be fractionally quantized in some cases
even when $c_2/2$ is an integer class,
as discussed further in \cref{sec:quantization}, although in such
situations integer values of $\phi$ are still possible.

To clarify some of the issues involved here, let us define\footnote{The expression
  $\httv(X_5,\Z)$ is defined in \cite{Jefferson:2021bid} as the
  first of these objects, but may in other places in the literature be
defined as the latter; we define these distinct symbols to avoid
confusion and ambiguity.}  two sublattices of $H_4 (X_5,\Z)$:
\begin{align}
 \Lambda_{\rm vert} & = {\rm span}_\Z (S_{IJ}) / \sim\label{eq:lv}\\
\bar{\Lambda}_{\rm vert} & =
H_4 (X_5,\Z) \cap \httv(X_5,\C) \,. \label{eq:lvb}
\end{align}
%In this discussion we assume that $c_2 (X_5)$ is even; suitable
%modifications must be made to the discussion when this condition does
%not hold.
The quotient in the first line of the above equation is by the
nullspace of $M$, i.e., by the equivalence relation
\begin{align}
	\phi \sim \phi' \Leftrightarrow M_{(IJ)(KL)} (\phi -
        \phi')^{KL} = 0 \text{ for all } (IJ)\,.
\end{align}
In general, $\bar{\Lambda}_{\rm vert}$ is an overlattice of
$\Lambda_{\rm vert}$.
While $\Lambda_{\rm vert}$ is easily computed from the set of integral
divisors, as far as we know there is not yet any systematic way of determining
$\blv$ for an arbitrary CY fourfold.
When $c_2$ is even, but $c_2/2$ is not contained in $\lv$, then integer
values of $\phi$ are still allowed.  When $c_2$
lies in $\lv$ but is not even in $H_4
(X_5,\Z)$, however, then $\phi$ must have some half-integer components.
Half-integer elements of $\phi$ can  modify
certain important aspects of the theory, such as the precise
quantization of the chiral indices or even the enforcement of the
symmetry constraints \labelcref{symconst},
although as discussed in more detail below
 we do not encounter the latter issue for the constructions
here. For this reason, it is important
to determine for what choices of characteristic data $c_2/2$ fails to
be an integer class.

One way to check if $c_2/2$ is not an integer class is to compute
$\chi /24$ where $\chi$ is the Euler characteristic of the smooth CY
fourfold $X_5$. If $\chi/24$ fails to be an integer, this indicates
that $c_2/2$ is not an integer class. In fact, from a physical
standpoint, in such situations it is important that
$c_2/2$ is not an integer class in order to maintain the integrality
of the M2-brane tadpole,
	\begin{equation}
	\label{tadpole}
		N_{\text{M2}}=  \frac{1}{24} \chi(X_5) - \frac{1}{2} \int_{X_5} G_4 \wedge G_4 \in \Z_{\geq 0}\,.
	\end{equation}
When $\chi/24$ fails to be an integer, the fractional part of
$\chi/24$ is precisely canceled by contributions to the integral
$\int_{X_5} G_4 \wedge G_4$ coming from the fractional part of $G_4$ (i.e., $c_2/2$), see, e.g., \cite{Collinucci:2010gz}.

We can use the pushforward technology described in the previous
subsection to evaluate the pushforward of the Euler characteristic to
$B$:\footnote{Note that the total Chern class $c(X_5)$ can be
  expressed as a formal power series in $\bm{D}_\alpha ,\bm{H} ,
  \bm{E}_{i}$---see Appendix E of \cite{Jefferson:2021bid}).}
	\begin{align}
	\begin{split}
		\chi(X_5) = &~ \int_{X_5} c_4\\
		 = &-12 c_2(B) \cdot K-144 K^3-96 K^2  \cdot \Sigma_2-168 K^2 \cdot \Sigma_3\\
		 &-24 K\cdot \Sigma_2^2-66 K \cdot \Sigma_3^2-75 K\cdot \Sigma_2\cdot \Sigma_3-6 \Sigma_3^3-15 \Sigma_2 \cdot\Sigma_3^2-9 \Sigma_2^2\cdot \Sigma_3\\
		 &+ \left(-144 K^2-51 K\cdot \Sigma_2-111 K\cdot \Sigma_3-3 \Sigma_2^2-21 \Sigma_3^2-18 \Sigma_2 \cdot \Sigma_3\right)\cdot Y \\
		 &+\left(-66 K-15 \Sigma_2-27 \Sigma_3\right)\cdot Y^2 -6 Y^3\,.
	\end{split}
	\end{align}
This expression for the Euler
characteristic  may be useful in various aspects of further analysis
of these constructions, for instance aspects involving the tadpole constraint (\cref{tadpole}).
It is also useful to have an explicit expression for $c_2$. For example, using the fact that the hyperplane class is
	\begin{equation}
		\boldsymbol{H} \cdot \boldsymbol{X}_5 = - \hat K -\hat \Sigma_3 + \hat D_0 + 2 \hat D_1 + \hat D_3 + \hat D_5
	\end{equation}
we can represent the image of $c_2$ of in the lattice $H_{2,2}^{\text{vert}}(X_5,\Z)$ as
	\begin{align}
	\begin{split}
\label{eq:c2}
		c_2(X_5) = &~\left(-6 K-2 \Sigma _2-\Sigma _3-Y\right)^\alpha S_{0\alpha } + \left(-6 K+2 \Sigma _2+\Sigma _3+Y\right)^\alpha S_{1\alpha }\\
		&+\left(c_2(B)+5 K^2+2 K\cdot \Sigma _2+K\cdot \Sigma _3+5 K \cdot Y+Y^2+\Sigma _2 \cdot Y\right)^{\alpha \beta} S_{\alpha \beta } \\
		&+ \left(-4 K-\Sigma _3-2 Y\right)^\alpha S_{\alpha 3}+ \left(4 K+\Sigma _2+\Sigma _3+2 Y\right)^\alpha S_{\alpha 4}+\left(\Sigma _2-2 K\right)^\alpha S_{\alpha 5}\\
		&+3 S_{03}+S_{05}-S_{34}\,.
	\end{split}
	\end{align}
Notice that the above expression for $c_2$ is expanded in the reduced
basis $S_{I\alpha}, S_{03}, S_{05}, S_{34}$ of
$H_{2,2}^{\text{vert}}(X_5,\Z)$.
From this expression we can immediately see that for all our
constructions, $c_2/2$ cannot be an element of $\lv$, since it
contains the components $(3 S_{03}+S_{05}-S_{34})/2$.  As discussed
above, however, this does not definitively indicate that $c_2/2$
cannot be an integral element of $H_4 (X_5,\Z)$, as it may live in $\blv$.

Another way to check if $c_2/2$ is \emph{not} an integer element of cohomology is
to compute $c_2^2/4$.  If this is not integer-valued, then $c_2/2$
cannot be an integer class, since the intersection product on integer
cohomology takes integer values.
The matrix $M_{\text{red}}$,
of intersection pairings of any two elements belonging to
$H_{2,2}^{\text{vert}}(X_5,\Z)$ is presented in
\cref{t:mr-alternate} using the basis relevant for
\cref{eq:c2}. Using the explicit expression given there for $M_{\text{red}}$, we
find that
%so that for any element $V = V^{IJ} S_{IJ} \in H_{2,2}^{\text{vert}}(X_5, \Z)$, we have
%	\begin{equation}
%		V \cdot c_2 = V^{IJ}  M_{\text{red}(IJ)(KL)}c_2^{KL},~~~~ IJ = I\alpha, 03,05,34.
%	\end{equation}
one-fourth of the square of the second Chern class is given by
	\begin{align}
	\begin{split}
		\frac{1}{4}	c_2(X_5)^2&=-6 c_2 \cdot K-8 K^2 \cdot \Sigma_2-14 K^2 \cdot \Sigma_3-12 K^2 \cdot Y-12 K^3-\frac{17}{4} K \cdot \Sigma_2 \cdot Y\\
		&-\frac{37}{4} K \cdot \Sigma_3 \cdot Y-2 K\cdot \Sigma_2^2-\frac{11}{2} K\cdot \Sigma_3^2-\frac{25}{4} K\cdot \Sigma_2\cdot \Sigma_3-\frac{11}{2} K\cdot Y^2\\
		&-\frac{5}{4} \Sigma_2 \cdot Y^2-\frac{9}{4} \Sigma_3 \cdot Y^2-\frac{1}{4} \Sigma_2^2\cdot Y-\frac{7}{4} \Sigma_3^2\cdot Y-\frac{3}{2} \Sigma_2\cdot \Sigma_3\cdot Y-\frac{5}{4} \Sigma_2 \cdot \Sigma_3^2\\
		&-\frac{3}{4} \Sigma_2^2 \cdot
                \Sigma_3-\frac{1}{2}\Sigma_3^3-\frac{1}{2} Y^3\,.
\label{eq:c2-squared}
	\end{split}
	\end{align}
Whenever the characteristic data $K, \Sigma_2, \Sigma_3, Y$ are such that the above expression is non-integer
valued, $c_2$ cannot be an even class.  In  \cref{sec:p3} we show
explicitly that there are many cases where this occurs, so that
integer flux parameters $\phi^{IJ}$ are not possible in those cases.

\begin{table}
\centerline{\includegraphics[scale=.27]{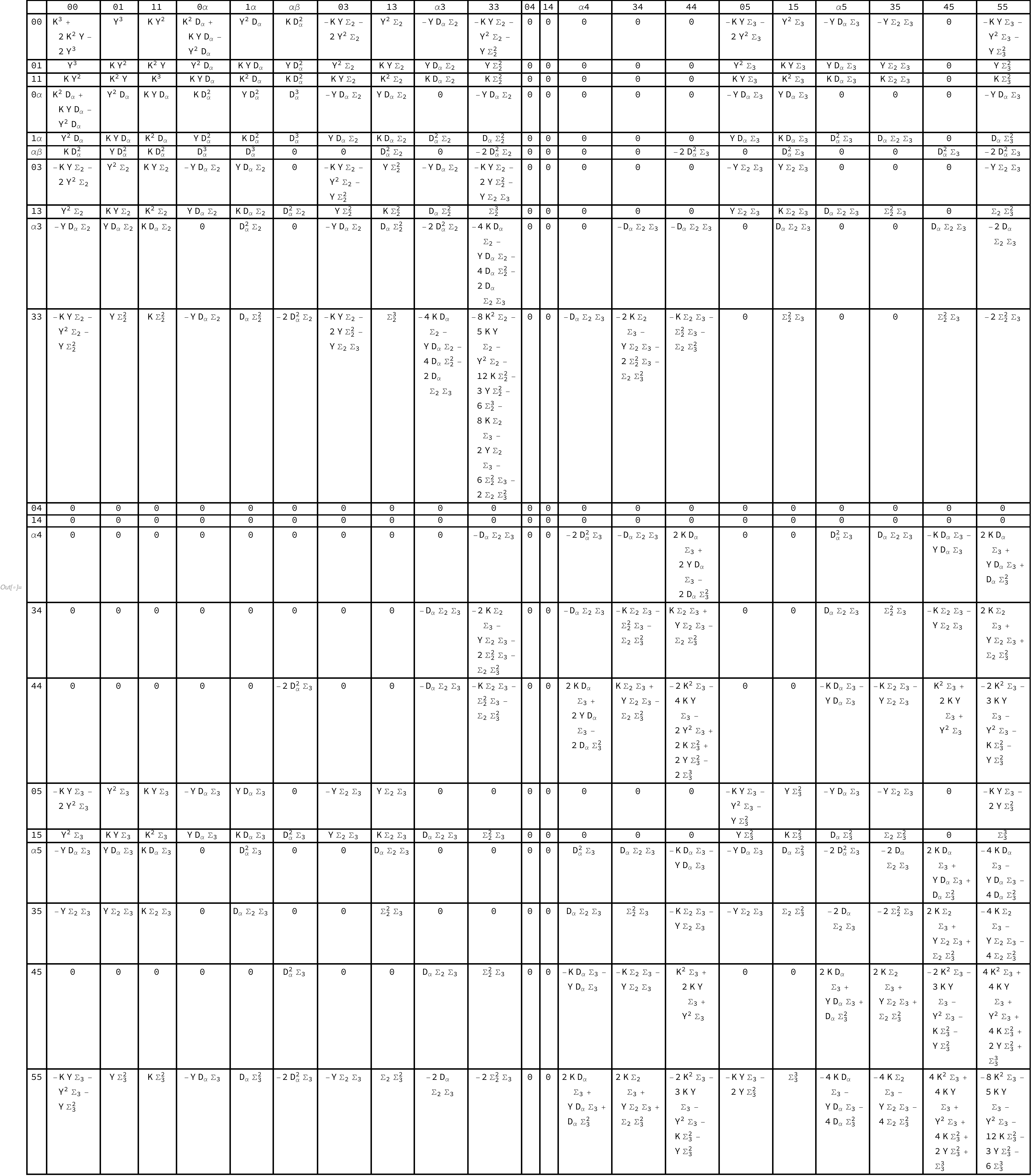}}
\caption{Matrix of pushforwards of quadruple intersection numbers $M_{(IJ)(KL)} = S_{IJ} \cdot S_{KL}$ (recall that $S_{IJ} := \hat D_I \cdot \hat D_J$) associated to the resolution $X_5$ of the $\SM$ model. The entries of the uppermost (leftmost) row (column) are the indices $IJ$ of the corresponding classes $S_{IJ}$. We work in the basis $\hat D_{I = 0,1,\alpha,i_s}$ where $\hat D_0, \hat D_1$ are the rational sections, $\hat D_\alpha$ is the pullback of a divisor in $B$, and $\hat D_{i_2 = 3},\hat D_{i_3=4,5}$ are (resp.) the $\asu(2), \asu(3)$ Cartan divisors. We omit the $\cdot$ notation indicating the intersection product for brevity and use the shorthand $D_\alpha^2 = D_\alpha \cdot D_\beta, D_\alpha^3 = D_\alpha \cdot D_\beta \cdot D_\gamma$.}
\label{fig:321intnum}
\end{table}

\begin{table}
\begin{center}
\includegraphics[scale=.36]{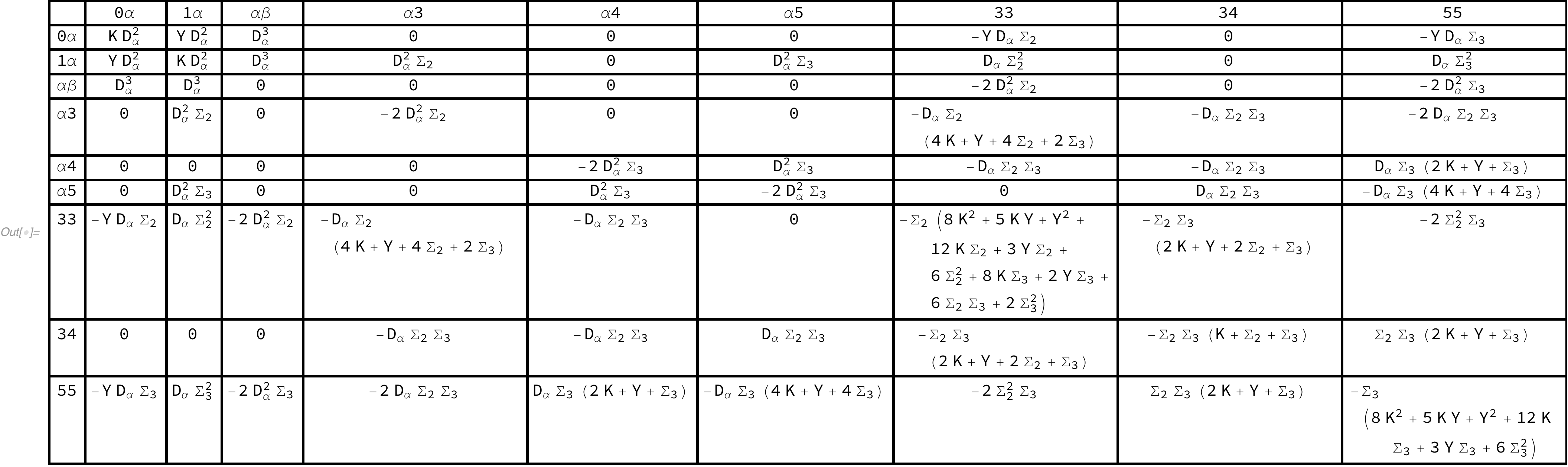}
\caption{Matrix elements $M_{\text{red}(IJ)(KL)}$ for the reduced intersection pairing $M_{\text{red}}$ of the lattice $H_{2,2}^{\text{vert}}(X_5,\Z)$. We compute the matrix elements $M_{\text{red}(IJ)(KL)}$ by quotienting by the equivalence relation $\sim$, where $\phi \sim \phi'$ iff $M_{(IJ)(KL)} (\phi-\phi')^{KL} = 0$, assuming additional null vectors have been removed to get reduced bases for $S_{\alpha \beta}, S_{I \alpha}, etc.$. Note that the indices used in this reduced basis are $IJ = I\alpha, 33, 34, 55$. We use the same notation as in \cref{fig:321intnum}, namely we omit the $\cdot$ notation indicating the intersection product for brevity and use the shorthand $D_\alpha^2 = D_\alpha \cdot D_\beta, D_\alpha^3 = D_\alpha \cdot D_\beta \cdot D_\gamma$.}
\label{Mred}
\end{center}
\end{table}

\begin{table}
\begin{center}
	\scalebox{.73}{$
\begin{array}{|c|c|c|c|c|c|c|c|c|c|}\hline
 \text{} & \text{0$\alpha $} & \text{1$\alpha $} & \alpha \beta  & \text{$\alpha $3} & \text{$\alpha $4} & \text{$\alpha $5} & 03 & 05 & 34 \\\hline
 \text{0$\alpha' $} & K D_{\alpha }^2 & Y D_{\alpha }^2 & D_{\alpha }^3 & 0 & 0 & 0 & -\Sigma _2 Y D_{\alpha } & -\Sigma _3 Y D_{\alpha } & 0 \\\hline
 \text{1$\alpha' $} & Y D_{\alpha }^2 & K D_{\alpha }^2 & D_{\alpha }^3 & \Sigma _2 D_{\alpha }^2 & 0 & \Sigma _3 D_{\alpha }^2 & \Sigma _2 Y D_{\alpha } & \Sigma _3 Y D_{\alpha } & 0 \\\hline
 \alpha' \beta'  & D_{\alpha }^3 & D_{\alpha }^3 & 0 & 0 & 0 & 0 & 0 & 0 & 0 \\\hline
 \text{$\alpha' $3} & 0 & \Sigma _2 D_{\alpha }^2 & 0 & -2 \Sigma _2 D_{\alpha }^2 & 0 & 0 & -\Sigma _2 Y D_{\alpha } & 0 & -\Sigma _2 \Sigma _3 D_{\alpha } \\\hline
 \text{$\alpha' $4} & 0 & 0 & 0 & 0 & -2 \Sigma _3 D_{\alpha }^2 & \Sigma _3 D_{\alpha }^2 & 0 & 0 & -\Sigma _2 \Sigma _3 D_{\alpha } \\\hline
 \text{$\alpha' $5} & 0 & \Sigma _3 D_{\alpha }^2 & 0 & 0 & \Sigma _3 D_{\alpha }^2 & -2 \Sigma _3 D_{\alpha }^2 & 0 & -\Sigma _3 Y D_{\alpha } & \Sigma _2 \Sigma _3 D_{\alpha } \\\hline
 03 & -\Sigma _2 Y D_{\alpha } & \Sigma _2 Y D_{\alpha } & 0 & -\Sigma _2 Y D_{\alpha } & 0 & 0 & -\Sigma _2 Y \left(K+\Sigma _2+Y\right) & -\Sigma _2 \Sigma _3 Y & 0 \\\hline
 05 & -\Sigma _3 Y D_{\alpha } & \Sigma _3 Y D_{\alpha } & 0 & 0 & 0 & -\Sigma _3 Y D_{\alpha } & -\Sigma _2 \Sigma _3 Y &- \Sigma _3 Y \left(K+\Sigma _3+Y\right) & 0 \\\hline
 34 & 0 & 0 & 0 & -\Sigma _2 \Sigma _3 D_{\alpha } & -\Sigma _2 \Sigma _3 D_{\alpha } & \Sigma _2 \Sigma _3 D_{\alpha } & 0 & 0 & -\Sigma _2 \Sigma _3 \left(K+\Sigma _2+\Sigma _3\right) \\\hline
\end{array}$}
\caption{Matrix elements $M_{\rm red (IJ)(KL)}$ in alternative
  basis. Compared to the basis used in \cref{Mred}, we find this basis
  to be more convenient
in some approaches for computing the chiral
  indices---see \cref{sec:321-spectrum}.}
\label{t:mr-alternate}
\end{center}
\end{table}

\subsection{3D Chern--Simons couplings}
As described in Section 5 of \cite{Jefferson:2021bid}, and following
\cite{Grimm:2011fx,Grimm:2011sk,Cvetic:2012xn,Cveti__2014}, the final step in our strategy
for computing the chiral indices $\chi_{\mathsf r} = n_{\mathsf r} - n_{\mathsf r^*}$ of the $\SM$ model is to match the vertical F-theory fluxes
$\Theta_{\bar I \bar J}$ (satisfying the symmetry constraints \cref{symconst}) with the one-loop Chern--Simons (CS) couplings
$\Theta^{\text{3D}}_{\bar I \bar J}$ characterizing
the low-energy 3D $\cN=2$ action describing M-theory
compactified on $X_5$.\footnote{The ``gauge'' basis
  indices $\bar I$ take the values $\bar I = \bar 0, \bar 1, \alpha,
  3,4 ,5$, where $\hat D_{\bar 0},\hat D_{\bar 1}$ are defined in
  \cref{U1div}. See also \cref{changebasis}.} We then use the fact that the 3D CS couplings can
be expressed as linear combinations of the chiral indices,
	\begin{align}
		 \Theta^{\text{3D}}_{\bar I \bar J} =- x_{\bar I \bar J}^{\mathsf r} \chi_{\mathsf r}\,,
	\end{align}
leading to the invertible\footnote{It is not entirely clear that every crepant resolution corresponds to a 3D phase in which the system \cref{linearsystem} is invertible---see Appendix G of \cite{Jefferson:2021bid}.} system
	\begin{align}
	\label{linearsystem}
		\Theta_{\bar I \bar J} =- x_{\bar I \bar J}^{\mathsf r} \chi_{\mathsf r}\,.
	\end{align}
Inverting the matrix of coefficients $x_{\bar I \bar J}^{\mathsf r}$ allows us to write the chiral indices as linear combinations of $\Theta_{\bar I \bar J}$.

Since we have already computed the F-theory fluxes through surfaces, what remains is to compute the coefficients $x_{\bar I \bar K}^{\textbf r}$. This can be accomplished with the help of 3D $\cN=2$ supersymmetry. From the point of view of the 3D effective field theory, the one-loop 3D CS couplings can be expressed as\footnote{Our conventions are as follows. Let $\mathfrak{g}$ be a semisimple Lie algebra. We may view a collection of real Coulomb branch moduli $\varphi^i$ as the components of a real 3D vector multiplet scalar expanded in a basis of simple coroots, $\varphi =\varphi^i \alpha^{\vee}_i$ where $i = 1, \dots, \text{rank}(\mathfrak{g})$. Similarly, a weight $w$ transforming in a representation $\mathsf r$ of $\mathfrak{g}$ may be expanded in a basis of fundamental weights, $w = w_i \omega^i$. The fundamental weights are canonically dual to the simple coroots and we denote their inner product by $\alpha^{\vee}_i \cdot \omega^j =\delta_i^j$, hence $\varphi \cdot w = \varphi^i w_i$. We define $\varphi^{\bar 0} = 1/r_{\text{KK}}$.}
	\begin{align}
	\begin{split}
	\label{qIJ1loop}
		\Theta_{i j}^\text{3D} &=   \sum_{w} ( \frac{1}{2} + \floor{|r_\text{KK}\varphi \cdot w| })  \,\sign(\varphi \cdot w)w_{i} w_{j}\,,\\
		\Theta_{ \bar 0i}^\text{3D}  &=\sum_{w}  (  \frac{1}{12} +\frac{1}{2} \floor{|r_\text{KK} \varphi \cdot w|} ( \floor{|r_\text{KK}\varphi \cdot w| } + 1) ) w_{i}\,,\\
		\Theta_{ \bar 0 \bar  0}^\text{3D}  &=\sum_{w}  \frac{1}{6} \floor{|r_\text{KK}\varphi \cdot w| } (\floor{|r_\text{KK}\varphi \cdot w|} + 1) (2 \floor{|r_\text{KK}\varphi \cdot w|} + 1)\,.
	\end{split}
	\end{align}
In order to match the above expressions with our M-theory vacuum, we must supply as input the signs of the central charges of 3D BPS particles in the specific phase of the vector multiplet moduli space described by the resolution $X_5$. These signs can in principle be computed geometrically by using the K\"ahler class $\hat J = \varphi^{\bar I} \hat D_{\bar I}$ to compute the volumes $\text{vol}(C_w) = \hat J \cdot C_w = \varphi^{\bar I} w_{\bar I}$ of all primitive holomorphic and anti-holomorphic curves $C_w$ in the Mori cone of $X_5$ (and its negative), as the 3D BPS particle spectrum descends from M2 branes wrapping these curves.\footnote{The coefficients $w_{\bar I}$ can either be interpreted as KK charges $w_{\bar 0}$, or as Cartan charges/Dynkin labels $w_{i}$ of weights transforming in some representation $\mathsf R$ of $\mathfrak{g}$.} Since the volumes $\text{vol}(C_w)$ are equal to the BPS central charges, partitioning the Mori cone generators $\{C_w\}$ into holomorphic and anti-holomorphic (based on whether $\text{vol}(C_w)$ is positive or negative) determines the full set of signs associated to $X_5$.

Rather than following the above approach, we use a shortcut. Namely,
we identify the signs appearing in \cref{qIJ1loop} by formally
matching intersection numbers of the form $W_{\bar I \bar J \bar K}$
with 5D CS couplings $k_{\bar I \bar J \bar K}$ in which the
coefficients have been appropriately ``promoted'' to the classes of
matter curves $C_{\mathsf R} \in B$; see Section 5 of
\cite{Jefferson:2021bid} for a more detailed discussion of this procedure.

To simplify the task of computing the 5D CS couplings, we first focus on the KK charges. To this end, we use the fact that $Y$ is the pushforward of the intersection of the zero section with the generating section to conclude that the only elementary BPS particles with nontrivial KK charge are those transforming in the representations $(\textbf{1},\textbf{1})_2,(\textbf{1},\textbf{2})_{\frac{3}{2}},(\textbf{3},\textbf{1})_{-\frac{4}{3}}$, as these are the only local matter representations whose associated matter curves in $B$ have the schematic form
	\begin{align}
		C_{\mathsf R} = Y \cdot (\cdots)\,.
	\end{align}
We then proceed to compute 5D Chern--Simons couplings according to the standard matching procedure detailed in Section 5.2 of \cite{Jefferson:2021bid}. Following this approach, we find that the resolution \cref{eq:321res} that leads to $X_5$ corresponds to the signs in \cref{321signstable} and KK masses in the parametric regime
	\begin{align}
	\label{eqn:KKcharge}
		\floor{|r_\text{KK} \varphi \cdot w^{(\textbf{1},\textbf{1})_2}|}  =  \floor{|r_\text{KK} \varphi \cdot w_+^{(\textbf{1},\textbf{2})_{\frac{3}{2}}}|} =\floor{|r_\text{KK} \varphi \cdot w_-^{(\textbf{3},\textbf{1})_{-\frac{4}{3}}}|} = 1\,,
	\end{align}
with all other primitive BPS particles having vanishing KK charge. This data can in turn be used to evaluate the one-loop exact expressions for the 3D Chern--Simons couplings presented in \cref{qIJ1loop}.

In the following subsection, we describe the results of inverting the linear system \cref{linearsystem}.

\begin{table}
	\begin{center}
	\scalebox{.9}{$
	\begin{array}{cccc}
	(\textbf{1},\textbf{1})_{w_{\bar 1}} &
	(\textbf{1},\textbf{2})_{w_{\bar 1}} &
	(\textbf{3},\textbf{1})_{w_{\bar 1}} &
	(\textbf{3},\textbf{2})_{w_{\bar 1}} \\\hline\hline
	\begin{array}{c}
		\left(	\begin{array}{c|cccc}
			\frac{\varphi \cdot w}{|\varphi \cdot w|} & w_{\bar 1} &w_3 & w_4 & w_5 \\\hline
			+&1&0&0&0\\\hline
			+&2&0&0&0
		\end{array}\right)
	\end{array}&\begin{array}{c}
\left(
\begin{array}{c|cccc}
	\frac{\varphi \cdot w}{|\varphi \cdot w|} & w_{\bar 1} &w_3 & w_4 & w_5 \\\hline
 +&\frac{1}{2} & 1 & 0 & 0 \\
 +&\frac{1}{2} & -1 & 0 & 0 \\\hline
 +&\frac{3}{2} & 1 & 0 & 0 \\
 +&\frac{3}{2} & -1 & 0 & 0 \\
\end{array}
\right)
	\end{array}&\begin{array}{c}\left(
\begin{array}{c|cccc}
	\frac{\varphi \cdot w}{|\varphi \cdot w|} & w_{\bar 1} &w_3 & w_4 & w_5 \\\hline
+& \frac{1}{3} & 0 & 1 & 0 \\
 -&\frac{1}{3} & 0 & -1 & 1 \\
 -&\frac{1}{3} & 0 & 0 & -1 \\\hline
-&- \frac{4}{3} & 0 & 1 & 0 \\
 -&-\frac{4}{3} & 0 & -1 & 1 \\
 -&-\frac{4}{3} & 0 & 0 & -1 \\\hline
 +&\frac{2}{3} & 0 & 1 & 0 \\
 +&\frac{2}{3} & 0 & -1 & 1 \\
 +&\frac{2}{3} & 0 & 0 & -1 \\
\end{array}
\right)\end{array}&\begin{array}{c}\left(
\begin{array}{c|cccc}
	\frac{\varphi \cdot w}{|\varphi \cdot w|} & w_{\bar 1} &w_3 & w_4 & w_5 \\\hline
+& \frac{1}{6} & 1 & 1 & 0 \\
 +&\frac{1}{6} & 1 & -1 & 1 \\
 +&\frac{1}{6} & 1 & 0 & -1 \\
 +&\frac{1}{6} & -1 & 1 & 0 \\
 -&\frac{1}{6} & -1 & -1 & 1 \\
 -&\frac{1}{6} & -1 & 0 & -1 \\
\end{array}
\right)\end{array}
	\end{array}
$}
\end{center}
\caption{Signs and Cartan charges associated to the BPS spectrum of the $\SM$ model resolution \cref{eq:321res}. The charges are the Dynkin coefficients $w_i$ of the weights $w$ transforming in various representations and the signs correspond to the signs of the BPS central charges $\varphi \cdot w$ for a given choice of Coulomb branch moduli $\varphi^i$. The indices $i$ of the Dynkin coefficients are chosen to match the indices of the Cartan divisors $\hat D_i$ in \cref{321Cartandiv} associated to the simple coroots of the gauge group. We adopt the convention that the Coulomb branch modulus $\varphi^{\bar 1}$ dual to the $\au(1)$ factor is non-positive in the Coulomb branch, in contrast to the Coulomb branch moduli $\varphi^{i_2=3}, \varphi^{i_3 =4,5}$ associated to nonabelian Cartan $\au(1)$s, which are non-negative.}
\label{321signstable}
\end{table}

\subsection{Chiral indices}
\label{chiralind}

Our goal in this section is to match the geometric fluxes with the 3D Chern--Simons (CS) terms in order to determine the chiral indices of the $\SM$ model. For ease of comparison, we first convert the fluxes to the gauge basis $\bar I = \bar 0, \bar 1, \alpha, i_s$ using \cref{geotogauge} (recall that the indices $\bar 0, \bar 1$ correspond, respectively, to the abelian gauge factors $\U(1)_{\text{KK}}, \U(1)$). In the gauge basis, we find that the linear relations \cref{geobasisnullfluxes} can be expressed as:
	\begin{align}
	\begin{split}
	\label{fluxconstraints}
		0&=\Theta_{ \bar 0 \bar  0} =\Theta_{ \bar 0 4 } =\Theta_{35} = \Theta_{44}\,, \\
		0&=\Theta_{45} + \Theta_{55}\,,\\
		0&=3\Theta_{\bar 1 5}- 4 \Theta_{55}\,,\\
		0&=\Theta_{ \bar 0 5} + \Theta_{55}\,, \\
		0&=6 \Theta_{ \bar 1 4} - 3 \Theta_{34}+ 4\Theta_{55}\,, \\
		0&=6\Theta_{\bar  1 3} - 9 \Theta_{33} - 2 \Theta_{34}\,, \\
		0&=\Theta_{ \bar 0 3} + \Theta_{33}\,, \\
		0&= 12 \Theta_{\bar 1  \bar 1} -15 \Theta_{33}- 4 \Theta_{34} - 16 \Theta_{55}\,, \\
		0&= 6 \Theta_{ \bar 0 \bar  1} + 3 \Theta_{33} + 4 \Theta_{55}\,.
	\end{split}
	\end{align}
Moreover, the one-loop 3D Chern--Simons couplings $\Theta_{\bar I \bar J}^\text{3D}$ are given by
	\begin{align}
	\begin{split}
		\Theta^\text{3D}_{\bar 0 \bar 1}&=\frac{1}{12} \chi _{(\textbf{1},\textbf{1})_{1}}-\frac{11}{6} \chi _{(\textbf{1},\textbf{1})_{2}}-\frac{1}{12} \chi _{(\textbf{1},\textbf{1})_{\frac{1}{2}}}-\frac{7}{4} \chi _{(\textbf{1},\textbf{1})_{\frac{3}{2}}}\\
		&+\frac{1}{6} \chi _{(\textbf{3},\textbf{1})_{\frac{2}{3}}}-\frac{1}{12} \chi _{(\textbf{3},\textbf{1})_{-\frac{1}{3}}}+\chi _{(\textbf{3},\textbf{1})_{-\frac{4}{3}}}-\frac{1}{12} \chi _{(\textbf{3},\textbf{2})_{\frac{1}{6}}}\,,\\
		\Theta^\text{3D}_{\bar 1\bar 1} &= \frac{1}{2} \chi _{(\textbf{1},\textbf{1})_1}+6 \chi _{(\textbf{1},\textbf{1})_2}+\frac{1}{4} \chi _{(\textbf{1},\textbf{2})_{\frac{1}{2}}}+\frac{9}{2} \chi _{(\textbf{1},\textbf{2})_{\frac{3}{2}}}\\
		&+\frac{2}{3} \chi _{(\textbf{3},\textbf{1})_{\frac{2}{3}}}-\frac{1}{18} \chi _{(\textbf{3},\textbf{1})_{-\frac{1}{3}}}-\frac{40}{9} \chi _{(\textbf{3},\textbf{1})_{-\frac{4}{3}}}+\frac{1}{36} \chi _{(\textbf{3},\textbf{2})_{\frac{1}{6}}}\,,\\
		\Theta^\text{3D}_{\bar 13} &=-\frac{3}{2}  \chi _{(\textbf{1},\textbf{2})_{\frac{3}{2}}}-\frac{1}{3} \chi _{(\textbf{3},\textbf{2})_{\frac{1}{6}}}\,,\\
		\Theta^\text{3D}_{\bar 14}&=\frac{1}{3} \chi _{(\textbf{3},\textbf{1})_{-\frac{1}{3}}}-\frac{1}{6} \chi _{(\textbf{3},\textbf{2})_{\frac{1}{6}}}\,,\\
		\Theta^\text{3D}_{\bar 15} &= \frac{4}{3} \chi _{(\textbf{3},\textbf{1})_{-\frac{4}{3}}}\,,\\
		\Theta^\text{3D}_{33} &=\chi _{(\textbf{1},\textbf{2})_{\frac{1}{2}}}+2 \chi _{(\textbf{1},\textbf{2})_{\frac{3}{2}}}+\chi _{(\textbf{3},\textbf{2})_{\frac{1}{6}}}\,,\\
		\Theta^\text{3D}_{34}&=-\chi _{(\textbf{3},\textbf{2})_{\frac{1}{6}}}\,,\\
		\Theta^\text{3D}_{35}&=0\,,\\
		\Theta^\text{3D}_{44}&=\chi _{(\textbf{3},\textbf{1})_{\frac{2}{3}}}-\chi _{(\textbf{3},\textbf{1})_{-\frac{4}{3}}}+\chi _{(\textbf{3},\textbf{2})_{\frac{1}{6}}}\,,\\
		\Theta^\text{3D}_{45}&=-\frac{1}{2} \chi _{(\textbf{3},\textbf{1})_{\frac{2}{3}}}+\frac{1}{2} \chi _{(\textbf{3},\textbf{1})_{-\frac{1}{3}}}+\frac{1}{2} \chi _{(\textbf{3},\textbf{1})_{-\frac{4}{3}}}\,,\\
		\Theta^\text{3D}_{55}&=\chi _{(\textbf{3},\textbf{1})_{\frac{2}{3}}}-\chi _{(\textbf{3},\textbf{1})_{-\frac{1}{3}}}-2 \chi _{(\textbf{3},\textbf{1})_{-\frac{4}{3}}}\,.
	\end{split}
	\end{align}
By identifying the 3D CS terms with the geometric fluxes in \cref{chiralind} (see \cref{3Dmatch}), we learn that the following constraints are satisfied by the chiral multiplicities in the $\SM$ model:
	\begin{align}
		\begin{split}
			\label{321anomaly}
				0 &=-\chIndex{(\textbf{1},\textbf{1})_2}\ -\chIndex{(\textbf{1},\textbf{2})_{\frac{3}{2}}}+\frac{1}{3} \chIndex{(\textbf{3},\textbf{1})_{\frac{2}{3}}}-\frac{1}{3} \chi_{(\textbf{3},\textbf{1})_{-\frac{1}{3}}}\,,\\
				0&=-\chIndex{(\textbf{1},\textbf{2})_{\frac{1}{2}}} -3 \chIndex{(\textbf{1},\textbf{2})_{\frac{3}{2}}}+\frac{2}{3} \chIndex{(\textbf{3},\textbf{1})_{\frac{2}{3}}}+\frac{1}{3} \chi_{(\textbf{3},\textbf{1})_{-\frac{1}{3}}}\,,\\
				0&=-\chIndex{(\textbf{3},\textbf{1})_{-\frac{4}{3}}}+ \frac{1}{3} \chi_{(\textbf{3},\textbf{1})_{\frac{2}{3}}}-\frac{1}{3} \chi_{(\textbf{3},\textbf{1})_{-\frac{1}{3}}}\,,\\
				0&=-\chIndex{(\textbf{1},\textbf{1})_1}+ 2 \chIndex{(\textbf{1},\textbf{2})_{\frac{3}{2}}}-\frac{4}{3} \chIndex{(\textbf{3},\textbf{1})_{\frac{2}{3}}}+\frac{1}{3} \chi_{(\textbf{3},\textbf{1})_{-\frac{1}{3}}}\,,\\
				0&=-\chIndex{(\textbf{3},\textbf{2})_{\frac{1}{6}}} -\frac{2}{3} \chIndex{(\textbf{3},\textbf{1})_{\frac{2}{3}}}-\frac{1}{3} \chi_{(\textbf{3},\textbf{1})_{-\frac{1}{3}}}\,.
		\end{split}
	\end{align}
After rearrangement, these constraints exactly match those in \cref{eq:321-AC-sols}. The constraints can also be used to write the chiral indices in terms of a minimal subset of the fluxes:
	\begin{align}
		\begin{split}
		\label{chiasphi}
			\chi _{(\textbf{1},\textbf{1})_1}&=2 \Theta _{33}+\Theta _{34}+2 \Theta _{55}\,,\\
			\chi _{(\textbf{1},\textbf{1})_2}&=- \Theta _{33}-\Theta _{55}\,,\\
			\chi _{(\textbf{1},\textbf{2})_{\frac{1}{2}}}&= -3 \Theta _{33}-\Theta _{34}\,,\\
			\chi _{(\textbf{1},\textbf{2})_{\frac{3}{2}}}&= \Theta _{33}\,,\\
			\chi _{(\textbf{3},\textbf{1})_{-\frac{4}{3}}}&=- \Theta _{55}\,,\\
			\chi _{(\textbf{3},\textbf{1})_{-\frac{1}{3}}}&= 2 \Theta _{55}-\Theta _{34}\,,\\
			\chi _{(\textbf{3},\textbf{1})_{\frac{2}{3}}}&=- \Theta _{34}-\Theta _{55}\,,\\
			\chi _{(\textbf{3},\textbf{2})_{\frac{1}{6}}}&=  \Theta _{34}\,.
		\end{split}
	\end{align}
In terms of the
parameterization in \cref{t:321-families}, we see that the multiplicities of the
first and second families are given by $-\Theta_{55}$ and
$\Theta_{33}$ respectively, and the number of Standard Model
generations is given by $\Theta_{34} + 2 (\Theta_{33} +
\Theta_{55})$.  When there is no exotic matter, the number of
generations of MSSM matter is then simply $\Theta_{34}$.

Note that in the chiral multiplicity formulae of \cref{chiasphi}, we
can freely replace the fluxes $\Theta_{33}, \Theta_{55}$ with
$-\Theta_{03}, - \Theta_{05}$, using \cref{fluxconstraints} and the
fact that $\Theta_{\bar{0}i}= \Theta_{0i}$ in the symmetry-constrained
space where $\Theta_{\alpha I}= 0$.  This allows us to easily move
back and forth between the bases of \cref{Mred} and
\cref{t:mr-alternate}.

\subsection{Chiral spectrum}
\label{sec:321-spectrum}

We now use the formulation of  $\mr$ in \cref{t:mr-alternate} to solve
the constraint equations $\Theta_{\alpha I} = 0$ and compute
formulae for the chiral multiplicities, where $\Theta
\sim\mr \phi$ is expressed in terms of flux parameters $\phi^{IJ}$.
This can be done in a systematic fashion even though, as mentioned
above, in some bases null vectors may remain for \cref{t:mr-alternate}.
However, unlike F-theory models without $\U(1)$ factors, it is not straightforward to
obtain a completely general solution by solving the constraint equations for all flux parameters
$\phi^{\alpha I}$. As discussed in \cite{Jefferson:2021bid}, the main obstruction to obtaining completely general solutions in models with $\U(1)$ factors is imposing the constraints
\begin{equation}
	\Theta_{1\alpha}=0\,.
\end{equation}
Despite these apparent obstructions, in the case of the $\SM$ model we nevertheless find that we are able to circumvent the difficulties associated with solving the above constraints in full generality. The approach described in this subsection thus leads to a general solution to the constraints $\Theta_{I\alpha}= 0$, along with general expressions for the associated fluxes and chiral
multiplicities, but at the cost of leaving the solution for the variable $\phi^{1
  \alpha}$ implicit.  In \cref{altsym}, we specialize to specific bases and follow an approach that more closely parallels the approach used in \cite{Jefferson:2021bid} for
purely nonabelian theories.

To illustrate our systematic solution, we spell out the
first constraints explicitly.  We begin by solving the constraint
$\Theta_{\alpha \beta} = 0$.  Note that the space of homologically
nontrivial $S_{\alpha \beta}$ has the same dimension $h^{1, 1}(B)$ as
the space of, e.g., $S_{0 \alpha}$ by Poincar\'{e} duality.  From
\cref{t:mr-alternate}, we see that this set of constraints simply
imply that
\begin{equation}
\phi^{1 \alpha} = -\phi^{0 \alpha} \,.
\label{eq:c1}
\end{equation}
We next solve the equations $\Theta_{0 \alpha} = 0$ for the flux
parameters $\phi^{\alpha \beta}$, giving
\begin{equation}
\phi^{\alpha \beta} = \phi^{0 \alpha} (Y-K)^\beta +
\phi^{03} (\Sigma_2)^\alpha Y^\beta +
\phi^{05} (\Sigma_3)^\alpha Y^\beta  \,.
\label{eq:cc2}
\end{equation}
Next, consider the equation for $\Theta_{\alpha 4}$:
\begin{equation}
 \phi^{\alpha 5} (\Sigma_3)^\beta =
2 \phi^{\alpha 4} (\Sigma_3)^\beta +
 \phi^{34} (\Sigma_2)^\alpha (\Sigma_3)^\beta  \,.
\label{eq:c3}
\end{equation}
Since $\phi^{\alpha 5} D_\alpha$ is always combined with a factor of
$\Sigma_3$, we can use this equation to replace $\phi^{\alpha 5}$
everywhere it appears, and although not all the parameters
$\phi^{\alpha 5}$ are fixed through this equation, those that are not
are null vectors.
Note that the constraint equations \labelcref{eq:c1,eq:cc2,eq:c3} are
all consistent even when $c_2$ is not an even class, so that the
associated symmetries can be preserved even in such cases.  For
example, from  \cref{eq:c2} we see that $\phi^{1 \alpha}$ and $\phi^{0
\alpha}$ each are half-integer flux parameters precisely when
$(\Sigma_3+ Y)^{\alpha}$ is odd
and $c_2$ is not even, so \cref{eq:c1} is consistent even in
these cases.  Similar arguments hold for the other symmetry
constraints, using in particular the fact that $c_2 (B) + K^2$ is
always even \cite{Collinucci:2010gz}.

Continuing in this fashion, we can solve the $\Theta_{\alpha 3} =
\Theta_{\alpha 5}= 0$ constraints, giving
\begin{equation}
\label{eq:a5}
\begin{aligned}
\phi^{\alpha 3} (\Sigma_2)^\beta  &=
\frac{1}{2} (-\phi^{0 \alpha} -\phi^{34} (\Sigma_3)^\alpha -\phi^{03}Y^\alpha)
(\Sigma_2)^\beta\,,
\\
 \phi^{\alpha  4} (\Sigma_3)^\beta  &=
\frac{1}{3} (-\phi^{0 \alpha} -\phi^{34} (\Sigma_2)^\alpha -\phi^{05}Y^\alpha)
(\Sigma_2)^\beta\,.
\end{aligned}
\end{equation}
Note that this implies some extra nontrivial integral constraints on
some of the flux parameters.

Putting this all together, the remaining constraint $\Theta_{1 \alpha}
= 0$ states that
\begin{equation}
A \cdot   \bar{H}
= \phi^{34} \Sigma_2 \cdot \Sigma_3 -9 \phi^{03} \Sigma_2 \cdot Y
-8 \phi^{05} \Sigma_3 \cdot Y
\label{eq:remaining-constraint}
\end{equation}
where
	\begin{align}
	A = \phi^{0 \alpha} D_\alpha =-\phi^{1 \alpha}
D_\alpha
\end{align} and
\begin{align}
	\bar{H} =
-12K -4 \Sigma_3-3 \Sigma_2+12Y
\end{align}
is the (scaled) height pairing
divisor.
Before imposing this set of constraints, the remaining fluxes are
given by
\begin{equation}
\label{eq:general-fluxes}
\begin{aligned}
\Theta_{03} & =
\frac{1}{2}\Sigma_2 \cdot Y \cdot
(-3A + \phi^{34} \Sigma_3 -2 \phi^{05} \Sigma_3
- \phi^{03} ( 2K + 2 \Sigma_2 + Y))\,,\\
\Theta_{05} & =
-\frac{1}{3}\Sigma_3 \cdot Y \cdot
(4A + \phi^{34} \Sigma_2  + 3 \phi^{03} \Sigma_2
+ \phi^{03} ( 3K + 3 \Sigma_2 + Y))\,,\\
\Theta_{34} & =
\frac{1}{6}\Sigma_2 \cdot  \Sigma_3 \cdot
(A +  (3\phi^{03} -2 \phi^{05}) Y
- \phi^{34} (6K + 2 \Sigma_2 + 3 \Sigma_3))\,.
\end{aligned}
\end{equation}
Note that this matches with the expectation that the exotic
matter multiplicities
$\chi_{(\textbf{1},\textbf{2})_{\frac{3}{2}}},
\chi_{(\textbf{3},\textbf{1})_{-\frac{4}{3}}}$
 associated with $\Theta_{03}, \Theta_{05}$ explicitly can only be
 nonzero when $\Sigma_2 \cdot Y, \Sigma_3 \cdot Y$ are non-vanishing
 as expected from  \cref{tab:codim12}, while the possibility of matter
 charged under both the $\SU(3)$ and $\SU(2)$ factors controlled by
 $\Theta_{34}$ can only be non-vanishing if $\Sigma_2 \cdot \Sigma_3$
 is a nontrivial curve in the base.

The constraints imposed by \cref{eq:remaining-constraint}
reduce the number of independent fluxes $\phi^{1 \alpha}$, $\phi^{03}$,
$\phi^{05}$, $\phi^{34}$ to three independent parameters
that give the fluxes \cref{eq:general-fluxes} and some null vectors
associated with parameters in $A$.
One way to think about these constraints is as a geometric condition
on curves.  The curves $\Sigma_2 \cdot \Sigma_3$, $\Sigma_2 \cdot Y$,
$\Sigma_3 \cdot Y$ span a space of dimension at most three in the
linear
space
of curves on the base.  We can think of $\bar{H}$ as a linear map from the
space of divisors on $B$ to the dual space of curves on $B$.  The set
of possible flux configurations can be determined from the
intersection of the image of this map $\bar{H}$ with the space of curves
spanned by the primary pair intersections listed above. These constraints can be thought of as constraints on the parameters
in $A$ or on the  $\phi^{03},
\phi^{05}, \phi^{34}$.  For example, if all three of the primary pair
intersections are independent and the image of $\bar{H}$ contains this full
space (in which case, the curves are all contained in the divisor
$\bar{H}$), then we can take as the basic set of three parameters any set of
all or some of the parameters $\phi^{03},
\phi^{05}, \phi^{34}$, and a complementary set of parameters in $A$.

In some sense,
\cref{eq:general-fluxes}
along with the constraint
\cref{eq:remaining-constraint} give the simplest general formulation of the
available fluxes for  an arbitrary base geometry.
We can proceed somewhat more explicitly, however, if we assume that the
primary pair intersection products appearing in the fluxes are
non-vanishing.  In particular, if $\Sigma_2 \cdot \Sigma_3$ is a
non-vanishing curve in the base, then we can simply solve
\cref{eq:remaining-constraint} for $\phi^{34} \Sigma_2 \cdot \Sigma_3$
and plug into \cref{eq:general-fluxes}.  This gives a general formula
for
the chiral multiplicities
%\wati{Patrick, please check relation of
  %multiplicities to fluxes in this basis is correct with proper signs.
%Do you see any significance to the shift on H appearing here?}\patrick{What is $N$ in the first line of (7.45)?}
\begin{equation}
\label{eq:general-multiplicities}
%\chi_{(\textbf{1},\textbf{2})_{\frac{3}{2}}}& = -\Theta_{03}=
%Y \cdot \left(
%\phi^{03}K \cdot \Sigma_2 + (\phi^{03} \Sigma_2 + \phi^{05} \Sigma_3) \cdot(N
%-4Y) - A\cdot (\bar{H} -3 \Sigma_2)/2 \right)
%\\
%\chi_{(\textbf{3},\textbf{1})_{-\frac{4}{3}}}& = \Theta_{05}
%=-Y \cdot \left(
%\phi^{03} \Sigma_2 \cdot (\Sigma_3  + 3 Y)
%+ \phi^{05} \Sigma_3 \cdot( K + 3 Y + \Sigma_3)
%+ A \cdot (\bar{H} -4 \Sigma_3)/3 \right)
%\\
%\chi_{(\textbf{3},\textbf{2})_{-\frac{1}{6}}}& = \Theta_{34}
%= -Y \cdot \left( K\cdot (9 \phi^{03} \Sigma_2 + 8 \phi^{05} \Sigma_3) +
%(4 \Sigma_3 + 3 \Sigma_2)\cdot
%(\phi^{03} \Sigma_2 +  \phi^{05} \Sigma_3) \right)\\
%&
%\hspace*{0.1in}
%-A \cdot (12K^2 + 2 \Sigma_3^2 + 3 \Sigma_3\cdot \Sigma_2 + \Sigma_2^2 + K\cdot (10
%\Sigma_3 + 7 \Sigma_2-12Y)-Y\cdot (6 \Sigma_3 + \Sigma_2)) \,.
%%%%%
\begin{aligned}
\chi_{(\textbf{1},\textbf{2})_{\frac{3}{2}}} = -\Theta_{03}&=\Sigma_2 \cdot Y \cdot  \left(K+\Sigma_2-4 Y\right)\phi^{03}+\Sigma_3 \cdot Y \cdot \left(\Sigma_2-4 Y\right)  \phi^{05}\\
& ~~~-\frac{1}{2} A \cdot Y \cdot \left( \bar H-3 \Sigma
_2\right)\,,\\
& =\Sigma_2 \cdot Y \cdot  \left(K+\Sigma_2-4
Y\right)\phi^{03}+\Sigma_3 \cdot Y \cdot \left(\Sigma_2 -4Y\right)  \phi^{05}\\
& ~~~ +A \cdot Y \cdot \left(6K + 2 \Sigma_3 + 3 \Sigma_2-6Y
\right)\,,\\
\chi_{(\textbf{3},\textbf{1})_{-\frac{4}{3}}}= \Theta_{05} &=
%-\frac{1}{3} Y \cdot \left(3 K\cdot \Sigma_3+6 \Sigma_3 \cdot
%\Sigma_2+9 \Sigma_2\cdot Y+\Sigma_3\cdot Y\right)\phi^{03}\\
-\Sigma_2 Y \cdot \left(\Sigma_3+ 3Y\right)\phi^{03}
-\Sigma_3 Y \cdot \left(\Sigma_3+ 3Y + K\right)\phi^{05}
\\
& ~~~-\frac{1}{3} A \cdot Y \cdot \left(\bar H+4 \Sigma_3\right)\,\\
& =-\Sigma_2 Y \cdot \left(\Sigma_3+ 3Y\right)\phi^{03}
-\Sigma_3 Y \cdot \left(\Sigma_3+ 3Y + K\right)\phi^{05}
\\
& ~~~+A \cdot Y \cdot \left( 4K + \Sigma_2 + 4Y \right)\,,\\\chi_{(\textbf{3},\textbf{2})_{-\frac{1}{6}}} = \Theta_{34} &=-\Sigma_2 \cdot Y \cdot \left(9 K+3 \Sigma_2+4 \Sigma_3\right) \phi^{03}-\Sigma_3 \cdot Y \cdot \left(8 K+3 \Sigma_2+4 \Sigma_3\right) \phi^{05}\\
& ~~~+\frac{1}{6} A  \cdot \left(\Sigma_2 \cdot \Sigma_3-\bar H \cdot \left(6 K+2 \Sigma_2+3 \Sigma_3\right)\right)\\
&=-\Sigma_2 \cdot Y \cdot \left(9 K+3 \Sigma_2+4 \Sigma_3\right)
\phi^{03}-\Sigma_3 \cdot Y \cdot \left(8 K+3 \Sigma_2+4 \Sigma_3\right) \phi^{05}\\
& ~~~-A \cdot Y  \cdot \left(12K + 6  \Sigma_3
+4\Sigma_2\right)
\\
& ~~~+ A \cdot K \cdot \left( 12K + 10 \Sigma_3 + 7 \Sigma_2\right)\\
&
 ~~~+ A \cdot (\Sigma_2 \cdot \Sigma_2 + 3 \Sigma_3 \cdot \Sigma_2 + 2 \Sigma_3 \cdot \Sigma_3)\,.
\end{aligned}
\end{equation}
Note that there appear to be more than three independent flux parameters here
since there are multiple parameters in $A$; these linearly independent
extra parameters, however, correspond to a combination of null vectors of the full
matrix $\mr$ and directions that are constrained by
\cref{eq:remaining-constraint} as discussed above.
For generic characteristic data, there are no further linear
dependencies on these multiplicities, and we can realize all three
linearly independent sets of chiral matter for many bases with arbitrary
characteristic data.  Note, however, that since
the flux parameters on the LHS of \cref{eq:a5} must be (half-)integer, and
there must be an integer (or half-integer) parameter $\phi^{05}$ satisfying
\cref{eq:remaining-constraint}, there are nontrivial
constraints on the choices of flux parameters appearing in
\cref{eq:general-multiplicities}.
We illustrate these features explicitly with some examples in the
following section.

The general formula for the multiplicities
\cref{eq:general-multiplicities} holds whenever $\Sigma_3 \cdot
\Sigma_2$ is a nontrivial curve, even if one of the other
primary pair intersection products vanishes such as $\Sigma_2 \cdot Y
= 0$, in which case $\Theta_{03} = 0$.
In the cases where $\Sigma_3 \cdot \Sigma_2 = 0$, we cannot have
jointly charged $\SU(3)\times \SU(2)$ matter, and the theory simplifies
significantly, essentially to the combinations of anomaly-free matter
expected for gauge groups $(\SU(2) \times \U(1))/\Z_2$ (as described in
more detail in \cite{Jefferson:2021bid}) and $(\SU(3) \times
\U(1))/\Z_3$.  The multiplicities in such a case can be similarly
computed assuming, e.g., that $\Sigma_2 \cdot Y$ is non-vanishing, in
which case we can solve \cref{eq:remaining-constraint} for $\phi^{03}
\Sigma_2 \cdot Y$ and compute the resulting multiplicities in a
similar fashion to the above analysis.

\subsection{Alternative treatment of the symmetry constraints}
\label{altsym}

As an alternative approach, closer in spirit to the direct approach
used for purely nonabelian theories in which the constraints are
solved immediately for all $\phi^{I \alpha}$, one can solve the symmetry constraints
$\Theta_{I\alpha}=0$ before restricting to the sublattice
$H_{2,2}^{\text{vert}}(X_5,\Z)$. We accomplish this by leaving
$\phi^{\hat K \hat L}$ free (note that the indices $\hat I \hat J$ are
still restricted to an appropriate subset of the indices $IJ$) and
instead replacing the intersection matrix $M_{(\hat I \hat J)(\hat K
  \hat L)}$ with the formal projection\footnote{In generic situations,
  we may solve the symmetry constraints $\Theta_{I\alpha} = 0$ by
  eliminating only independent flux parameters of the form
  $\phi^{I\alpha}$ (in fact, this can always be done for resolutions
  of F-theory models that exhibit strictly nonabelian gauge symmetry
  and admit a holomorphic zero section). In such situations, pairs of
  hatted indices $\hat I \hat J$, for which $\hat I, \hat J = 0,1,
  i_s$, are used to label the remaining unconstrained flux
  parameters. However, when the F-theory gauge algebra $\mathfrak g$
  includes $\au(1)$ factors, in some cases imposing the
  symmetry constraints $\Theta_{I\alpha} =0$ cannot be accomplished
  without eliminating some independent parameters $\phi^{IJ}$ for
  which $I, J \ne \alpha$ (and consequently leaving a subset of
  parameters of the form $\phi^{I\alpha}$ unconstrained.) In these
  somewhat special cases, we abuse notation and use $\hat I \hat J$, more generally, to denote the parameters that remain unconstrained after
  solving $\Theta_{I\alpha} = 0$, keeping in mind that the definition
  $\hat I \ne \alpha$ does not apply. \label{hat}}
(see Appendix C of \cite{Jefferson:2021bid} for further
details in a more general context)
	\begin{align}
	\label{geotheta1}
		{M}_{C( \hat I \hat J) (\hat K \hat L) }  = M_{C_{\text{na}}( \hat I \hat J) (\hat K  \hat L) } - M_{C_{\text{na}}( \hat I \hat  J)(1\alpha)} M_{C_{\text{na}}}^{+(1\alpha)(1\beta)}  M_{C_{\text{na}}(1\beta)( \hat K \hat L)}\,.
	\end{align}
In the above expression, $M_{C(IJ)(KL)}$ is the restriction of
$M_{(IJ)(KL)}$ to $\Lambda_C$ and similarly $ {
  M_{C_{\text{na}}(IJ)(KL)}}$ is the restriction of $M_{(IJ)(KL)}$ to
the sublattice $\Lambda_{ C_{\text{na}}} \subset \Lambda_S$ (note
$\Lambda_C \subset \Lambda_{ C_{\text{na}}}$) of backgrounds
preserving local Lorentz symmetry and at least the nonabelian (in this
case, $\asu(3) \oplus \asu(2)$) gauge symmetry. The
symmetric matrix $M_{C_{\text{na}}}$ can be obtained from $M$ by
acting on the right with a projection matrix
$P_{\text{na}}$\footnote{The right-acting projection $P_{\text{na}}$
  always exists for the class of resolutions of F-theory
  models analyzed in \cite{Jefferson:2021bid}.}
	\begin{equation}
		M_{C_{\text{na}}} = MP_{\text{na}} = P^{\text{t}}_{\text{na}} M
	\end{equation}
and has the following nontrivial matrix elements:
	\begin{align}
	\begin{split}
	\label{omegabar}
		 { M_{C_{\text{na}}}}_{ ({I} {J}) (K L)}&= W_{  I  J K L}- W_{  I  J|i_{s}} \cdot W^{i_{s}| j_{s'}} W_{  K Lj_{s'}} - W_{0  I  J} \cdot W_{  K L} - W_{  I  J} \cdot  W_{0  K L}\\
		&~~~ + W_{00}\cdot W_{  I  J} \cdot W_{  K L}
	\end{split}\\
	\begin{split}
	\label{omegabar2}
		 { M_{C_{\text{na}}}}_{ (1 \alpha) (KL)} &= D_\alpha \cdot W_{\bar 1 KL}\\
		 &=D_\alpha \cdot (-W_{1|k_{s''}}  W^{k_{s''}| i_s} W_{i_s  KL} + W_{1  IJ}-W_{0 KL}+(W_{00}-W_{01}) \cdot W_{ KL})
	\end{split}\\
	\begin{split}
	\label{omegabar3}
		{ M_{C_{\text{na}}}}_{(1\alpha)(1\beta)}&= D_\alpha \cdot D_\beta \cdot W_{\bar 1 \bar 1}\\
		&= D_\alpha \cdot D_\beta \cdot (- W_{1|k_{s''}} W^{k_{s''}|i_s} W_{1 i_s }+2(W_{00} - W_{01}) )\,.
	\end{split}
	\end{align}
Note that in contrast to $M_{C_{\text{na}}}$, whose components can be
expressed solely in terms of the triple intersections of the
characteristic data $K, \Sigma_2, \Sigma_3, Y$, the components of the
matrix $M^{+}_{C_{\text{na}}}$
(which is the inverse of ${ M_{C_{\text{na}}}}_{(1\alpha)(1\beta)}$,
assuming it exists)
generically depend on more intersection
numbers of $B$ as they are determined by the intersections of the
height pairing divisor $W_{\bar 1 \bar q}$ with the vertical curves of
$B$.
Without a more general approach to the analysis, $M^{+}_{C_{\text{na}}}$ can only be computed
explicitly for a specific choice of base, and thus has to be worked
out on a case-by-case basis. We can plug in the values of $W_{IJKL}$
presented in \cref{fig:321intnum} into the above expressions in order
to explicitly evaluate the fluxes $\Theta_{\hat I \hat J} = M_{C(\hat
  I \hat J)(\hat K \hat L) } \phi^{\hat K \hat L}$.
We perform this analysis explicitly for the base $\bP^3$ in the
next section.

\section{Examples}
\label{sec:specific}

We next turn our attention to some examples with constrained choices
of characteristic data and/or a specific choice of base $B$.

\subsection{\texorpdfstring{$B = \bP^3$}{}}
\label{sec:p3}
We describe the specific case $B= \bP^3, K= - 4 H, \Sigma_2 =
n_2 H, \Sigma_3 = n_3 H, Y = y H$
to illustrate features of the general theory, which we analyze in several
complementary ways.

\subsubsection{\texorpdfstring{$B = \bP^3$}{} as a special case
  of the general formalism}

The allowed values of the parameters $n_2, n_3, y$ must
satisfy the inequalities
\begin{equation}
n_2 > 0, \; \; n_3 > 0, \; \; [s_1] = 4 + y-n_2-n_3  \geq 0, \; \;
[d_2] = 16-2y-n_2-2n_3 \geq 0 \,,
\end{equation}
or
\begin{equation}
y = 0, \; \; n_2 > 0, \; \; n_3 > 0, \; \; [d_2] = 16-n_2-2n_3 \geq 0 \,,
\end{equation}
for a good Weierstrass model to exist with the gauge group $\SM$ and
no further enhancement of the group \cite{Raghuram:2019efb,
Raghuram:2020vxm}. Note that in this example there is a single base index $\alpha = H$.

Note that for this class of examples, we have
	\begin{align}
	\begin{split}
		\frac{1}{4} c_2^2&=\left(17 n_2+37 n_3-192\right) y+22 n_3^2+8 \left(n_2-16\right) n_2+25 n_2 n_3-224 n_3+22 y^2+912 \\
		&~~~-\frac{1}{4} [ \left(5 n_2+9 n_3\right)
                  y^2+\left(n_2^2+6 n_3 n_2+7 n_3^2\right) y+n_3
                  \left(n_2+n_3\right) \left(3 n_2+2 n_3\right)+2
                  y^3]\,.
\label{eq:c2squared}
	\end{split}
	\end{align}
In order for $c_2/2$ to be an integer class (and note that this is a necessary, but not sufficient condition), we at least require that the term in square brackets in the second line of the above expression is a multiple of 4. Even for the $F_{11}$ model, there are many examples where this condition cannot be satisfied and hence the flux must be half-integer quantized. In the case of the $F_{11}$ model over $B = \bP^3$, we must set $y = 0$, so that the term in brackets reduces to
	\begin{align}
		n_3 \left(n_2+n_3\right) \left(3 n_2+2 n_3\right)\,.
	\end{align}
By scanning over various small integer values of $n_2, n_3$, one can
verify %(at least qualitatively)
 that choices of characteristic data
within this class of examples for which $c_2/2$ is not integer appear
to be quite common (for example, $n_2 = 4, n_3 = 1$), and hence the
subtleties surrounding half-integer quantization of the flux cannot
generically be ignored when studying these models.

As mentioned earlier, a complication that appears to plague F-theory
models with $\U(1)$ gauge factors such as the $\SM$ model is that, unlike
in purely nonabelian theories, it is not straightforward to solve the
constraints $\Theta_{I\alpha} = 0$ directly for the full set of
variables $\phi^{I \alpha}$.  The approach of \cref{sec:321-spectrum}
gives a general, but somewhat more indirect solution of this problem,
while the more direct approach taken in \cref{altsym} involves
inversion of the matrix of intersections of height pairing divisor
with other base divisors, typically giving a rational form of the
solution. Fortunately, for the latter approach, in this specific
example, the matrix of intersections of the height pairing divisor
with other base divisors is rather simple, namely
	\begin{equation}
		M_{C(1H)(1H)} =: h/6\,,
	\end{equation}
where $h$ is the (rescaled) height pairing divisor
	\begin{equation}
	h	 =48 + 12y-4n_3-3n_2
= 12[s_1] + 8n_3 + 9n_2\,.
	\end{equation}
From the constraints on the characteristic data described at the beginning of this subsection, we know that $h$ is invertible (in fact, positive)
in the region of allowed values for $n_2, n_3, y$.
Thus, we can proceed by solving the
constraint equations $\Theta_{IH} = 0$ for the corresponding
$\phi^{IH}$, leading to expressions with $h$ in the denominator.
Alternatively, we can solve
for other variables to get polynomial expressions, following the lines
of \cref{sec:321-spectrum}.
The following two subsections give the details of the solution using
these two approaches.
%, but these expressions are not particularly illuminating, so we pursue the former strategy.) In the next subsection, we describe these solutions in more detail.
We recall that the three
independent non-trivial fluxes can alternatively be taken to be $ \Theta_{33}, \Theta_{34},
\Theta_{55}$ or $\Theta_{03}, \Theta_{05}, \Theta_{34}$.  We use these
two different basis sets in the two different analyses that follow,
recalling that they can be related through
\cref{geobasisnullfluxes}.

\subsubsection{Rational solution}
\label{rational}

Solving the constraint equations $\Theta_{IH} = 0$ for  all $\phi^{IH}$ gives
	\begin{align}
%	\begin{split}
		\Theta_{\hat I \hat J} &= M_{C(\hat I \hat J)(33)} (-\phi^{03} + \phi^{33} )+ M_{C(\hat I \hat J)(34)} \phi^{34} + M_{C(\hat I \hat J)(55)} (-\phi^{05} -  \phi^{45} + \phi^{55})\,,
%	\end{split}
\label{eq:rational-fluxes}
	\end{align}
where the coefficients for $\Theta_{33}$ are
	\begin{align}
		\begin{split}
		\label{Theta33}
		%c_{33} &= \frac{3 n_2+4 n_3-12 y-48}{n_2 y}\\
%			 \widetilde{MP}_{(33)(03)} &= c_{33} ( 3 n_2^2+4 n_2 n_3+3 n_2 Y-60 n_2+2 n_3 Y-16 n_3-6 Y^2+24 Y+192 ) \\
%			 \widetilde{MP}_{(33)(05)}&=  c_{33}n_3 (3 n_2+4 n_3-48) \\
			 M_{C(33)(33)} &= -\frac{n_2 y(-192-3 n_2^2-4 n_2 n_3-3 n_2 y+60 n_2-2 n_3 y+16 n_3+6 y^2-24 y)}{h}\,,\\
			  M_{C(33)(34)} &=-\frac{n_2 n_3y (24-3 n_2-2 n_3+6 y)}{h}\,,\\
			 %  \widetilde{MP}_{(33)(45)} & =c_{33}(n_3 (3 n_2+4 n_3-48))\\
			    M_{C(33)(55)} &=-\frac{n_2 n_3
                              y(48-3 n_2-4 n_3)}{h}\,,
%{3 n_2+4 n_3-12 y-48}\,,
		\end{split}
	\end{align}
the coefficients for $\Theta_{34}$ are
		\begin{align}
		\begin{split}
	%	c_{34} &=\frac{3 n_2+4 n_3-12 y-48}{n_2 n_3}\\
%			 \widetilde{MP}_{(34)(03)} &= c_{34} ( -Y (-3 n_2-2 n_3+6 Y+24) ) \\
%			 \widetilde{MP}_{(34)(05)}&=  c_{34}Y (-n_2+4 Y+16) \\
			 M_{C(34)(33)} &= -\frac{n_2 n_3 y (24-3 n_2-2 n_3+6 y)}{h}\,,\\
			  M_{C(34)(34)} &=-\frac{  n_2 n_3(-192 -n_2^2-3 n_2 n_3+4 n_2 y+28 n_2-2 n_3^2+6 n_3 y+40 n_3-48 y)}{h}\,,\\
		%	   \widetilde{MP}_{(34)(45)} & =c_{34}Y (-n_2+4 Y+16)\\
			    M_{C(34)(55)} &=\frac{n_2 n_3
                              y ( 16-n_2 +4 y)}{h}\,,
		\end{split}
	\end{align}
and the coefficients for $\Theta_{55}$ are
		\begin{align}
		\begin{split}
	%	c_{55} &=\frac{3 n_2+4 n_3-12 y-48}{n_3 y}\\
%			 \widetilde{MP}_{(55)(03)} &= c_{55} n_2 (3 n_2+4 n_3-48) \\
%			 \widetilde{MP}_{(55)(05)}&=  c_{55}(3 n_2 n_3+n_2 Y-12 n_2+4 n_3^2-64 n_3-4 Y^2+32 Y+192) \\
			 M_{C(55)(33)} &= -\frac{n_2 n_3 y (48-3 n_2-4 n_3)}{h}\,,\\
			  M_{C(55)(34)} &=\frac{n_2 n_3 y
                            (16-n_2 +4 y)}{h}\,,\\
			%   \widetilde{MP}_{(55)(45)} & =c_{55}(3 n_2 n_3+n_2 Y-12 n_2+4 n_3^2-64 n_3-4 Y^2+32 Y+192)\\
			    M_{C(55)(55)} &=\frac{n_3 y(-192-3 n_2 n_3-n_2 y+12 n_2-4 n_3^2+64 n_3+4 y^2-32 y)}{h}\,.
		\end{split}
	\end{align}
From these expressions it is clear that the chiral multiplicities are
parameterized by the three independent linear combinations of fluxes
appearing in \cref{eq:rational-fluxes}.
While these expressions all have a factor of $h$ in the denominator,
the fact that all flux backgrounds $\phi^{\hat I\hat J}$
must be (half-)integral
guarantees that this factor will be cancelled
properly for any set of allowed integer fluxes, such that
 corresponding fluxes $\Theta_{\hat I \hat J}$
are then integral.

%As described in \cref{sec:21-polynomial}, one can alternatively solve
%the constraint equation $\Theta_{1H}$ for a variable other than
%$\phi_{1H}$ and get polynomial expressions for the fluxes
%$\Theta_{\hat{I}\hat{J}}$.  For example this can be done solving the
%last equation for $\phi^{05}$.  The expressions are not particularly
%illuminating, however, so we omit them here.
%%\subsection{Polynomial solution}

\subsubsection{Polynomial expressions and
chiral multiplicities}

In the base $\bP^3$, the triple intersection products in the base
simply become products of numerical integer factors.  Thus,
from the general form of $\mr$, in a basis of fluxes (we use the index $2$ here for the
divisor $H$)
\begin{align}
\phi^{02}, \quad \phi^{22}, \quad \phi^{23}, \quad \phi^{24}, \quad \phi^{25}, \quad \phi^{12},
\quad \phi^{03}, \quad \phi^{05}, \quad \phi^{34}
\end{align}
the reduced matrix of \cref{t:mr-alternate} is simply
	\begin{align}
	\label{321mr}
	M_{\text{red}} = \begin{pmatrix}
-4& 1& 0& 0& 0& y& -n_2 y& -n_3 y& 0\\
 1& 0& 0& 0& 0& 1& 0& 0& 0\\
 0& 0& -2 n_2& 0& 0& n_2& -n_2 y& 0& -n_2 n_3\\
 0& 0& 0& -2 n_3& n_3& 0& 0& 0&
  -n_2 n_3\\
 0& 0& 0& n_3& -2 n_3& n_3& 0& -n_3 y& n_2 n_3\\
 y& 1& n_2& 0& n_3& -4& n_2 y& n_3 y& 0\\
 -n_2 y& 0& -n_2 y& 0& 0& n_2 y& (4 -y - n_2) n_2 y&
  -n_2 n_3 y& 0\\
 -n_3 y& 0& 0& 0& -n_3 y& n_3 y& -n_2 n_3 y&(4 - y - n_3) n_3 y& 0\\
 0& 0& -n_2 n_3& -n_2 n_3& n_2 n_3& 0& 0& 0& n_2 n_3 (4 -n_2 -n_3)
\end{pmatrix}\,,
	\end{align}
Note that with this ordering we see the characteristic structure of
the upper left 5 $\times$ 5 block, which is resolution-independent and
encodes the basic structure of the base geometry and the nonabelian
gauge factors.

%We now specialize to the special case $y = 1$ on the base $\bP^3$, for a simple illustration.
In this case the analysis of \cref{sec:321-spectrum}
applies, and the space of curves on $B$ is one-dimensional (spanned by
$H^2$), so all curves in \cref{eq:remaining-constraint}  lie in the
same one-dimensional space and we can solve the constraint for any of
the parameters involved.  For variety, we solve here explicitly for the
$\phi^{2I}$ variables except $\phi^{12}$ and use $\phi^{05}$ instead
to solve \cref{eq:remaining-constraint}, giving
% WT: changed from -33, 34, 55.
\begin{equation}
\label{eq:redsol}
\begin{aligned}
\Theta_{03} & = \frac{n_2}{8} \left[  (n_2 + 32-4y) y \phi^{03}
- (48-3n_2-4n_3) \phi^{12} + n_3 (4y-n_2) \phi^{34} \right]\,,\\
\Theta_{34} & =  \frac{n_2}{8} \left[(3 n_2 +  4n_3)y \phi^{03}
+ (n_2- 16-4y) \phi^{12} + n_3  (32-3n_2-4n_3) \phi^{34} \right]\,,\\
\Theta_{05} & =  - \frac{n_2}{8} \left[ (36-3y-n_3)y \phi^{03}
 + n_3 (n_3  + 3y-4) \phi^{34}\right.\\
& \hspace*{0.5in} \left.
-(192 + 32y -4y^2-(12-y)n_2-64n_3 + 3n_2n_3 + 4n_3^2) \phi^{12}
 \right] \,.
\end{aligned}
\end{equation}
The requirement that all $\phi^{IJ}$ take integer (or, appropriately,
half-integer) values provides
additional constraints that guarantee that these expressions related
to chiral multiplicities are all
integers.
These constraints
% are
all follow from the condition
\begin{align}
%\phi^{12} & \equiv y \phi^{03} + n_3 \phi^{34}\ {\rm (mod\ 2)}\\
%\phi^{12} & \equiv y \phi^{05} + n_2 \phi^{34}\ {\rm (mod\ 3)}\\
8n_3y \phi^{05}  & = -9n_2y \phi^{03}
 + h \phi^{12}
+ n_2n_3 \phi^{34} \,, \label{eq:05-12-constraint}
\end{align}
%The first and second of these constraints are automatically implied by
%the third,
which is \cref{eq:remaining-constraint}.  Generally, the
resulting chiral multiplicities are all nonzero, so that all three
independent families of chiral matter are allowable, though the
multiplicities only take certain combinations of integer values
when the flux parameters $\phi^{IJ}$ are (half-)integral.

In order to make a direct comparison with the rational solutions
described in \cref{rational}, we need to use the constraint
\cref{eq:05-12-constraint} to eliminate the parameter $\phi^{12}$ in
the expressions for $\Theta_{03}, \Theta_{34},\Theta_{05}$ given in
\cref{eq:redsol}. For generic choices of characteristic data, the scaled
height pairing divisor $h \ne 0$, so in any such case we impose the
constraint
	\begin{align}
	\label{solvedphi12}
		\phi^{12} = \frac{1}{h} ( 9 n_2 y \phi^{03} + 8 n_3 y \phi^{05} - n_2 n_3 \phi^{34} )\,.
	\end{align}
We also make use of the relations implied by the restriction of the
null vectors \cref{nullspace} to the subspace $\Lambda_C$ defined by
the condition $\Theta_{1\alpha} =0$. These relations enable us to
account for the redundancy in the set of parameters $\phi^{03},
\phi^{33}, \phi^{34}, \phi^{05}, \phi^{45}, \phi^{55}$ used to
parametrize the solutions of \cref{rational}. Specifically, we find that before eliminating the null directions, we may make the identification
	\begin{align}
	\begin{split}
		&\phi^{03} S_{03} + \phi^{33} S_{33} + \phi^{34} S_{34} + \phi^{05} S_{05} + \phi^{45} S_{45} + \phi^{55} S_{55}\\
		&~~~~ \cong ( \phi^{03} - \phi^{33} )S_{03} + \phi^{34} S_{34} + (\phi^{05} + \phi^{45} - \phi^{55} ) S_{05}\,.
	\end{split}
	\end{align}
The constrained null vectors also imply relations among the fluxes,
in particular the last 3 relations of \cref{geobasisnullfluxes}.
%in
%particular
%	\begin{align}
%	\begin{split}
%		0&=\Theta_{33} +\Theta_{03}\\
%		0& = \Theta_{05} - \Theta_{45} \\
%		0&= \Theta_{05} + \Theta_{55}.
%	\end{split}
%	\end{align}
This in turn implies that when comparing the solutions in
\cref{rational} to \cref{eq:redsol}, we should identify, e.g., the
coefficient of $(\phi^{03}- \phi^{33})$ in $\Theta_{33}$ in
\cref{rational} with the coefficient of $\phi^{03}$ in $\Theta_{03}$
of \cref{eq:redsol}, and so on. As an explicit illustration of this
method of comparison, we impose the condition \cref{solvedphi12} in
the expression for $\Theta_{03}$ in \cref{eq:redsol} to obtain
	\begin{align}
	\begin{split}
		\Theta_{03} &=-\frac{n_2 y \phi _{03} \left(3 n_2 y+2 n_3 y+3 n_2^2-60 n_2+4 n_2 n_3-16 n_3-6 y^2+24 y+192\right)}{3 n_2+4 n_3-12 y-48}\\
		&~~~+\frac{n_2 n_3 y \phi _{34} \left(3 n_2+2 n_3-6
                  y-24\right)}{3 n_2+4 n_3-12 y-48}-\frac{n_2 n_3
                  \left(3 n_2+4 n_3-48\right) y \phi _{05}}{3 n_2+4
                  n_3-12 y-48}\,,
\label{eq:t03-relation}
	\end{split}
	\end{align}
which is precisely (minus) the expression for $\Theta_{33}$ whose coefficients are presented in \cref{Theta33}.

As an example of applying these results,
consider the case with
$n_2 = n_3 = y = 1$.  In this case \cref{eq:05-12-constraint} becomes
$\phi^{34} = -53 \phi^{12} + 9 \phi^{03} + 8 \phi^{34}$.
While $\phi^{03}, \phi^{05}, \phi^{34}$ can be half-integer valued
(from \cref{eq:c2}), $\phi^{12}$ is integer-valued.
This gives the fluxes
\begin{equation}
\Theta_{03} =
-25 \phi^{12} + 7 \phi^{03} + 3 \phi^{05},  \hspace*{0.1in}
\Theta_{05}=
19 \phi^{12} - 4 \phi^{03},  \hspace*{0.1in}
  \Theta_{34} =    -168 \phi^{12} + 29 \phi^{03} + 25 \phi^{05} \,,
\label{eq:111-spectrum}
\end{equation}
which are all clearly integer-valued, even when $\phi^{03}, \phi^{05}$
are both half-integers.

As another class of examples,
we can look at the chiral
multiplicities that are possible without exotic matter
representations.
Setting $\Theta_{03} =\Theta_{05} = 0$ in
\cref{321mr}, the expression for $\Theta_{34}$ simplifies to
\begin{equation}
\Theta_{34} = \frac{1}{2}(12-n_2-n_3 -y)  (4-m-n + y) \phi^{12} \,.
\end{equation}
For example with $n_2 = n_3 = y = 1$, the number of generations of
Standard Model matter is given by $ -27  \phi^{12}/2$.
The condition from \cref{eq:111-spectrum} that $\Theta_{05} = 0$
implies that $\phi^{12}$ is a multiple of 2 (whether $\phi^{03}$ is an
integer or of the form $(2k +1)/2$), so the number of generations is a
multiple of 27.  Imposing also the condition $\Theta_{03} = 0$ gives
the general solution
\begin{equation}
\phi^{03} =\frac{19}{2} \beta,~~~~ \phi^{05} =-\frac{11}{2} \beta,~~~~ \phi^{12} = 2\beta,~~~~
\Theta_{34} = -198 \beta \,,
\end{equation}
so in general the multiplicity can be any multiple of 198, when
$c_2/2$ is an even homology class, and any odd multiple of 198 when
$c_2/2$ is not even (in this case
\cref{eq:c2squared} is integral, which does not answer the question of
evenness ---
for further discussion of this distinction see
\cref{sec:quantization}).
Note, as a cross-check on the equivalence of the
different methods of analysis used above, that plugging these values into
 \cref{eq:t03-relation}, the numerators of the first and third terms
 add to $3225\, \beta/2$, while the numerator of the second term is
 $25$, so indeed (half-)integer quantization of $\phi_{3 4}$ is
 compatible with the integrality of $\Theta_{34}$
%\footnote{The condition $\Theta_{03} = 0$ implies
%  $\phi^{03} \equiv \phi^{12}$ (mod\ 3) is needed for $\phi^{05}$ to
%  be integral, but this congruence is automatically implied by
%  $\Theta_{55} = 0$.}, so in fact the chiral multiplicity can be an
%arbitrary multiple of 396.\wati{[pass 1] someone should check these examples.}

\subsection{Special cases \texorpdfstring{$Y = 0$ ($F_{11}$}{} model)}
\label{sec:f11}

\subsubsection{General $F_{11}$ models}

In the special case $Y=0$ for which the $\SM$ model reduces to the
$F_{11}$ model \cite{KleversEtAlToric}, the chiral indices of the exotic 4D matter must vanish and
we are left with a one-dimensional family of non-trivial fluxes.
$\Theta_{34}$.
%	\begin{align}
%	\begin{split}
%	\label{F11flux}
%		\Theta_{34} = &~~( 4 K + \Sigma_2 + 2 \Sigma_3) \cdot (3 K + \Sigma_2 + \Sigma_3) \\
%		&\cdot ( - K \phi^{11} + \Sigma_2 (-\phi^{13} + 2 \phi^{33} ) + \Sigma_3 (- \phi^{15} + \phi^{44} - \phi^{45} + 2 \phi^{55} ) ).
%	\end{split}
%	\end{align}
In particular, in this particular class of models,
the flux parameters $\phi^{03}, \phi^{05}$ become null vectors of
$\mr$, the multiplicity formulae
\cref{eq:general-multiplicities} reduce to
\begin{equation}
\label{eq:f11-fluxes}
\begin{aligned}
\Theta_{03} & =
0\,,\\
\Theta_{05} & =
0\,,\\
\Theta_{34} & =
\frac{1}{6}\Sigma_2 \cdot  \Sigma_3 \cdot
(A
- \phi^{34} (6K + 2 \Sigma_2 + 3 \Sigma_3))\,,
\end{aligned}
\end{equation}
 and the constraint \cref{eq:remaining-constraint} reduces to
\begin{equation}
A \cdot  \bar{H}
= \phi^{34} \Sigma_2 \cdot \Sigma_3 \,.
\label{eq:p3-constraint}
\end{equation}
Imposing the constraint we can alternatively write
\begin{equation}
\chi_{(\textbf{3},\textbf{2})_{-\frac{1}{6}}} = \Theta_{34}
=
-A (12K^2 + 2 \Sigma_3^2 + 3 \Sigma_3 \Sigma_2 + \Sigma_2^2 + K (10
\Sigma_3 + 7 \Sigma_2)) \,,
\end{equation}
where the parameters in $A$ are constrained such that
\cref{eq:p3-constraint} can be satisfied for (half-)integer $\phi^{34}$; in
particular, parameters in $A$ with a nonzero image under  the map
induced by $\bar{H}$ from divisors to curves in any direction other than
$\Sigma_2 \cdot \Sigma_3$ are fixed to vanish, while parameters in $A$
associated with divisors such that $D \cdot \bar{H} = 0, D \cdot
\Sigma_2\cdot \Sigma_3 \neq 0$ can contribute.

\subsubsection{Specialized ``$\SU(5)$'' $F_{11}$ models}
\label{sec:su5}

Specializing further to the case $\Sigma_2 = \Sigma_3 =- K$,
as studied in  \cite{CveticEtAlQuadrillion},
which are associated with a hidden broken $\SU(5)$ unification group
\cite{Raghuram:2019efb} as all the gauge branes lie on the same divisor,
we have
$\bar{H} = -5K, \Sigma_2 \cdot \Sigma_3 = K^2$, and
\begin{equation}
\Theta_{34} = -AK^2 = \frac{\phi^{34}}{5}K^3 \,,
\label{eq:theta-su5}
\end{equation}
with the constraint $5AK = -\phi^{34}K^2$, in agreement with the
results of  \cite{CveticEtAlQuadrillion}.  For example, if $K = k
\tilde{K}$ with $k \in\Z$ and $\tilde{K}, \tilde{K}^2$ a primitive divisor
and primitive curve respectively, then we can parameterize $A = a
\tilde{A}$ where $\tilde{A}$ is the  Poincar\'{e} dual divisor to $\tilde{K}^2$,
and we have $\Theta_{34} = -a k^2$, with the quantization condition
$5a = -\phi^{34} k\tilde{K}^3$.  Taking the special case of the base $\bP^3$
(i.e., $n_2
= n_3 = k =4, \tilde{K}^3 = 1$), we have, as in  \cite{CveticEtAlQuadrillion}:
	\begin{align}
		\Theta_{34} = -16a = \frac{64}{5} \phi^{34} \,.
\label{eq:p3-multiplicity}
	\end{align}
In this case, all parts of $c_2$ are manifestly even except for a
shift by $S_{03} + S_{05} + S_{34}$, so if $c_2$ is an even class  the
flux parameter $\phi^{34}$ is integral, while if $c_2$ is an odd
class, $\phi^{34}$ is of the form $(2k + 1)/2$.
The quantization constraint shows that $4 | a$, and
$ 5 | \phi^{34}$ when $\phi^{34}$ is an integer, and $2 | a, 5 | 2
\phi^{34}$ when $\phi^{34}$ is a half-integer.
We discuss the question of whether $c_2$ is even in the next section.

As another example, for the base $B =\bP^2 \times\bP^1$, we have $K
=\tilde{K}, K^3 = 18$, and  $\Theta_{34} = -a = 18 \phi^{34}/5$, with
similar quantization conditions.

\section{Quantization of multiplicities}
\label{sec:quantization}

The question of precisely which chiral multiplicities are allowed for
a given model is somewhat subtle.  We give a few brief comments here
that address some further aspects of which fluxes and corresponding
chiral multiplicities may be possible in a broad class of flux
compactifications.  For convenience here, in parts of the discussion
we take $G_{\Z} = G_4 + c_2 (X_5)/2 \in H^4 (X_5,\Z)$ to be an
integer class, even when $c_2$ is not even; when $c_2$ is not even,
fluxes must be suitably shifted by a half-integer class to $G_4$.  The
analysis here expands in particular on related discussions in
\cite{CveticEtAlQuadrillion, Jefferson:2021bid}.  Note that aside from
the specific examples at the end of this section, most of the
discussion of quantization here is relevant for general classes of
F-theory models, independent of the specific gauge group and matter
structure.

Recall from \cref{eq:lv} and \cref{eq:lvb} the definitions of the
lattices $\lv,\blv$.  The homology lattice $H_4 (X_5,\Z)$ has a
unimodular intersection pairing.  Thus, for any surface $S$
that is a primitive\footnote{A lattice element $S \in H_4(X_5,\Z)$
is primitive iff there does not exist a distinct non-zero element $S'
\in H_4(X_5,\Z)$ such that $S = m S'$ for $m \in
\Z_{>1}$.} element of this lattice,
% (i.e., $\nexists S' \in
%H_4 (X_5,\Z): S = m S', m \in\Z)$,
in particular for any primitive
surface in $\blv$, there exists some integer flux $G_{\Z}\in H^4
(X_5,\Z)$ such that $\int_SG_{\Z}$ = 1.  This would suggest that
chiral multiplicities can always be quantized in arbitrary integer
units, and that essentially any model would allow three generations of
anomaly-free chiral matter (leaving aside the issue, for the moment, of
non-even $c_2$).  There are several reasons, however, why such a flux
may not be possible or may not
 correspond to a physical vacuum of interest.  First, it may
%not
 be the case %that the flux through every primitive surface $S \in
%\bar \Lambda_{\text{vert}}$ corresponds to a well-defined physical
%quantity\wati{I don't understand what this means and I don't think I
%  agree with it.  Explain/discuss?}. Suppose, for example,
that a vertical surface $S_{IJ} = D_I \cap D_J \in\Lambda_{\rm vert}$
is primitive in $\lv$ but is an integer multiple of an element of $\blv$ and $H_4
(X_5,\Z)$. In particular,
 a (vertical) matter surface $S_{\mathsf r}$  can be
primitive in $\lv$ but not primitive in $\blv$; in such a situation,
since $S_{\mathsf r} = m
S',$ with $S'$ primitive in $\blv$,  the associated chiral
multiplicities $\chi_{\mathsf r} =\int_{S_{\mathsf r}} G_{\Z}$
will always be multiples of $m$, and unit multiplicity, for example,
(or multiplicity 3 if $m \neq 3$)
is not possible.
%\footnote{For example, one could imagine a situation in which the
%identification with 3D CS couplings implies that the flux $\int_{S'}
%G_{\Z}$ is equal to $1/m$ times $\chi_{\mathsf r}$.}
Second, it is
possible that the specific $G_{\Z}$ that is dual to a surface $S$
that is primitive in $\blv$ and $\hfz$ breaks Poincar\'{e} or gauge
invariance by giving a nonzero $\int_{S_{I \alpha}}G_{\Z}$.  If
neither of the conditions just summarized are the case, then,
recalling the decomposition \cref{eq:orthogonal-decomposition}, the
flux $G_{\Z}$ that is dual to a primitive vertical matter surface $S$
may have a fractional component in the vertical cohomology and a
fractional part in the horizontal or remainder cohomology, and satisfy
the desired properties.
Completely determining which of these scenarios is realized depends on
a full understanding of $ H_4 (X_5,\Z)$ and its intersection form.
Note also that while in many models the matter surfaces lie in $\lv$,
this need not be the case in general; in any specific model, a clear
analysis of matter surfaces and/or the relationship between chiral
multiplicities and integrated fluxes $\Theta_{IJ}$ is needed for
precise statements regarding allowed multiplicities.
Finally, we must incorporate
the possibility of non-even values of $c_2$ into the preceding
discussion.  Note that when $c_2$ is not even, as in many examples
discussed in the preceding section, the chiral multiplicities
associated with integrating over matter curves are still  integral, $\int_S
c_2 \in\Z$, in which case the allowed multiplicities are odd multiples
of some basic factor.

Because of the orthogonal decomposition
\labelcref{eq:orthogonal-decomposition}, and the fact that the intersection
product on $H_4 (X_5,\Z)$ is unimodular, the set of possible
fluxes  $\int_S G_{\Z}$ through vertical surfaces, and hence the set
of chiral multiplicities, can be characterized by considering
$\phi'$ in the projection of $H_4 (X_5,\Z)$ to
$\bar{\Lambda}_{\rm vert}$, which  is the dual lattice ${\rm
  dual}(\bar{\Lambda}_{\rm vert})$ (i.e., the set of points in
$H_{2,2}^{\rm vert}(X_5,\C)$ with integer inner product under
$\mr$ with all points in $\bar{\Lambda}_{\rm vert}$).
When
$\bar{\Lambda}_{\rm vert}={\Lambda}_{\rm vert}$, then this dual space
is characterized by the set of necessary conditions stated in
\cite{CveticEtAlQuadrillion}, which are that $\int_S (G_4 + c_2 (X_5)/2)$ must be
integer-valued for any allowed flux $G_4$ and vertical surface
$S \in\lv$.\footnote{Note that when $c_2$ is not even, in principle it is
  possible that $\int_S (G_4 +c_2/2)$ may be half-integral for some
  vertical surfaces, although physical consistency dictates that
it must be integral for any vertical matter surface, as in the
examples studied in the previous section.  In such situations, $G_{\Z}$,
but not necessarily $G$,
must project to dual($\blv$).}
However,
$\bar{\Lambda}_{\rm vert}$ could contain extra integral points not realized
in ${\Lambda}_{\rm vert}$, in which case
these necessary conditions
would not be sufficient as there would be points in dual($\lv$) that are not in dual($\blv$) and hence do not correspond to the set of physically-allowed fluxes.  Furthermore, fluxes associated with some
points in ${\rm dual}(\bar{\Lambda}_{\rm vert})$ may break gauge or
Poincar\'{e} invariance.  The information of
${\Lambda}_{\rm vert}$ and $\mr$ is not generally sufficient to uniquely determine
$\bar{\Lambda}_{\rm vert}$.  For example, the lattice defined by the
inner product ${\rm diag} (4, 4)$ can alternatively be embedded in a
self-dual Euclidean 2D lattice by extending to an overlattice that
includes one half of each generator, or can represent the complete
integral 2D subspace of the unimodular lattice $E_8$ along a pair of
orthogonal primitive vectors of length 4.

The upshot of this discussion is that to determine the possible chiral
multiplicities for a given model, one must consider the possible
integer overlattices $\bar \Lambda_{\text{vert}}$ of ${\Lambda}_{\rm vert}$; for each such
overlattice the quantization condition is then determined by
identifying all elements of the dual space dual($\bar{\Lambda}_{\rm vert}$) that leave
Poincar\'{e} and gauge invariance unbroken.  Determining which is the
proper overlattice requires a complete determination of $H_4
(X_5,\Z)$, which is beyond the scope of this paper but will be
addressed in future work. Note also that consistent chiral multiplicities
associated with points $\phi_v$ in ${\rm dual}(\bar{\Lambda}_{\rm
  vert})$ that are not in $\bar{\Lambda}_{\rm vert}$ can be allowed
only if non-vertical components of flux are turned on.  In such cases,
the integrality of the D3-brane tadpole from $\phi_v$ can only
be violated when the vertical component of flux is considered;
thus, that condition (which was used in \cite{CveticEtAlQuadrillion})
is not actually necessary for an allowed multiplicity, but rather
amounts to the condition that the flux $\phi_v$ lies in an allowed
overlattice $\blv$. Moreover, when flux is quantized properly
with $G_{\Z} \in H^4 (X_5,\Z)$, the D3-brane quantization
condition is automatically satisfied.
The uncertainty of which overlattice is realized by $\blv$ also means
that unless there is a clear signal such as non-integrality of
\cref{eq:c2-squared} indicating that $c_2/2$ is not even, there may be
some overlattices $\blv$ that contain $c_2/2$, in which case $c_2$ is
even, and other overlattices that do not, in which case $c_2$ is not even---this further complicates the issue of precisely determining the flux quantization and associated chiral multiplicities.

As a specific example of the kinds of overlattice conditions that may
arise, let us consider the quantization of chiral multiplicities in
the simplest model considered above in \cref{sec:su5}, with base
$\bP^3$, and $\Sigma_2 = \Sigma_3 = -K = 4H, Y = 0$.  In this case, $\mr$ from \cref{321mr} simplifies to
(dropping the null directions $\phi^{03}, \phi^{05}$, and moving
$\phi^{12}$ to the second slot)
	\begin{align}
	\label{321mr-simplest}
	M_{\text{red}} = \begin{pmatrix}
 4& 0& 1& 0& 0& 0& 0\\
 0& 4& 1& -4& 0& -4& 0\\
 1& 1& 0& 0& 0& 0& 0\\
 0& -4& 0& 8& 0& 0& -16\\
 0& 0& 0& 0& 8& -4& -16\\
 0& -4& 0& 0& -4& 8& 16\\
 0& 0& 0& -16& -16& 16& 64
\end{pmatrix}\,.
	\end{align}
If we choose integer fluxes in $\Lambda_{\rm vert}$, as discussed in
\cref{sec:su5}, the chiral multiplicity is an integer multiple of 64
through \cref{eq:p3-multiplicity}.
The possible fluxes that preserve Poincar\'{e} and gauge invariance
are of the form
\begin{equation}
\phi = p/16 (1, -1, -4, -3, -2, 1, -5/4) \;\,.
\label{eq:phi-unit}
\end{equation}
When such a flux is allowed, it gives a chiral multiplicity
$\Theta_{34} = p$.  The flux in \cref{eq:phi-unit} is in the dual
lattice ${\rm dual}(\bar{\Lambda}_{\rm vert})$ whenever $p$ is
integral.

We now consider the issues of even/non-even $c_2$ and possible
overlattices for $\blv$ in this specific case.  First, note that
\begin{equation}
c_2/2 = c' + (0, 0, 0, 0, 0, 0, 1/2) \,,
\end{equation}
where $c' \in\lv$.  Thus, whether $c_2$ is even depends upon whether
$S_{34}/2\in\blv$.
Note that \cref{eq:c2-squared} is integral in this case, so we cannot
rule out the possibility that $c_2$ is even.  Furthermore, there are
several overlattices of $\lv$ that contain $c_2/2$.
There is a maximal overlattice $\Lambda^{\rm max}_{\rm over}$ that includes
the generators $(S_{01} -S_{12})/2, S_{23}/2, S_{24}/2, S_{25}/2,
S_{34}/8$, and which is self-dual.  If $\bar{\Lambda}_{\rm
  vert}={\Lambda}^{\rm max}_{\rm over}$ then $c_2$ is even,
 the minimum allowed $p$ in
\cref{eq:phi-unit} is $p = 16$,  the quantization is controlled by
integer values of $a$ in \cref{eq:p3-multiplicity} and the chiral
multiplicity must be a multiple of 16.
There is another overlattice of $\lv$ that contains only the
additional vector $c_2/2$.  The dual of this overlattice consists of
all vectors of the form  \cref{eq:phi-unit} with even $p$.  If this
overlattice is $\blv$, then the chiral multiplicity must be a multiple
of 2, but non-vertical flux must be included for any multiplicity that
is not a multiple of 32.  Finally, if
$\bar{\Lambda}_{\rm vert}={\Lambda}_{\rm vert}$ then
$c_2$ is odd, since $c_2/2$ is not in $\blv$.  Because $c_2 \in$
dual($\lv$) in this case,
the flux associated with chiral multiplicity is
quantized in units of 1,
but non-vertical flux must be included for
any chiral multiplicity that is not of the form $\chi = 32 (2k + 1)$, and, e.g., a model
with $\chi = 3$ will appear to
violate the D3-brane tadpole integrality
condition unless the non-vertical flux is properly included).
There are a variety of other possible overlattices, such as other
sublattices of $\Lambda_{\rm over}^{\rm max}$ that contain $\lv$, some
of which contain $c_2/2$ and others of which do not; we do not go
through all the possibilities here.
Thus, even in this simple case there are many possibilities for
$\blv$, each of which gives distinct quantization conditions when
restricted to vertical or more general integral fluxes.  It would be
interesting to gain further detailed information about $H_4
(X_5,\Z)$ to determine the precise form of $\blv$ and definitively
determine the precise quantization condition on chiral
multiplicities for
these models.

\section{Conclusions}
\label{sec:discussion}

In this paper we have performed a detailed analysis of the geometry
and flux-induced chiral matter of the universal tuned $G_{\rm SM}=\SM$
Weierstrass model identified in \cite{Raghuram:2019efb}.
This model admits three linearly independent families of chiral matter
transforming under the Standard Model gauge group $G_{\rm SM}$.  All
three families can be produced by F-theory fluxes, so the observed
Standard Model chiral matter structure is realized in a subset of
models where one or two discrete parameters are tuned to vanish.
Tuning the parameter $Y = 0$ gives the class of ``$F_{11}$'' models
identified and studied in
\cite{KleversEtAlToric,CveticEtAlThreeParam,CveticEtAlQuadrillion} and related works.

The analysis of this paper provides a framework for more detailed
study of further phenomenological features of these tuned Standard
Model-like models, including aspects of Yukawa interactions, chiral and
non-chiral spectra, and moduli stabilization.  This class of models
complements other types of F-theory constructions such as the tuned
$\SU(5)$ GUT models
\cite{Donagi:2008ca,BeasleyHeckmanVafaI,BeasleyHeckmanVafaII,DonagiWijnholtGUTs}
that have been studied for several decades, or the more recently
constructed models that arise from flux breaking of a rigid $E_6$ or $E_7$
symmetry \cite{Li:2021eyn,Li:2022aek}.  Comparing how these different
classes of F-theory Standard Model constructions differ in detailed
phenomenology and in frequency in the landscape may provide insights
into what features of beyond the
Standard Model physics are correlated with
the most typical or natural realizations of the Standard
Model in F-theory.

While in this paper we have focused on the specific class of models
with tuned gauge group $G_{\rm SM}$, we have also developed here
further the theoretical toolkit for analyzing the geometric
structure and chiral matter content of broad classes of F-theory
compactifications, extending the framework developed in
\cite{Jefferson:2021bid}.  In particular, we have extended the
analysis of the middle intersection form on vertical cohomology to
include theories with a $\U(1)$ factor, which complicates the structure
of the matrix $\mr$ substantially, and we have given a more thorough
analysis of quantization questions related to the multiplicity of
chiral matter and allowed fluxes for large classes of F-theory
models.  One interesting aspect of this is that in the particular
class of models studied here we were able to completely solve the flux
constrained equations associated with Poincar\'{e} and gauge
symmetries, although part of the solution was implicit.  It would be
interesting to investigate whether this kind of solution is possible
for more general models with $\U(1)$ gauge factors.  Regarding the
quantization of chiral multiplicities, the analysis here highlights the importance of attaining
a full understanding of the structure of $H_4 (X_5,\Z)$ for
F-theory models of interest.

\acknowledgments
We would like to thank Mboyo Esole, Manki Kim, Shing Yan Li, Nikhil Raghuram, and Timo Weigand for helpful discussions. This work
was supported by DOE grant DE-SC00012567. AT is also supported by DOE (HEP) Award DE-SC001352 and the Tushar Shah and Sara Zion Fellowship. WT would like to thank the Aspen
Center for Physics (ACP) for hospitality during part of this work. The
authors would all like to thank the Witwatersrand (Wits) rural
facility and the MIT International Science and Technology Initiatives
(MISTI) MIT--Africa--Imperial College seed fund program for
hospitality and support during some stages of this project.

\appendix

\section{Converting between the geometric and gauge bases}
\label{changebasis}

Let $X \rightarrow B$ denote a smooth CY manifold
that is elliptically fibered
over a smooth base $B$. There is a natural relationship between
divisors $\hat D_{\bar I} \subset X$ and gauge fields appearing in the
low-energy description of M-theory compactified on $X$. In particular,
the divisors $\hat D_{\bar I} $ are Poincar\'e dual to harmonic
$(1,1)$-forms $\omega_{\bar I}$ that appear in a local expansion of
the M-theory 3-form $C_3$ in the vicinity of the 7-brane loci, i.e.,
$C_3= A^{\bar I} \wedge \omega_{\bar I} + \cdots$. Thus, to each gauge
field $A^I$, there corresponds a particular choice of divisor $\hat
D_{\bar I}$, and the gauge charges $q_{\bar I}$ of BPS particles
corresponding to M2 branes wrapping holomorphic curves $C \subset X$
can be computed in terms of their intersection product with $\hat
D_{\bar I}$, i.e., $q_{\bar I} = \int_{C} \omega_{\bar I}= \hat
D_{\bar I} \cdot C$. This suggests that the low-energy gauge-theoretic
description of M-theory compactified on $X$ defines a canonical basis
of divisors $\hat D_{\bar I}$---we refer to this basis as the
``gauge'' basis of divisors.

Despite the fact that the gauge basis is natural from the point of
view of the low-energy effective theory, in practice, it tends to be
simpler to expand divisors of $X$ in terms of the ``geometric'' basis
$\hat D_I$, where $I = 0$ corresponds to a (non-unique) choice of zero
section and $I = a$ corresponds to a generator of the Mordell--Weil
group of sections. As we describe below, the divisors $\hat D_0, \hat
D_a$ are linearly related to the divisors $\hat D_{\bar 0}, \hat
D_{\bar a}$, which correspond (respectively) to Kaluza--Klein
$\U(1)_{\text{KK}}$ and $\U(1)$ factors belonging to the free abelian
part of the F-theory gauge group.

In this Appendix, we describe how to convert between the gauge and geometric bases. The change of basis from the geometric basis to the gauge basis is given by (see Appendix B in \cite{Jefferson:2021bid})
	\begin{align}
		\hat D_{\bar I} = \sigma^J_{\bar I} \hat D_J
	\end{align}
where in the case of the $\SM$ model, the matrix elements $\sigma^{J}_{\bar I}$ are:
	\begin{align}
		\begin{split}
			\sigma_{\bar 0}^I &= (1,0,-\frac{1}{2} K^\alpha ,0,0,0)^I\\
			\sigma_{\bar 1}^I &= (-1,1,K ^\alpha - Y^\alpha ,\frac{1}{2},\frac{1}{3},\frac{2}{3})^I\\
			\sigma_{\bar J}^I &= \delta^I_{\bar J},~~~~ \bar J \ne 0,1\,.
		\end{split}
	\end{align}
Note that $\hat D_{\bar 1} = \sigma_{\bar 1}^I \hat D_I$ is the image of the Shioda map described in \cite{MorrisonParkU1}. Using
\begin{align}
    \begin{split}
        (\sigma^{-1})_0^{\bar I} &= ( 1,0, \frac{1}{2} K^\alpha , 0,0,0 )^{\bar I} \\
        (\sigma^{-1})_{1}^{\bar I } &= (1,1, -\frac{1}{2}K^{\alpha} + Y^\alpha, - \frac{1}{2}, - \frac{1}{3}, - \frac{2}{3})^{\bar I}\\
        (\sigma^{-1})_{J}^{\bar I} &= \delta_{J}^{\bar I} ~~~~\text{for $J \ne 0,1$}\,,
    \end{split}
    \end{align}
one can invert the above linear transformation:
    \begin{align}
        (\sigma^{-1})_{J}^{\bar I} \sigma_{\bar I}^{K} = \delta_{J}^{K},~~~~ \sigma^{J}_{\bar K} (\sigma^{-1})_{J}^{\bar I} = \delta_{\bar K}^{\bar I}\,.
    \end{align}
 Thus, e.g., the change of basis from geometric fluxes $\Theta_{IJ}$ to gauge basis fluxes $\Theta_{\bar I \bar J}$ relevant for computing the image of the nullspace of $M$ in the constrained sublattice $\Lambda_C \subset \Lambda$ is
\begin{align}
	\begin{split}
	\label{geotogauge}
		\Theta_{\bar 1 I} &=\sigma_{\bar 1}^{J} \Theta_{JI}=  -\Theta_{0I} + \Theta_{1I} + \frac{\Theta_{3I}}{2} + \frac{\Theta_{4I}}{3} + \frac{2 \Theta_{5I}}{3},~~~~ I \ne 1\\
		\Theta_{\bar 1\bar 1}&=\sigma_{\bar 1}^J \sigma_{\bar 1}^K \Theta_{JK} =\Theta _{00}-2 \Theta _{01}-\Theta _{03}+\Theta _{11}+\Theta _{13}+\frac{2 \Theta _{14}}{3}+\frac{4 \Theta _{15}}{3}\\
		&+\frac{\Theta _{33}}{4}+\frac{\Theta _{34}}{3}+\frac{2 \Theta _{35}}{3}+\frac{\Theta _{44}}{9}+\frac{4 \Theta _{45}}{9}+\frac{4 \Theta _{55}}{9}-\frac{2 \Theta _{04}}{3}-\frac{4 \Theta _{05}}{3}\,.
	\end{split}
	\end{align}
Note that $\Theta_{I\bar 0} = \Theta_{I0}$ on the constrained sublattice, hence one can freely exchange the indices $0$ and $\bar 0$.

\bibliographystyle{JHEP}
\bibliography{references}

\providecommand{\href}[2]{#2}\begingroup\raggedright\begin{thebibliography}{10}

\bibitem{Cvetic:2022fnv}
M.~Cvetic, J.~Halverson, G.~Shiu and W.~Taylor, \emph{{Snowmass White Paper:
  String Theory and Particle Physics}},
  \href{https://arxiv.org/abs/2204.01742}{{\ttfamily 2204.01742}}.

\bibitem{Raghuram:2019efb}
N.~Raghuram, W.~Taylor and A.~P. Turner, \emph{{General F-theory models with
  tuned $(\operatorname{SU}(3) \times \operatorname{SU}(2) \times
  \operatorname{U}(1)) / \mathbb{Z}_6$ symmetry}},
  \href{https://doi.org/10.1007/JHEP04(2020)008}{\emph{JHEP} {\bfseries 04}
  (2020) 008} [\href{https://arxiv.org/abs/1912.10991}{{\ttfamily
  1912.10991}}].

\bibitem{VafaSwamp}
C.~Vafa, \emph{{The String landscape and the swampland}},
  \href{https://arxiv.org/abs/[hep-th/0509212]}{{\ttfamily [hep-th/0509212]}}.

\bibitem{OoguriVafaSwamp}
H.~Ooguri and C.~Vafa, \emph{{On the Geometry of the String Landscape and the
  Swampland}},
  \href{https://doi.org/10.1016/j.nuclphysb.2006.10.033}{\emph{Nucl. Phys.}
  {\bfseries B766} (2007) 21}
  [\href{https://arxiv.org/abs/hep-th/0605264}{{\ttfamily hep-th/0605264}}].

\bibitem{VafaF-theory}
C.~Vafa, \emph{{Evidence for F theory}},
  \href{https://doi.org/10.1016/0550-3213(96)00172-1}{\emph{Nucl. Phys.}
  {\bfseries B469} (1996) 403}
  [\href{https://arxiv.org/abs/hep-th/9602022}{{\ttfamily hep-th/9602022}}].

\bibitem{MorrisonVafaI}
D.~R. Morrison and C.~Vafa, \emph{{Compactifications of F theory on Calabi--Yau
  threefolds --- I}},
  \href{https://doi.org/10.1016/0550-3213(96)00242-8}{\emph{Nucl. Phys.}
  {\bfseries B473} (1996) 74}
  [\href{https://arxiv.org/abs/hep-th/9602114}{{\ttfamily hep-th/9602114}}].

\bibitem{MorrisonVafaII}
D.~R. Morrison and C.~Vafa, \emph{{Compactifications of F theory on Calabi--Yau
  threefolds --- II}},
  \href{https://doi.org/10.1016/0550-3213(96)00369-0}{\emph{Nucl. Phys.}
  {\bfseries B476} (1996) 437}
  [\href{https://arxiv.org/abs/hep-th/9603161}{{\ttfamily hep-th/9603161}}].

\bibitem{TaylorWangMC}
W.~Taylor and Y.-N. Wang, \emph{{A Monte Carlo exploration of threefold base
  geometries for 4d F-theory vacua}},
  \href{https://doi.org/10.1007/JHEP01(2016)137}{\emph{JHEP} {\bfseries 01}
  (2016) 137} [\href{https://arxiv.org/abs/1510.04978}{{\ttfamily
  1510.04978}}].

\bibitem{HalversonLongSungAlg}
J.~Halverson, C.~Long and B.~Sung, \emph{{Algorithmic universality in F-theory
  compactifications}},
  \href{https://doi.org/10.1103/PhysRevD.96.126006}{\emph{Phys. Rev.}
  {\bfseries D96} (2017) 126006}
  [\href{https://arxiv.org/abs/1706.02299}{{\ttfamily 1706.02299}}].

\bibitem{TaylorWangLandscape}
W.~Taylor and Y.-N. Wang, \emph{{Scanning the skeleton of the 4D F-theory
  landscape}}, \href{https://doi.org/10.1007/JHEP01(2018)111}{\emph{JHEP}
  {\bfseries 01} (2018) 111}
  [\href{https://arxiv.org/abs/1710.11235}{{\ttfamily 1710.11235}}].

\bibitem{TaylorWangVacua}
W.~Taylor and Y.-N. Wang, \emph{{The F-theory geometry with most flux vacua}},
  \href{https://doi.org/10.1007/JHEP12(2015)164}{\emph{JHEP} {\bfseries 12}
  (2015) 164} [\href{https://arxiv.org/abs/1511.03209}{{\ttfamily
  1511.03209}}].

\bibitem{Donagi:2008ca}
R.~Donagi and M.~Wijnholt, \emph{{Model Building with F-Theory}},
  \href{https://doi.org/10.4310/ATMP.2011.v15.n5.a2}{\emph{Adv. Theor. Math.
  Phys.} {\bfseries 15} (2011) 1237}
  [\href{https://arxiv.org/abs/0802.2969}{{\ttfamily 0802.2969}}].

\bibitem{BeasleyHeckmanVafaI}
C.~Beasley, J.~J. Heckman and C.~Vafa, \emph{{GUTs and Exceptional Branes in
  F-theory - I}},
  \href{https://doi.org/10.1088/1126-6708/2009/01/058}{\emph{JHEP} {\bfseries
  01} (2009) 058} [\href{https://arxiv.org/abs/0802.3391}{{\ttfamily
  0802.3391}}].

\bibitem{BeasleyHeckmanVafaII}
C.~Beasley, J.~J. Heckman and C.~Vafa, \emph{{GUTs and Exceptional Branes in
  F-theory - II: Experimental Predictions}},
  \href{https://doi.org/10.1088/1126-6708/2009/01/059}{\emph{JHEP} {\bfseries
  01} (2009) 059} [\href{https://arxiv.org/abs/0806.0102}{{\ttfamily
  0806.0102}}].

\bibitem{DonagiWijnholtGUTs}
R.~Donagi and M.~Wijnholt, \emph{{Breaking GUT Groups in F-Theory}},
  \href{https://doi.org/10.4310/ATMP.2011.v15.n6.a1}{\emph{Adv. Theor. Math.
  Phys.} {\bfseries 15} (2011) 1523}
  [\href{https://arxiv.org/abs/0808.2223}{{\ttfamily 0808.2223}}].

\bibitem{Chen:2009me}
C.-M. Chen and Y.-C. Chung, \emph{{A Note on Local GUT Models in F-Theory}},
  \href{https://doi.org/10.1016/j.nuclphysb.2009.09.008}{\emph{Nucl. Phys. B}
  {\bfseries 824} (2010) 273}
  [\href{https://arxiv.org/abs/0903.3009}{{\ttfamily 0903.3009}}].

\bibitem{Heckman:2009mn}
J.~J. Heckman, A.~Tavanfar and C.~Vafa, \emph{{The Point of E(8) in F-theory
  GUTs}}, \href{https://doi.org/10.1007/JHEP08(2010)040}{\emph{JHEP} {\bfseries
  08} (2010) 040} [\href{https://arxiv.org/abs/0906.0581}{{\ttfamily
  0906.0581}}].

\bibitem{Blumenhagen:2009yv}
R.~Blumenhagen, T.~W. Grimm, B.~Jurke and T.~Weigand, \emph{{Global F-theory
  GUTs}}, \href{https://doi.org/10.1016/j.nuclphysb.2009.12.013}{\emph{Nucl.
  Phys. B} {\bfseries 829} (2010) 325}
  [\href{https://arxiv.org/abs/0908.1784}{{\ttfamily 0908.1784}}].

\bibitem{Dudas:2009hu}
E.~Dudas and E.~Palti, \emph{{Froggatt-Nielsen models from E(8) in F-theory
  GUTs}}, \href{https://doi.org/10.1007/JHEP01(2010)127}{\emph{JHEP} {\bfseries
  01} (2010) 127} [\href{https://arxiv.org/abs/0912.0853}{{\ttfamily
  0912.0853}}].

\bibitem{Grimm:2009yu}
T.~W. Grimm, S.~Krause and T.~Weigand, \emph{{F-Theory GUT Vacua on Compact
  Calabi-Yau Fourfolds}},
  \href{https://doi.org/10.1007/JHEP07(2010)037}{\emph{JHEP} {\bfseries 07}
  (2010) 037} [\href{https://arxiv.org/abs/0912.3524}{{\ttfamily 0912.3524}}].

\bibitem{Marsano:2009wr}
J.~Marsano, N.~Saulina and S.~Schafer-Nameki, \emph{{Compact F-theory GUTs with
  U(1) (PQ)}}, \href{https://doi.org/10.1007/JHEP04(2010)095}{\emph{JHEP}
  {\bfseries 04} (2010) 095} [\href{https://arxiv.org/abs/0912.0272}{{\ttfamily
  0912.0272}}].

\bibitem{Leontaris:2011wtt}
G.~K. Leontaris, \emph{{Aspects of F-Theory GUTs}},
  \href{https://doi.org/10.22323/1.155.0095}{\emph{PoS} {\bfseries CORFU2011}
  (2011) 095} [\href{https://arxiv.org/abs/1203.6277}{{\ttfamily 1203.6277}}].

\bibitem{Callaghan:2012rv}
J.~C. Callaghan and S.~F. King, \emph{{E6 Models from F-theory}},
  \href{https://doi.org/10.1007/JHEP04(2013)034}{\emph{JHEP} {\bfseries 04}
  (2013) 034} [\href{https://arxiv.org/abs/1210.6913}{{\ttfamily 1210.6913}}].

\bibitem{Mayrhofer:2013ara}
C.~Mayrhofer, E.~Palti and T.~Weigand, \emph{{Hypercharge Flux in IIB and
  F-theory: Anomalies and Gauge Coupling Unification}},
  \href{https://doi.org/10.1007/JHEP09(2013)082}{\emph{JHEP} {\bfseries 09}
  (2013) 082} [\href{https://arxiv.org/abs/1303.3589}{{\ttfamily 1303.3589}}].

\bibitem{Callaghan:2013kaa}
J.~C. Callaghan, S.~F. King and G.~K. Leontaris, \emph{{Gauge coupling
  unification in $E_6$ F-theory GUTs with matter and bulk exotics from flux
  breaking}}, \href{https://doi.org/10.1007/JHEP12(2013)037}{\emph{JHEP}
  {\bfseries 12} (2013) 037} [\href{https://arxiv.org/abs/1307.4593}{{\ttfamily
  1307.4593}}].

\bibitem{Braun:2014pva}
A.~P. Braun, A.~Collinucci and R.~Valandro, \emph{{Hypercharge flux in F-theory
  and the stable Sen limit}},
  \href{https://doi.org/10.1007/JHEP07(2014)121}{\emph{JHEP} {\bfseries 07}
  (2014) 121} [\href{https://arxiv.org/abs/1402.4096}{{\ttfamily 1402.4096}}].

\bibitem{Li:2021eyn}
S.~Y. Li and W.~Taylor, \emph{{Natural F-theory constructions of Standard Model
  structure from $E_7$ flux breaking}},
  \href{https://arxiv.org/abs/2112.03947}{{\ttfamily 2112.03947}}.

\bibitem{Li:2022aek}
S.~Y. Li and W.~Taylor, \emph{{Gauge symmetry breaking with fluxes and natural
  Standard Model structure from exceptional GUTs in F-theory}},
  \href{https://arxiv.org/abs/2207.14319}{{\ttfamily 2207.14319}}.

\bibitem{HeckmanReview}
J.~J. Heckman, \emph{{Particle Physics Implications of F-theory}},
  \href{https://doi.org/10.1146/annurev.nucl.012809.104532}{\emph{Ann. Rev.
  Nucl. Part. Sci.} {\bfseries 60} (2010) 237}
  [\href{https://arxiv.org/abs/1001.0577}{{\ttfamily 1001.0577}}].

\bibitem{GrassiHalversonShanesonTaylor}
A.~Grassi, J.~Halverson, J.~Shaneson and W.~Taylor, \emph{{Non-Higgsable QCD
  and the Standard Model Spectrum in F-theory}},
  \href{https://doi.org/10.1007/JHEP01(2015)086}{\emph{JHEP} {\bfseries 01}
  (2015) 086} [\href{https://arxiv.org/abs/1409.8295}{{\ttfamily 1409.8295}}].

\bibitem{TaylorTurnerGeneric}
W.~Taylor and A.~P. Turner, \emph{{Generic matter representations in 6D
  supergravity theories}},
  \href{https://doi.org/10.1007/JHEP05(2019)081}{\emph{JHEP} {\bfseries 05}
  (2019) 081} [\href{https://arxiv.org/abs/1901.02012}{{\ttfamily
  1901.02012}}].

\bibitem{KleversEtAlToric}
D.~Klevers, D.~K. Mayorga~Pena, P.-K. Oehlmann, H.~Piragua and J.~Reuter,
  \emph{{F-Theory on all Toric Hypersurface Fibrations and its Higgs
  Branches}}, \href{https://doi.org/10.1007/JHEP01(2015)142}{\emph{JHEP}
  {\bfseries 01} (2015) 142} [\href{https://arxiv.org/abs/1408.4808}{{\ttfamily
  1408.4808}}].

\bibitem{CveticEtAlThreeParam}
M.~Cveti\v{c}, D.~Klevers, D.~K.~M. Pe\~na, P.-K. Oehlmann and J.~Reuter,
  \emph{{Three-Family Particle Physics Models from Global F-theory
  Compactifications}},
  \href{https://doi.org/10.1007/JHEP08(2015)087}{\emph{JHEP} {\bfseries 08}
  (2015) 087} [\href{https://arxiv.org/abs/1503.02068}{{\ttfamily
  1503.02068}}].

\bibitem{CveticEtAlQuadrillion}
M.~Cveti\v{c}, J.~Halverson, L.~Lin, M.~Liu and J.~Tian, \emph{{Quadrillion
  $F$-Theory Compactifications with the Exact Chiral Spectrum of the Standard
  Model}}, \href{https://doi.org/10.1103/PhysRevLett.123.101601}{\emph{Phys.
  Rev. Lett.} {\bfseries 123} (2019) 101601}
  [\href{https://arxiv.org/abs/1903.00009}{{\ttfamily 1903.00009}}].

\bibitem{Bies:2014sra}
M.~Bies, C.~Mayrhofer, C.~Pehle and T.~Weigand, \emph{{Chow groups, Deligne
  cohomology and massless matter in F-theory}},
  \href{https://arxiv.org/abs/1402.5144}{{\ttfamily 1402.5144}}.

\bibitem{Bies:2021nje}
M.~Bies, M.~Cveti\v{c}, R.~Donagi, M.~Liu and M.~Ong, \emph{{Root Bundles and
  Towards Exact Matter Spectra of F-theory MSSMs}},
  \href{https://arxiv.org/abs/2102.10115}{{\ttfamily 2102.10115}}.

\bibitem{Bies:2021xfh}
M.~Bies, M.~Cveti\v{c} and M.~Liu, \emph{{Statistics of Limit Root Bundles
  Relevant for Exact Matter Spectra of F-Theory MSSMs}},
  \href{https://arxiv.org/abs/2104.08297}{{\ttfamily 2104.08297}}.

\bibitem{Bies:2022wvj}
M.~Bies, M.~Cveti\v{c}, R.~Donagi and M.~Ong, \emph{{Brill-Noether-general
  Limit Root Bundles: Absence of vector-like Exotics in F-theory Standard
  Models}},  \href{https://arxiv.org/abs/2205.00008}{{\ttfamily 2205.00008}}.

\bibitem{Jefferson:2021bid}
P.~Jefferson, W.~Taylor and A.~P. Turner, \emph{{Chiral matter multiplicities
  and resolution-independent structure in 4D F-theory models}},
  \href{https://arxiv.org/abs/2108.07810}{{\ttfamily 2108.07810}}.

\bibitem{Grimm:2013oga}
T.~W. Grimm, A.~Kapfer and J.~Keitel, \emph{{Effective action of 6D F-Theory
  with U(1) factors: Rational sections make Chern-Simons terms jump}},
  \href{https://doi.org/10.1007/JHEP07(2013)115}{\emph{JHEP} {\bfseries 07}
  (2013) 115} [\href{https://arxiv.org/abs/1305.1929}{{\ttfamily 1305.1929}}].

\bibitem{LawrieEtAlRational}
C.~Lawrie, S.~Schafer-Nameki and J.-M. Wong, \emph{{F-theory and All Things
  Rational: Surveying U(1) Symmetries with Rational Sections}},
  \href{https://doi.org/10.1007/JHEP09(2015)144}{\emph{JHEP} {\bfseries 09}
  (2015) 144} [\href{https://arxiv.org/abs/1504.05593}{{\ttfamily
  1504.05593}}].

\bibitem{WeigandTASI}
T.~Weigand, \emph{{TASI Lectures on F-theory}},
  \href{https://arxiv.org/abs/[1806.01854]}{{\ttfamily [1806.01854]}}.

\bibitem{Marsano_2011}
J.~Marsano and S.~Sch{\"a}fer-Nameki, \emph{Yukawas, g-flux, and spectral
  covers from resolved calabi-yau's},
  \href{https://doi.org/10.1007/jhep11(2011)098}{\emph{Journal of High Energy
  Physics} {\bfseries 2011} (2011) }
  [\href{https://arxiv.org/abs/1108.1794}{{\ttfamily 1108.1794}}].

\bibitem{Kuntzler:2012bu}
M.~Kuntzler and S.~Schafer-Nameki, \emph{{G-flux and Spectral Divisors}},
  \href{https://doi.org/10.1007/JHEP11(2012)025}{\emph{JHEP} {\bfseries 11}
  (2012) 025} [\href{https://arxiv.org/abs/1205.5688}{{\ttfamily 1205.5688}}].

\bibitem{Cveti__2014}
M.~Cveti\v{c}, A.~Grassi, D.~Klevers and H.~Piragua, \emph{Chiral
  four-dimensional f-theory compactifications with su(5) and multiple
  u(1)-factors}, \href{https://doi.org/10.1007/jhep04(2014)010}{\emph{Journal
  of High Energy Physics} {\bfseries 2014} (2014) }
  [\href{https://arxiv.org/abs/1306.3987}{{\ttfamily 1306.3987}}].

\bibitem{LinWeigandG4}
L.~Lin and T.~Weigand, \emph{{G4-flux and standard model vacua in F-theory}},
  \href{https://doi.org/10.1016/j.nuclphysb.2016.09.008}{\emph{Nucl. Phys.}
  {\bfseries B913} (2016) 209}
  [\href{https://arxiv.org/abs/1604.04292}{{\ttfamily 1604.04292}}].

\bibitem{Grassi:2018wfy}
A.~Grassi, J.~Halverson, C.~Long, J.~L. Shaneson and J.~Tian,
  \emph{{Non-simply-laced Symmetry Algebras in F-theory on Singular Spaces}},
  \href{https://doi.org/10.1007/JHEP09(2018)129}{\emph{JHEP} {\bfseries 09}
  (2018) 129} [\href{https://arxiv.org/abs/1805.06949}{{\ttfamily
  1805.06949}}].

\bibitem{Grassi:2021ptc}
A.~Grassi, J.~Halverson, C.~Long, J.~L. Shaneson, B.~Sung and J.~Tian,
  \emph{{$6$D Anomaly-Free Matter Spectrum in F-theory on Singular Spaces}},
  \href{https://arxiv.org/abs/2110.06943}{{\ttfamily 2110.06943}}.

\bibitem{Katz:2022vwe}
S.~Katz and W.~Taylor, \emph{{Dimensional Reduction of B-Fields in F-theory}},
  \href{https://arxiv.org/abs/2204.13146}{{\ttfamily 2204.13146}}.

\bibitem{Bies_2017}
M.~Bies, C.~Mayrhofer and T.~Weigand, \emph{Algebraic cycles and local
  anomalies in f-theory},
  \href{https://doi.org/10.1007/jhep11(2017)100}{\emph{Journal of High Energy
  Physics} {\bfseries 2017} (2017) }
  [\href{https://arxiv.org/abs/1706.08528}{{\ttfamily 1706.08528}}].

\bibitem{Corvilain:2017luj}
P.~Corvilain, T.~W. Grimm and D.~Regalado, \emph{{Chiral anomalies on a circle
  and their cancellation in F-theory}},
  \href{https://doi.org/10.1007/JHEP04(2018)020}{\emph{JHEP} {\bfseries 04}
  (2018) 020} [\href{https://arxiv.org/abs/1710.07626}{{\ttfamily
  1710.07626}}].

\bibitem{Cheng:2021zjh}
P.~Cheng, R.~Minasian and S.~Theisen, \emph{{Anomalies as Obstructions: from
  Dimensional Lifts to Swampland}},
  \href{https://arxiv.org/abs/2106.14912}{{\ttfamily 2106.14912}}.

\bibitem{Raghuram34}
N.~Raghuram, \emph{{Abelian F-theory Models with Charge-3 and Charge-4
  Matter}}, \href{https://doi.org/10.1007/JHEP05(2018)050}{\emph{JHEP}
  {\bfseries 05} (2018) 050}
  [\href{https://arxiv.org/abs/1711.03210}{{\ttfamily 1711.03210}}].

\bibitem{CveticKleversPiraguaMultU1}
M.~Cveti\v{c}, D.~Klevers and H.~Piragua, \emph{{F-Theory Compactifications
  with Multiple U(1)-Factors: Constructing Elliptic Fibrations with Rational
  Sections}}, \href{https://doi.org/10.1007/JHEP06(2013)067}{\emph{JHEP}
  {\bfseries 06} (2013) 067} [\href{https://arxiv.org/abs/1303.6970}{{\ttfamily
  1303.6970}}].

\bibitem{Cvetic:2015ioa}
M.~Cveti\v{c}, D.~Klevers, H.~Piragua and W.~Taylor, \emph{{General U(1) x U(1)
  F-theory compactifications and beyond: geometry of unHiggsings and novel
  matter structure}},
  \href{https://doi.org/10.1007/JHEP11(2015)204}{\emph{JHEP} {\bfseries 11}
  (2015) 204} [\href{https://arxiv.org/abs/1507.05954}{{\ttfamily
  1507.05954}}].

\bibitem{Aspinwall:2000kf}
P.~S. Aspinwall, S.~H. Katz and D.~R. Morrison, \emph{{Lie groups, Calabi-Yau
  threefolds, and F theory}},
  \href{https://doi.org/10.4310/ATMP.2000.v4.n1.a2}{\emph{Adv. Theor. Math.
  Phys.} {\bfseries 4} (2000) 95}
  [\href{https://arxiv.org/abs/hep-th/0002012}{{\ttfamily hep-th/0002012}}].

\bibitem{AspinwallMorrisonNonsimply}
P.~S. Aspinwall and D.~R. Morrison, \emph{{Nonsimply connected gauge groups and
  rational points on elliptic curves}},
  \href{https://doi.org/10.1088/1126-6708/1998/07/012}{\emph{JHEP} {\bfseries
  07} (1998) 012} [\href{https://arxiv.org/abs/hep-th/9805206}{{\ttfamily
  hep-th/9805206}}].

\bibitem{Grimm:2010ez}
T.~W. Grimm and T.~Weigand, \emph{{On Abelian Gauge Symmetries and Proton Decay
  in Global F-theory GUTs}},
  \href{https://doi.org/10.1103/PhysRevD.82.086009}{\emph{Phys. Rev. D}
  {\bfseries 82} (2010) 086009}
  [\href{https://arxiv.org/abs/1006.0226}{{\ttfamily 1006.0226}}].

\bibitem{KatzEtAlTate}
S.~Katz, D.~R. Morrison, S.~Schafer-Nameki and J.~Sully, \emph{{Tate's
  algorithm and F-theory}},
  \href{https://doi.org/10.1007/JHEP08(2011)094}{\emph{JHEP} {\bfseries 08}
  (2011) 094} [\href{https://arxiv.org/abs/1106.3854}{{\ttfamily 1106.3854}}].

\bibitem{Cvetic:2012xn}
M.~Cvetic, T.~W. Grimm and D.~Klevers, \emph{{Anomaly Cancellation And Abelian
  Gauge Symmetries In F-theory}},
  \href{https://doi.org/10.1007/JHEP02(2013)101}{\emph{JHEP} {\bfseries 02}
  (2013) 101} [\href{https://arxiv.org/abs/1210.6034}{{\ttfamily 1210.6034}}].

\bibitem{MorrisonParkU1}
D.~R. Morrison and D.~S. Park, \emph{{F-Theory and the Mordell--Weil Group of
  Elliptically-Fibered Calabi--Yau Threefolds}},
  \href{https://doi.org/10.1007/JHEP10(2012)128}{\emph{JHEP} {\bfseries 10}
  (2012) 128} [\href{https://arxiv.org/abs/1208.2695}{{\ttfamily 1208.2695}}].

\bibitem{Esole:2017kyr}
M.~Esole, P.~Jefferson and M.~J. Kang, \emph{{Euler Characteristics of Crepant
  Resolutions of Weierstrass Models}},
  \href{https://doi.org/10.1007/s00220-019-03517-1}{\emph{Commun. Math. Phys.}
  {\bfseries 371} (2019) 99}
  [\href{https://arxiv.org/abs/1703.00905}{{\ttfamily 1703.00905}}].

\bibitem{genfun}
P.~Jefferson and A.~P. Turner, \emph{{Generating functions for intersection
  products of divisors in resolved F-theory models}},
  \href{https://arxiv.org/abs/2206.11527}{{\ttfamily 2206.11527}}.

\bibitem{Witten:1996md}
E.~Witten, \emph{{On flux quantization in M theory and the effective action}},
  \href{https://doi.org/10.1016/S0393-0440(96)00042-3}{\emph{J. Geom. Phys.}
  {\bfseries 22} (1997) 1}
  [\href{https://arxiv.org/abs/hep-th/9609122}{{\ttfamily hep-th/9609122}}].

\bibitem{Greene:1993vm}
B.~R. Greene, D.~R. Morrison and M.~Plesser, \emph{{Mirror manifolds in higher
  dimension}}, \href{https://doi.org/10.1007/BF02101657}{\emph{Commun. Math.
  Phys.} {\bfseries 173} (1995) 559}
  [\href{https://arxiv.org/abs/hep-th/9402119}{{\ttfamily hep-th/9402119}}].

\bibitem{Braun:2014xka}
A.~P. Braun and T.~Watari, \emph{{The Vertical, the Horizontal and the Rest:
  anatomy of the middle cohomology of Calabi-Yau fourfolds and F-theory
  applications}}, \href{https://doi.org/10.1007/JHEP01(2015)047}{\emph{JHEP}
  {\bfseries 01} (2015) 047} [\href{https://arxiv.org/abs/1408.6167}{{\ttfamily
  1408.6167}}].

\bibitem{Dasgupta:1999ss}
K.~Dasgupta, G.~Rajesh and S.~Sethi, \emph{{M theory, orientifolds and G -
  flux}}, \href{https://doi.org/10.1088/1126-6708/1999/08/023}{\emph{JHEP}
  {\bfseries 08} (1999) 023}
  [\href{https://arxiv.org/abs/hep-th/9908088}{{\ttfamily hep-th/9908088}}].

\bibitem{Collinucci:2010gz}
A.~Collinucci and R.~Savelli, \emph{{On Flux Quantization in F-Theory}},
  \href{https://doi.org/10.1007/JHEP02(2012)015}{\emph{JHEP} {\bfseries 02}
  (2012) 015} [\href{https://arxiv.org/abs/1011.6388}{{\ttfamily 1011.6388}}].

\bibitem{Grimm:2011fx}
T.~W. Grimm and H.~Hayashi, \emph{{F-theory fluxes, Chirality and Chern-Simons
  theories}}, \href{https://doi.org/10.1007/JHEP03(2012)027}{\emph{JHEP}
  {\bfseries 03} (2012) 027} [\href{https://arxiv.org/abs/1111.1232}{{\ttfamily
  1111.1232}}].

\bibitem{Grimm:2011sk}
T.~W. Grimm and R.~Savelli, \emph{{Gravitational Instantons and Fluxes from
  M/F-theory on Calabi-Yau fourfolds}},
  \href{https://doi.org/10.1103/PhysRevD.85.026003}{\emph{Phys. Rev. D}
  {\bfseries 85} (2012) 026003}
  [\href{https://arxiv.org/abs/1109.3191}{{\ttfamily 1109.3191}}].

\bibitem{Raghuram:2020vxm}
N.~Raghuram, W.~Taylor and A.~P. Turner, \emph{{Automatic enhancement in 6D
  supergravity and F-theory models}},
  \href{https://doi.org/10.1007/JHEP07(2021)048}{\emph{JHEP} {\bfseries 07}
  (2021) 048} [\href{https://arxiv.org/abs/2012.01437}{{\ttfamily
  2012.01437}}].

\end{thebibliography}\endgroup

\end{document}